%% file: main.tex
\documentclass[acmlarge]{acmart}
\makeatletter
\newcommand{\myconfshort}{\acmConference@shortname}
\newcommand{\myconffull}{\acmConference@name}
\newcommand{\myconfdate}{\acmConference@date}
\newcommand{\myconfloc}{\acmConference@venue}
\AtBeginDocument{
  \fancypagestyle{firstpagestyle}{
    \fancyhead{}%
    \fancyfoot[C]{}%
  }
  \fancyhf{}
  \fancyhead[LO]{\@headfootfont\shorttitle}%
  \fancyhead[RE]{\@headfootfont\@shortauthors}%
  \fancyhead[LE]{\@headfootfont\footnotesize \myconfshort, \myconfdate, \myconfloc}%
  \fancyhead[RO]{\@headfootfont\footnotesize \myconfshort, \myconfdate, \myconfloc}%
  \fancyfoot[C]{}%
}
\makeatother
\acmBooktitle{\conffull\@ (\confshort), \confdate, \confloc}

\usepackage{todonotes}
\makeatletter
\newcommand*\iftodonotes{\if@todonotes@disabled\expandafter\@secondoftwo\else\expandafter\@firstoftwo\fi} 
\makeatother

\usepackage{cleveref}
\crefname{section}{\S}{\S\S}
\Crefname{section}{\S}{\S\S}
\crefname{table}{Table}{Tables}
\crefname{figure}{Figure}{Figures}
\crefname{algorithm}{Algorithm}{}
\crefname{equation}{eq.}{}
\crefname{appendix}{App.}{}
\crefformat{section}{\S#2#1#3} 
\crefformat{subsection}{\S#2#1#3}
\crefformat{subsubsection}{\S#2#1#3}

\usepackage{tabularx}
\usepackage{booktabs}
\usepackage{multirow}
\usepackage{rotating}
\usepackage{nicematrix}

\usepackage[dvipsnames]{xcolor}

\definecolor{mypink1}{rgb}{0.858, 0.188, 0.478}
\definecolor{mypink2}{RGB}{219, 48, 122}
\definecolor{mypink3}{cmyk}{0, 0.7808, 0.4429, 0.1412}
\definecolor{mygray}{RGB}{245,245,245}
\definecolor{myorange}{RGB}{255,230,204}%{237, 157, 72}
\definecolor{mylighterorange}{RGB}{255,230,204}%{255,206,154}%{255, 187, 117}
\definecolor{myblue}{RGB}{218,232,252}%{77,146,169}%{48, 125, 147}
\AtBeginDocument{%
  }

\setcopyright{none}
\copyrightyear{2026}
\acmYear{2026}
\setcopyright{cc}
\setcctype{by-nc-nd}
\acmConference[FAccT '26]{The 2026 ACM Conference on Fairness, Accountability, and Transparency}{June 25--28, 2026}{Montreal, QC, Canada}
\acmBooktitle{The 2026 ACM Conference on Fairness, Accountability, and Transparency (FAccT '26), June 25--28, 2026, Montreal, QC, Canada}
\acmDOI{10.1145/3805689.3812236}
\acmISBN{979-8-4007-2596-8/2026/06}

\usepackage{fancybox}
\usepackage{hyperref}

\begin{document}

%%
%% The "title" command has an optional parameter,
%% allowing the author to define a "short title" to be used in page headers.
\title{A Human-Centric Framework for Data Attribution in Large Language Models}

%%
%% The "author" command and its associated commands are used to define
%% the authors and their affiliations.
%% Of note is the shared affiliation of the first two authors, and the
%% "authornote" and "authornotemark" commands
%% used to denote shared contribution to the research.

\author{Amelie W{\"u}hrl}
\email{amwy@itu.dk}
\orcid{0009-0008-3382-0423}
\affiliation{%
  \institution{IT University of Copenhagen}
  \city{Copenhagen}
  \country{Denmark}
}

\author{Mattes Ruckdeschel}
%\authornote{Corresponding Author.}
\email{mruc@itu.dk}
\orcid{0009-0000-1888-9408}
\affiliation{%
  \institution{IT University of Copenhagen}
  \city{Copenhagen}
  \country{Denmark}
}

\author{Kyle Lo}
%\authornote{Corresponding Author.}
\email{kyleclo@cs.washington.edu}
\orcid{0000-0002-1804-2853}
\affiliation{%
  \institution{University of Washington, Allen Institute for AI}
  \city{Seattle}
  \country{USA}
}

\author{Anna Rogers}
%\authornote{Corresponding Author.}
\email{arog@itu.dk}
\orcid{0000-0002-4845-4023}
\affiliation{%
  \institution{IT University of Copenhagen}
  \city{Copenhagen}
  \country{Denmark}
}

%%
%% By default, the full list of authors will be used in the page
%% headers. Often, this list is too long, and will overlap
%% other information printed in the page headers. This command allows
%% the author to define a more concise list
%% of authors' names for this purpose.
\renewcommand{\shortauthors}{W{\"u}hrl et al.}

%%
%% The abstract is a short summary of the work to be presented in the
%% article.
\input{sections/0_abstract}

%%
%% The code below is generated by the tool at http://dl.acm.org/ccs.cfm.
%% Please copy and paste the code instead of the example below.
%%
\begin{CCSXML}
<ccs2012>
   <concept>
       <concept_id>10003456.10003462.10003463</concept_id>
       <concept_desc>Social and professional topics~Intellectual property</concept_desc>
       <concept_significance>500</concept_significance>
       </concept>
   <concept>
       <concept_id>10010147.10010178.10010179.10010182</concept_id>
       <concept_desc>Computing methodologies~Natural language generation</concept_desc>
       <concept_significance>300</concept_significance>
       </concept>
   <concept>
       <concept_id>10003456.10003457.10003567.10010990</concept_id>
       <concept_desc>Social and professional topics~Socio-technical systems</concept_desc>
       <concept_significance>500</concept_significance>
       </concept>
   <concept>
       <concept_id>10003456.10003462.10003544.10003589</concept_id>
       <concept_desc>Social and professional topics~Governmental regulations</concept_desc>
       <concept_significance>300</concept_significance>
       </concept>
   <concept>
       <concept_id>10003456.10003462.10003588.10003589</concept_id>
       <concept_desc>Social and professional topics~Governmental regulations</concept_desc>
       <concept_significance>300</concept_significance>
       </concept>
 </ccs2012>
\end{CCSXML}

\ccsdesc[500]{Social and professional topics~Intellectual property}
\ccsdesc[300]{Computing methodologies~Natural language generation}
\ccsdesc[500]{Social and professional topics~Socio-technical systems}
\ccsdesc[300]{Social and professional topics~Governmental regulations}

%%
%% Keywords. The author(s) should pick words that accurately describe
%% the work being presented. Separate the keywords with commas.
\keywords{Data attribution, large language models, human-centric, data economy, policy making}
%% A "teaser" image appears between the author and affiliation
%% information and the body of the document, and typically spans the
%% page.
%\begin{teaserfigure}
%  \includegraphics[width=\textwidth]{sampleteaser}
%  \caption{Seattle Mariners at Spring Training, 2010.}
%  \Description{Enjoying the baseball game from the third-base
%  seats. Ichiro Suzuki preparing to bat.}
%  \label{fig:teaser}
%\end{teaserfigure}

%\received{20 February 2007}
%\received[revised]{12 March 2009}
%\received[accepted]{5 June 2009}

%%
%% This command processes the author and affiliation and title
%% information and builds the first part of the formatted document.

% cover letter to highlight the changes made during revisions
%\input{sections/cover-letter}
%\newpage
%\setcounter{page}{1}
\maketitle

% introduction
\input{sections/0_introduction}

\input{sections/1_related_work}

% The current LLM ecosystem
\input{sections/2_llm-ecosystem}

% attribition-worthiness framework
\input{sections/4_attribution-worthiness}

% emerging practices - tda in practice
%\input{sections/4.5_practice}

% moonshot - tda in practice
\input{sections/5_attribution-in-practice}

% conclusion / discussion
\input{sections/6_conclusion}

\bibliographystyle{ACM-Reference-Format}
\bibliography{literature,zotero_related_work}

\appendix
\input{sections/2.2_concepts}

\end{document}

%% file: sections/0_abstract.tex
\begin{abstract}
In the current Large Language Model (LLM) ecosystem, creators have little agency over how their data is used, and LLM users may find themselves unknowingly plagiarizing existing sources. Attribution of LLM-generated text to LLM input data could help with these challenges, but so far we have more questions than answers: what elements of LLM outputs require attribution, what goals should it serve, how should it be implemented? 

We contribute a human-centric data attribution framework, which situates the attribution problem within the broader data economy. Specific use cases for attribution, such as creative writing assistance or fact-checking, can be specified via a set of parameters (including stakeholder objectives and implementation criteria). These criteria are up for negotiation by the relevant stakeholder groups: creators, LLM users, and their intermediaries (publishers, platforms, AI companies). The outcome of domain-specific negotiations can be implemented and tested for whether the stakeholder goals are achieved. The proposed approach provides a bridge between methodological NLP work on data attribution, governance work on policy interventions, and economic analysis of creator incentives for a sustainable equilibrium in the data economy.

\end{abstract}

%% file: sections/0_introduction.tex
\section{Introduction}
The current Large Language Models (LLM) ecosystem is socially unsustainable. LLM performance is predicated on high-quality training data~\cite{Sankaran_2024_OpenAI_says_it_is_impossible_to_train_AI_without_using_copyrighted_works_for_free,KandpalRaffel_2025_Position_Most_Expensive_Part_of_LLM_should_be_its_Training_Data}, but its creators currently have little agency over the use of their data, or economic benefits from it. Instead, they witness their data being scraped or watch their publishers license their content without their consent~\cite{AIwatchdog_2025_Content_Licensing_Deals}. At the same time, LLM users cannot tell either whether the outputs are trustworthy~\cite{JazwinskaChandrasekar_AI_Search_Has_Citation_Problem}, and when they infringe on someone's rights~\cite{gupta-pruthi-2025-glitters}. The LLM industry currently benefits from the status quo, but long-term it also depends on a healthy data economy: if creators do not have meaningful incentives to participate with new high-quality data, their contributions will decline ~\cite{longpreConsentCrisisRapid2024,ZhangJiaoEtAl_2025_Fairshare_Data_Pricing_via_Data_Valuation_for_Large_Language_Models,PeukertAbeillonEtAl_2025_AI_and_Dynamic_Supply_of_Training_Data}. 

In theory, data attribution presents a solution.
We consider the data attribution task broadly as providing acknowledgment that a certain work was in some way used by the model to produce an output. In Natural Language Processing (NLP) research, attribution is typically\footnote{In NLP research, even instance-level attribution has been criticized as an overloaded term describing two families of methods~\cite{worledge2024unifying}. Still, we deliberately take a broad view of the attribution problem to encompass both instance and dataset-level attribution, because the latter is prerequisite for the former (see \cref{sec:ecosystem-disruptions}), and this perspective helps to place attribution into the socioeconomic context.} discussed at the level of a specific output instance (e.g. a reference of a claim to its source document), but it can also be considered at the level of datasets included in model training, and broad model functionality (e.g. a model X has improved performance on task Y after it was trained on dataset Z). Data attribution could help to sustain a healthy data economy by incentivizing the creators to contribute to the data available to LLMs. It could also help LLM users to avoid plagiarism and check whether the provided sources are trustworthy.

Data attribution for LLMs is an active research area, but it is still largely unsolved. Besides major technical challenges, it requires transparency about the training data, which commercial services currently do not even disclose. 
More fundamentally, what do we even need attribution for? Common facts like ``grass is green'' do not need it, whereas newsworthy claims clearly do. But there is a lot of gray area, and many use cases require different solutions (e.g. some facts may be attributable via string matching, but not stylistic features in creative writing). Technical solutions can only be defined by practical goals, but they are external to NLP research.

This work situates the data attribution problem in the LLM data ecosystem (\cref{sec:ecosystem}). We characterize the variety and conflicts of goals of different stakeholders, which make a single data attribution solution impossible. To address that, we contribute a human-centric\footnote{We use the term 'human-centric' to indicate the goal of data attribution, rather than the use of human-computer interaction (HCI) methods to achieve it.} data attribution framework (\cref{sec:framework}), which grounds the implementation of data attribution in case-by-case negotiations between stakeholders. 
We then present a moonshot for attribution-backed LLM  services (\cref{sec:moonshot}). Overall, our goal is to bridge NLP research with the broader socioeconomic context of data attribution, and to support economic and governance work, providing a blueprint for how the data attribution task could be specified for an implementation.

%% file: sections/1_related_work.tex
\section{Related Work}
\label{sec:related}

In modern LLM systems, data can shape model behavior in two ways: as \emph{training data} which determines model parameters, and as \emph{input context} provided to the model at inference time. 
Most existing methods for attribution to training data are post hoc, aiming to associate a given LLM generation with training examples using techniques such as influence functions~\citep{10.5555/3305381.3305576}, Shapley-value-based data attribution~\cite{wang2024economicsolutioncopyrightchallenges}, text similarity functions~\citep{akyurek-etal-2022-towards}, among others.
There are also approaches relevant to attribution that focus on different roles of subsets of training data during the model development phase: 
this includes methods like explicit training data weighting~\citep{yoon2019datavaluationusingreinforcement} and modular model architectures with respect to data~\citep{shi2025flexolmoopenlanguagemodels}. A complementary line of work focuses on attribution-aware system design, most notably retrieval-augmented generation (RAG) \citep{lewis2020retrieval} that conditions the model on specific input data, which is then provided as sources~\citep{menick2022teachinglanguagemodelssupport}. \citet{LivniMoranEtAl_2024_Credit_attribution_and_stable_compression} propose a notion of attribution based on differential privacy.  
Finally, data valuation research~\cite{SimXuEtAl_2022_Data_Valuation_in_Machine_Learning_Ingredients_Strategies_and_Open_Challenges,ChoeAhnEtAl_2024_What_is_Your_Data_Worth_to_GPT_LLM-Scale_Data_Valuation_with_Influence_Functions} aims to quantify the value of specific data points or subsets for model training. See~\cite{worledge2024unifying,dengSurveyDataAttribution2025} for methodological surveys of attribution methods.

The NLP community has also contributed to relevant discussions of LLM data governance~\cite{JerniteNguyenEtAl_2022_Data_Governance_in_Age_of_Large-Scale_Data-Driven_Language_Technology}, data documentation~\cite{BenderFriedman_2018_Data_Statements_for_Natural_Language_Processing_Toward_Mitigating_System_Bias_and_Enabling_Better_Science,GebruMorgensternEtAl_2020_Datasheets_for_Datasets,HutchinsonSmartEtAl_2021_Towards_Accountability_for_Machine_Learning_Datasets_Practices_from_Software_Engineering_and_Infrastructure,iiMeasuringData2023}, and data markets~\cite{AzcoitiaIordanouEtAl_2023_Understanding_Price_of_Data_in_Commercial_Data_Marketplaces,FallahJordanEtAl_2024_On_Three-Layer_Data_Markets}. 
With respect to acknowledgment that certain data was used, there are multiple non-profit-driven LLM development projects with fully open training data (e.g.~\cite{ScaoFanEtAl_2022_BLOOM_176BParameter_OpenAccess_Multilingual_Language_Model,BidermanSchoelkopfEtAl_2023_Pythia_Suite_for_Analyzing_Large_Language_Models_Across_Training_and_Scaling,groeneveldOLMoAcceleratingScience2024}). There is also work on providing searchable interfaces for LLM-scale corpora\cite{PiktusAkikiEtAl_2023_ROOTS_Search_Tool_Data_Transparency_for_LLMs}. Projects with open LLM training data enabled work on its analysis and description~\cite{DodgeSapEtAl_2021_Documenting_Large_Webtext_Corpora_Case_Study_on_Colossal_Clean_Crawled_Corpus,KreutzerCaswellEtAl_2022_Quality_at_Glance_Audit_of_WebCrawled_Multilingual_Datasets,ElazarBhagiaEtAl_2023_Whats_In_My_Big_Data}.
Another relevant research direction is data extraction, indicating that certain data was used for training~\cite{HaimVardiEtAl_2022_Reconstructing_Training_Data_From_Trained_Neural_Networks,CooperGokaslanEtAl_2025_Extracting_memorized_pieces_of_copyrighted_books_from_open-weight_language_models}. Dataset contamination research~\cite{deng-etal-2024-unveiling} is similarly related as it studies if particular evaluation data was part of the training process.

In economic terms, there have long been calls to recognize data as a kind of labor~\cite{Lanier_2014_Who_owns_future,ArrietaIbarraGoffEtAl_2017_Should_We_Treat_Data_as_Labor_Moving_Beyond_Free}. There is a recent discussion of AI systems as data markets~\cite{Jordan_2025_Collectivist_Economic_Perspective_on_AI}, and game-theoretic analysis of consequences of unfair data pricing~\cite{ZhangJiaoEtAl_2025_Fairshare_Data_Pricing_via_Data_Valuation_for_Large_Language_Models}. Warning of the dangers for the information economy, \citet{vincentCollectiveBargainingInformation2025} argue for collective bargaining by data creator coalitions.  
Outside computer science, relevant discussions are ongoing in economics and business studies \cite[e.g.][]{PeukertAbeillonEtAl_2025_AI_and_Dynamic_Supply_of_Training_Data,AgarwalSen_2026_Google_AI_Overviews_and_Publisher_Traffic_Evidence_from_Field_Experiment}, copyright law \cite[e.g.][]{2025_Report_on_Copyright_and_Artificial_Intelligence_Part_3_Generative_AI_Training_pre-publication_version,HendersonLiEtAl_2023_Foundation_Models_and_Fair_Use}, and political economy \cite[e.g.][]{TanThelen_2026_Cloud_Capitalism_and_AI_Transition,BurkhardtRieder_2024_Foundation_models_are_platform_models_Prompting_and_political_economy_of_AI}, inter alia. 

To the best of our knowledge, this is the first work attempting to provide a practical bridge between the NLP work on data governance and attribution, and socioeconomic work on LLMs in the data economy. For the latter, it outlines the main aspects of NLP research on data attribution and assessing the capabilities of current technology. For the former, this work explicitly brings in the stakeholders and their varying goals into the definition of the data attribution problem. \citet{wang2024economicsolutioncopyrightchallenges} propose a specific solution for compensating creators (based on Shapley-values), and call for cooperation between creators and LLM developers. We contribute a broader framework for similar contributions across many attribution use cases, which supports the overall collective bargaining approach advocated by \cite{vincentCollectiveBargainingInformation2025}.

%% file: sections/2_llm-ecosystem.tex
\section{The Current LLM Ecosystem: Creators, Distributors and Recipients of Data}
\label{sec:ecosystem}

We consider the \textit{LLM data ecosystem} as the set of main stakeholders who directly or indirectly participate in LLM service development and use (\cref{sec:stakeholders}). 
Historically, the relations between them have been mediated by legal and professional norms such as copyright, plagiarism and authorship, which we assume the reader to be broadly familiar with. For reference, we outline them in~\cref{sec:concepts}. We then summarize how this data ecosystem and its norms were disrupted when LLMs became a new information distribution channel (\ref{sec:ecosystem-disruptions}). Finally, we outline what role data attribution research could play in the re-negotiation of stakeholder roles for a sustainable data economy. 

\input{sections/2.1_stakeholders}

\input{sections/2.3_disruptions}

\input{sections/2.4_desiderata}

%% file: sections/2.1_stakeholders.tex
\subsection{Stakeholders}
\label{sec:stakeholders}

In scope of this work, we consider five groups of stakeholders in the LLM data ecosystem, defined as follows:

\begin{itemize}
\item \textit{Creators:} the individuals who satisfy the authorship criteria (\cref{sec:authorship}) relevant for their field of activity.
\item \textit{Publishers:} the organizations disseminating the creators' work, typically with at least a partial transfer of rights (see \cref{sec:copyright}) to the publisher. Publishers may add value through e.g. editorial services (though this is not always the case~\cite{Phelps_2022_Challenging_Academic_Publisher_Oligopoly}) and prestige~\cite{Palmer_2024_Prestige_Factor_Propping_Up_Academic_Publishers}.
\item \textit{Platforms:} the third-party services providing the creators with means of dissemination and/or monetization of their work. Examples include Kindle self-publishing, social media platforms like Reddit, blog platforms like Medium. Unlike publishers, platforms typically do not add editorial value. They also historically differ from publishers in terms of liability for content they provide~\cite{SkorupHuddleston_2019_Erosion_of_Publisher_Liability_in_American_Law_Section_230_and_Future_of_Online_Curation}. 
\item \textit{AI industry:} organizations that commercially provide LLM-based services, typically branded as `AI', and derived from the creators' data \cite{Sankaran_2024_OpenAI_says_it_is_impossible_to_train_AI_without_using_copyrighted_works_for_free}. 
\item \textit{Readers/Users:} the individuals acting as consumers of the creators' work.
\end{itemize}

We consider the  
individuals as \textit{primary stakeholders}: the creators and their intended audience. This is motivated by the fact that they are the main sources of value in the ecosystem (producers and consumers). Three other groups represent organizations acting as \textit{intermediaries} between the creators and the audience: publishers, platforms and AI industry. For example, a blogger is a creator, who may distribute their work to readers via a platform (e.g. Medium). The same creator may also be a reader of other blogs, now as a member of their audience.

\input{sections/tab-stakeholders}

Like all other spheres of human activity, data economy is affected by the existing incentives for the participants. \cref{tab:stakeholders} presents examples\footnote{We intentionally do not provide a quantitative analysis, because the interests are too heterogeneous even within one group, and the current literature likely over-represents the perspectives of those stakeholders who have the means to express them, especially in courts.} of such incentives. 
We group the incentives into extrinsic and intrinsic, per classic high-level distinction of motivation types in psychology~\cite{MorrisGrehlEtAl_2022_On_what_motivates_us_detailed_review_of_intrinsic_extrinsic_motivation}. We further distinguish between financial and social extrinsic incentives, since they can have different implications for the attribution problem: some kinds of financial incentives may be achievable via blanket license agreements, not necessarily requiring fine-grained attribution. 
In contrast, social incentives heavily depend on attribution: e.g. a content creator can only establish themselves as an expert if their contributions are attributed to them. Some types of intrinsic motivation may also rely on attribution: for example, LLM users may expect citations because that is required by their epistemic goals \citep{clark2025epistemicalignmentmediatingframework}, or because they prefer to support ethically sourced technology. Note that if an attribution-enabled service is more valuable to the users for social or intrinsic reasons, the industry has a financial incentive to provide it.

Our literature review suggests the following observations relevant to the attribution problem.

\begin{itemize}

\item \textit{Motivations within stakeholder groups are highly heterogeneous.} Even within a single sub-domain such as travel blogs, writers may work under different mixes of economic, social and intrinsic incentives, which would impact their attribution needs. Intermediaries also  vary in how much precedence their non-financial goals may have.
\item \textit{Conflicting interests.} The clearest example is that the working authors may prefer to maximize their earnings, while the intermediaries who distribute their work or its derivatives may prefer to maximize their share of that transaction (see \cref{sec:ecosystem-disruptions}). Intermediaries may have conflicting interests between themselves, e.g. in news organizations trying to get platforms to pay for their content~\cite{Nicholls_2024_Facebook_wont_keep_paying_Australian_media_outlets_for_their_content_Are_we_about_to_get_another_news_ban}. Stakeholders may differ in their position toward other stakeholders: e.g. publishers largely oppose the existence of the Internet Archive as a non-profit digital library for humans~\cite{2022_Statement_From_Terrence_Hart_General_Counsel_Association_of_American_Publishers_on_Disinformation_in_Internet_Archive_Case_AAP}, while many authors support it~\cite{1000_Authors_for_Libraries}.   
\item \textit{The stakeholder roles may be mixed.} For example, creators may also act as their own publishers (providing content for free, for subscriptions or donations, or ad-supported via their own websites or platforms). They may also themselves be users of LLM services, which results in an interesting conflict: creators may prefer to indirectly use other creators' work for free, while getting paid for their own work (e.g. some writers use generated art~\cite{Robertson_2025_How_Authors_Are_Thinking_About_AI_Survey_of_1200_Authors}). 
\item \textit{Structural power asymmetries.} Major publishers and platforms are large organizations, and as such, they had a big advantage in legal resources over individuals (both creators and users) even before LLMs. They may also enjoy a dominant market position~\cite{Manning_2005_Monopsony_in_Motion_Imperfect_Competition_in_Labor_Markets,Brooks_2020_Dilemma_of_Free_Facebooks_Monopsony_Power_and_Need_For_Antitrust_Renaissance}, i.e. become relatively free from competition to which the individuals could easily move. For-profit academic publishers are a prominent example~\cite{Phelps_2022_Challenging_Academic_Publisher_Oligopoly,Palmer_2024_Prestige_Factor_Propping_Up_Academic_Publishers}. Big tech platforms show a similar trend: once it is costly or impossible for individuals to switch, they can unilaterally change the terms in their favor~\cite{Doctorow_2025_Enshittification}, and refuse to negotiate even in response to legislation~\cite{Zandbergen_2023_Canadian_media_trained_audiences_to_use_Facebook_With_Meta_blocking_news_whats_next,Nicholls_2024_Facebook_wont_keep_paying_Australian_media_outlets_for_their_content_Are_we_about_to_get_another_news_bana}. 
\end{itemize}

To conclude, the LLM data ecosystem is characterized by a heterogeneous landscape of incentives for various stakeholders, which may have different and even conflicting interests, mixed roles, and unequal market power.

%% file: sections/tab-stakeholders.tex
\begin{table*}
\footnotesize
    \centering
\begin{tabular}{p{0.2cm} p{1.1cm} | p{4.3cm} p{3.3cm} p{5.1cm}}
\toprule

& & \multicolumn{3}{c}{Examples of incentives for participation in data economy} \\
\cmidrule(lr){3-5}
&  &  Extrinsic (financial)&  Extrinsic (social)& Intrinsic \\

\cmidrule(lr){3-3} \cmidrule(lr){4-4} \cmidrule(lr){5-5}

\multirow{2}{*}[6pt]{\rotatebox{90}{Primary stakeholders}}
&  Creators & Book commissions, salary (for staff writers), freelance payments, ad-supported monetization, subscriptions, donations, licensing \cite{DatasetProvidersAlliance_2024_Machine_Learning_AI_Data_Licensing}  
& Displaying one's knowledge, developing a `personal brand', building up a following, community-building 
& Advancing a cause/agenda, `making a difference', learning/exercising one's skills, contributing to public knowledge, self-realization, work for one's own use \cite{ForteBruckman_2005_Why_Do_People_Write_for_Wikipedia_Incentives_to_Contribute_to_Open-Content_Publishing,RafaeliAriel_2008_Online_Motivational_Factors_Incentives_for_Participation_and_Contribution_in_Wikipedia,AnthonySmithEtAl_2009_Reputation_and_Reliability_in_Collective_Goods_Case_of_Online_Encyclopedia_Wikipedia,Florida_2022_The_Rise_of_the_Creator_Economy} \\

%%% USERS %%%%

\cmidrule(lr){2-2} \cmidrule(lr){3-3} \cmidrule(lr){4-4} \cmidrule(lr){5-5}

&  Readers / users & Content access while optimizing their costs \cite{RingstadLoyland_2006_Demand_for_Books_Estimated_by_Means_of_Consumer_Survey_Data}, including indirect costs (e.g. collection of user data, exposure to ads \cite{Zuboff_2019_age_of_surveillance_capitalism_fight_for_human_future_at_new_frontier_of_power})
& Connecting with their favorite creators, social signals in sharing or otherwise interacting with creator content & Convenience \cite{LaiChang_2011_User_attitudes_toward_dedicated_e-book_readers_for_reading_effects_of_convenience_compatibility_and_media_richness}, ethical/legal principles they may wish to uphold \cite{HashimKannanEtAl_2018_Central_Role_of_Moral_Obligations_in_Determining_Intentions_to_Engage_in_Digital_Piracy,Jayasundara_2022_Study_on_Risk_of_Prosecution_and_Perceived_Proximity_on_State_University_Undergraduates_Behavioural_Intention_for_e-Book_Piracy,Robertson_2025_How_Authors_Are_Thinking_About_AI_Survey_of_1200_Authors}, willingness to support good creators \cite{LiuJiangEtAl_2021_Why_Audiences_Donate_Money_to_Content_Creators_Uses_and_Gratifications_Perspective} \\

%%%% INTERMEDIARIES %%%%
\cmidrule(lr){1-2} \cmidrule(lr){3-5}

\multirow{3}{*}[-12pt]{\rotatebox{90}{Intermediary parties}}
&  Publishers & Reader/institutional subscriptions, direct sales, publication fees \cite{VanNoorden_2013_Open_access_true_cost_of_science_publishing}, licensing \cite{Palmer_Taylor_Francis_AI_Deal_Sets_Worrying_Precedent_for_Academic_Publishing}, monetization of reader data \cite{Wu_2016_attention_merchants_from_daily_newspaper_to_social_media_how_our_time_and_attention_is_harvested_and_sold,Eiko_2022_Welcome_to_Hotel_Elsevier_you_can_check-out_any_time_you_like_not_Eiko_Fried} 
& Institutional reputation \cite{NelsonKim_2021_Improve_Trust_Increase_Loyalty_Analyzing_Relationship_Between_News_Credibility_and_Consumption,Palmer_2024_Prestige_Factor_Propping_Up_Academic_Publishers}, growing the audience 
& \multirow{3}{5cm}{Many non-profits and companies have social or scientific mission statements (e.g. `ensuring that AGI benefits all of humanity' (OpenAI)). It may be unclear how they interact with business goals, especially over time \cite{LoveSolonEtAl_2025_Google_Removes_Language_on_Weapons_From_Public_AI_Principles}. 
Non-profit status may change (e.g. OpenAI \cite{Landymore_2025_OpenAI_Successfully_Sheds_Its_Roots_as_Ethical_Non-Profit}). For-profit organizations may contribute to open-source to undercut their competition~\cite{WidderWhittakerEtAl_2024_Why_open_AI_systems_are_actually_closed_and_why_this_matters}.}
\\

\cmidrule(lr){2-2} \cmidrule(lr){3-3} \cmidrule(lr){4-4} 

&  Platforms & Subscriptions, ads  \cite{Wu_2016_attention_merchants_from_daily_newspaper_to_social_media_how_our_time_and_attention_is_harvested_and_sold,Zuboff_2019_age_of_surveillance_capitalism_fight_for_human_future_at_new_frontier_of_power}, licensing \cite{Wiggers_2024_OpenAI_inks_deal_to_train_AI_on_Reddit_data} 
& Growing the user base \cite{Murphy_2021_Facebook_confronts_growth_problems_as_number_of_young_users_in_US_declines}, brand reputation  \cite{Brown_2020_Should_Stay_or_Should_Leave_Exploring_Discontinued_Facebook_Use_After_Cambridge_Analytica_Scandal} 
 &
\\

\cmidrule(lr){2-2} \cmidrule(lr){3-3} \cmidrule(lr){4-4} 
&  AI industry & Investments \cite{MercedKaye_2025_Exclusive_OpenAI_Secures_Another_Giant_Funding_Deal}, subscriptions, ads \cite{Bort_2025_Perplexity_CEO_says_its_browser_will_track_everything_users_do_online_to_sell_hyper_personalized_ads}, stock valuations
& Growing the user base, brand reputation, attracting talent 
         & \\

\bottomrule

    \end{tabular}
    \caption{Examples of diverse interests of stakeholders in the LLM data ecosystem (not exhaustive and not representing any specific sector or demographic group).
    }
    \label{tab:stakeholders}

\end{table*}

%% file: sections/2.3_disruptions.tex
\subsection{LLMs as a New Information Distribution Channel}
\label{sec:ecosystem-disruptions}

As discussed in \cref{sec:stakeholders}, the data economy involves various intermediaries between the primary stakeholders (creators and readers/users). In such a situation, one could expect conflicts of interests based on how the intermediaries mediate the transactions 
and what would be a fair share for them. From that point of view, the generative chatbots could be described as a new distribution channel through which information moves from creators to readers. This disrupts the original flow of value, and does not provide equivalent incentives for the creators. 

\begin{figure}[t]
\includegraphics[width=\textwidth]{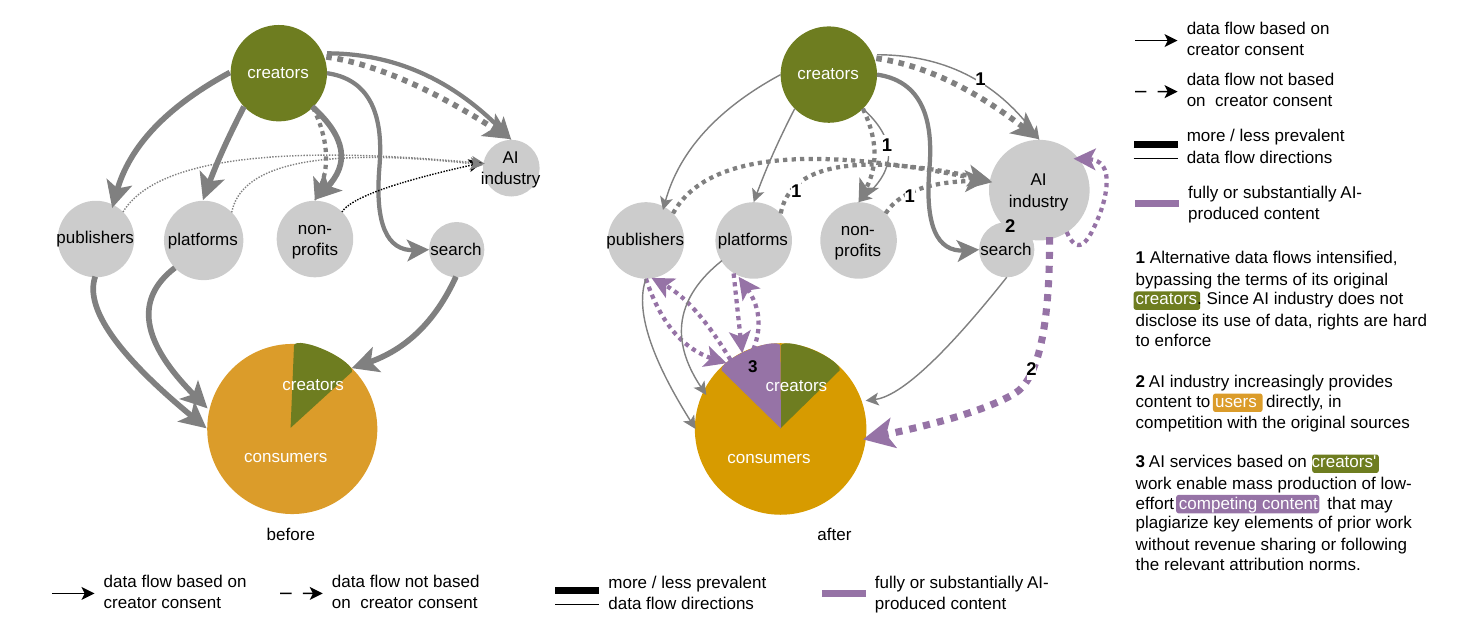}
\caption{Major changes in information flow from the creators to readers/users when LLMs started serving as providers of content, including for assisted production of new materials. In addition to stakeholders discussed in \cref{sec:stakeholders}, a relevant group is non-profit organizations who do not serve as intermediaries, but whose outputs (including datasets) may be used by AI industry.
}
\Description[Major information flow changes in the LLM ecosystem]{The major changes in information flow from the creators to readers/users when LLMs started serving as providers of content, including for assisted production of new materials.}
\label{fig:stakeholders}
\end{figure}

This process is schematically described in \autoref{fig:stakeholders}.  
It starts when the materials used by LLM services (training documents, RAG sources) may be obtained by the AI industry without knowledge or consent of the creators.\footnote{\label{footnote:abuse} The most straightforward cases are training on pirated resources~\cite{Wiggers_2025_Mark_Zuckerberg_gave_Metas_Llama_team_OK_to_train_on_copyrighted_works_filing_claims}, paywalled content~\cite{rosenblat2025beyond}, not respecting the robots.txt specifications~\cite{MEHROTRA-MARCHMAN-perplexity-bullshit-machine,Franceschi-Bicchierai2025_PerplexityScraping} and terms of other services~\cite{Belanger_2025_Lawsuit_Reddit_caught_Perplexity_red-handed_stealing_data_from_Google_results}. There is also the practice of `data laundering': commercial use of  data collected under non-profit auspices, which are themselves more likely to be considered `fair use' \cite[p.192-193]{Kim_2024_Data_Scraping_for_Generative_AI_To_What_Extent}. Academic institutions in particular may have various research exemptions for data collection, (e.g.~\cite{Hansen_2024_Text_Data_Mining_Research_DMCA_Exemption_Renewed_and_Expanded}), although these practices are themselves ethically complicated~\cite{KlassenFiesler_2022_This_Isnt_Your_Data_Friend_Black_Twitter_as_Case_Study_on_Research_Ethics_for_Public_Data,KairamBernsteinEtAl_2024_Community-Driven_Models_for_Research_on_Social_Platforms} and have their own attribution issues~\cite{Bruckman_2002_Studying_amateur_artist_perspective_on_disguising_data_collected_in_human_subjects_research_on_Internet}. Crucially, such exemptions are meant for public-interest research rather than development of commercial products, and the non-profit organizations are not meant to serve as intermediaries in the creator-consumer distribution chain. But in machine learning, academic research is increasingly mixed with commercial conflicts-of-interest 
\cite{AbdallaWahleEtAl_2023_Elephant_in_Room_Analyzing_Presence_of_Big_Tech_in_Natural_Language_Processing_Researcha,GnewuchWahleEtAl_2026_Big_tech-funded_AI_papers_have_higher_citation_impact_greater_insularity_and_larger_recency_bias}. Some academics closely collaborate with industry teams, subsequently using that data for commercial models ~\cite{Baio_2022_AI_Data_Laundering_How_Academic_and_Nonprofit_Researchers_Shield_Tech_Companies_from_Accountability}.  LLMs owe much to Common Crawl, a non-profit explicitly opting not to require attribution for the data they collected~\cite{Reisner_2025_Company_Quietly_Funneling_Paywalled_Articles_to_AI_Developers}.} That in itself is not unprecedented: e.g. the BERT model~\cite{DevlinChangEtAl_2019_BERT_Pre-training_of_Deep_Bidirectional_Transformers_for_Language_Understanding}, highly successful and even deployed into Google Search~\cite{Nayak_2019_Understanding_searches_better_than_ever_before}, was trained on a corpus of copyrighted fiction works, collected without author consent~\cite{BandyVincent_2021_Addressing_Documentation_Debt_in_Machine_Learning_Research_Retrospective_Datasheet_for_BookCorpus}. That fact did raise concerns, and in 2021 that corpus was no longer distributed. But the current LLMs have received a lot more objections than BERT, because they are generative, and they are deployed to function as standalone sources  for information seeking and production of new texts. The generated responses may combine public knowledge (e.g. English syntax rules, names of presidents) with elements that may be protected (e.g. characters of a fictional universe, facts in investigative journalism). 

Consequently, the creators may feel that this practice results in competing products derived from their work~\cite{Belanger_2023_Grisham_Martin_join_authors_suing_OpenAI_There_is_nothing_fair_about_this_Updated}, and that they should be compensated~\cite{Survey_Reveals_90_Percent_of_Writers_Believe_Authors_Should_Be_Compensated_for_Use_of_Their_Books_in_Training_Generative_AI}. As for RAG, even with a link to the original source, the LLM service user may feel that their information need is sufficiently addressed, and they do not need to visit the original source. Indeed, there are reports of decline in human web traffic to the original sources (compared to web searches)~\cite{Germain_2025_Is_Google_about_to_destroy_web,Newman_2026_Journalism_media_and_technology_trends_and_predictions_2026_Reuters_Institute_for_Study_of_Journalism}, and news publishers consider the practice as a kind of unsanctioned re-publishing~\cite{CaseC250_25_LikeCompanyvGoogle,Foxglove2025_GoogleNewsChallenge}. 

In both LLM training and RAG, the current creator protests are fundamentally about non-disclosure\footnote{
Historical note: before HTML was widely adopted, there was also a proposal for an alternative structure for the web, in which links would be bi-directional. This structure would enable (and even encourage) reuse and mixing of available data, but the original authors would be able to see each reuse (\cite{Nelson_1999_Xanalogical_structure_needed_now_more_than_ever_parallel_documents_deep_links_to_content_deep_versioning_and_deep_re-use}, see also discussion in \cite[ch.18]{Lanier_2014_Who_owns_future}). Besides factuality issues~\cite{LiuZhangEtAl_2023_Evaluating_Verifiability_in_Generative_Search_Engines}, one could see the current dissatisfaction of creators with RAG solutions as the result of lacking this \textit{second} link: the kind of meta-attribution which would allow them to see who uses their data, and set the terms for that. Such links would also support moral rights in the way that the unidirectional citations cannot, e.g. enabling the authors to object to misrepresentation.}, which prevents the creators from exercising their rights (economic and/or moral, see \cref{sec:ecosystem-concepts-moral-rights}). The same non-disclosure leads to the risk of plagiarism for the users (see below). The platforms and publishers themselves are also impacted in various ways.\footnote{As shown in \autoref{fig:stakeholders}, both platforms and publishers struggle with the influx of AI-generated content (e.g. \cite{Acovino_2023_Sci-Fi_magazine_stops_submissions_after_flood_of_AI_generated_stories,Knibbs_2024_Scammy_AI-Generated_Books_Are_Flooding_Amazon,JonesNewmanEtAl_2025_AI-Generated_Slop_in_Online_Biomedical_Science_Educational_Videos_Mixed_Methods_Study_of_Prevalence_Characteristics_and_Hazards_to_Learners_and_Teachers,LinShanEtAl_2025_Stop_DDoS_Attacking_Research_Community_with_AI-Generated_Survey_Papers}), which is problematic exactly because its quality and originality is hard to assess. Some platforms also experience conflicts with their creator communities \cite{Edwards_2024_Stack_Overflow_users_sabotage_their_posts_after_OpenAI_deal}. Some platforms, notably Quora and Stackoverflow, declined in popularity as similar functionality was provided by LLMs, possibly trained on their own data (although both platforms were reported to have systemic issues pre-dating ChatGPT \cite{Pahwa_2024_How_Quora_Died,Orosz_2025_Stack_overflow_is_almost_dead}.} 

The (mostly U.S.-based) AI industry argues that its data practices are covered by the U.S. `fair use' doctrine, although that is U.S.-specific and far from settled even there (see \cref{sec:fair-use}). OpenAI additionally argued that the `less innovative' countries cannot be allowed to enforce their own laws on the U.S. companies ~\cite[p.11]{OpenAI_2025_OpenAI_Response_OSTP_NSF_RFI_Notice_Request_for_Information_on_Development_of_Artificial_Intelligence_AI_Action_Plan}, in particular for the sake of U.S.-China `AI race' \cite{Belanger_2025_OpenAI_declares_AI_race_over_if_training_on_copyrighted_works_isnt_fair_use,Lorenz_2026_Dark-Money_Campaign_Is_Paying_Influencers_to_Frame_Chinese_AI_as_Threat}. According to OpenAI, in the long run, AI will ensure `universal prosperity' \cite{Altman_2026_Our_principles} backed up by some version of universal basic income \cite{HolderGhaffary_2024_Sam_Altman-Backed_Group_Completes_Largest_US_Study_on_Basic_Income,Mak_2026_OpenAIs_spin_on_universal_basic_income}. This would presumably render financial concerns obsolete, as the creators would be `unconstrained by the need to make a living'~\cite{Khosla_2024_Roadmap_to_AI_Utopia}.
\footnote{Since attribution serves not only economic, but also epistemic and social purposes, we note that it is still needed even on this view.} 

We note that the revenue sharing problems between creators, platforms and big publishers are serious, but not fundamentally new: e.g. news aggregators have long been in a conflict with news publishers~\cite{Colangelo_2022_Enforcing_copyright_through_antitrust_strange_case_of_news_publishers_against_digital_platforms}, and creators in conflict with platforms \cite{Smith_2021_13400_Artists_Out_of_7_Million_Earn_50k_or_More_From_Spotify_Yearly}. What is new and specific to generative tools is that their users both consume and produce new texts: about a quarter of ChatGPT queries are for information-seeking and writing assistance each~\cite{ChatterjiCunninghamEtAl_2025_How_People_Use_ChatGPT}, and there are now more generated texts than human-written \cite{ParedesSmithEtAl_2024_More_Articles_Are_Now_Created_by_AI_Than_Humans}. These texts mix user contributions, common knowledge and original elements from past work provided to the model, on a scale that would be impossible for a human learner (a fact recognized as relevant by the U.S. copyright office, see \cref{sec:fair-use}). Some abuse LLMs to flood the ecosystem with generated texts in unprecedented volume, competing with the original creators~\cite{Sommer_2025_AI-Generated_Books_on_Amazon_Are_Hurting_Authors_and_Publishing_Industry,Knibbs_2024_Scammy_AI-Generated_Books_Are_Flooding_Amazon} and flooding their infrastructure~\cite{Acovino_2023_Sci-Fi_magazine_stops_submissions_after_flood_of_AI_generated_stories}. The majority are presumably good-faith users who would like to avoid plagiarism, but that is currently impossible due to the lack of instance-level attribution for information retrieved from parametric memory.\footnote{The RAG setting is supposed to address exactly this problem, but the generated text may still misrepresent the RAG source~\cite{LiuZhangEtAl_2023_Evaluating_Verifiability_in_Generative_Search_Engines,JazwinskaChandrasekar_AI_Search_Has_Citation_Problem}, which in practice puts the onus on the user to either verify all citations manually or accept responsibility for errors.} 

For such good-faith LLM users, the LLM services are provided at `use at your own risk' basis with no quality assurances~\cite{pandit2026termsabuseanalysisgenai}, and \textit{the responsibility for copyright violations is on the users}. While AI industry takes some measures to avoid obvious plagiarism,\footnote{E.g. OpenAI stated that it has measures to limit `inadvertent memorization', and it defends in court the position that verbatim copying is a `rare bug'~\cite{Morrone_2024_New_report_60_of_OpenAI_models_responses_contain_plagiarism}. Note that avoiding verbatim copies does not necessarily ensure that the output is not plagiarizing, see \cref{sec:plagiarism} for background on plagiarism and~\cite{GuptaPruthi_2025_All_That_Glitters_is_Not_Novel_Plagiarism_in_AI_Generated_Researcha} for evidence of well-masked plagiarism in generated research proposals.} it leaves the users bearing any reputational and legal risks. E.g. the terms of service of MS Copilot insist that this product is `for entertainment purposes only',\footnote{In early April 2026 this text drew much public attention. In response to that, Microsoft stated it was a `legacy' description that would be updated \cite{Griffiths_2026_Microsoft_says_Copilot_isnt_just_for_entertainment_purposes_after_its_terms_of_service_language_goes_viral}. As of May 5 2026, the text remains the same.} and explicitly rejects responsibility for any copyright infringement that the users of their models may commit by using the model outputs (while reserving the right for MS to train on the data provided by those users)~\cite{2025_Microsoft_Copilot_Terms_of_Use}. 
Yet such violations are anticipated, which is suggested both by research~\cite{liu-et-al-2024,GuptaPruthi_2025_All_That_Glitters_is_Not_Novel_Plagiarism_in_AI_Generated_Researcha} and the fact that MS is offering to cover the legal fees of their enterprise clients~\cite{Nowbar_2023_Microsoft_announces_new_Copilot_Copyright_Commitment_for_customers}.  This reasoning goes all the way to the court: e.g. OpenAI (unsuccessfully) argued in a copyright dispute with GEMA, a German association of composers, lyricists and music publishers, that the users are responsible for any copyright infringement of song lyrics `caused' by their prompts~\cite{Ulea_2025_OpenAI_cannot_use_song_lyrics_without_paying_German_court_rules}.

Instance-level attribution could address the above problems, but this requires both further research and transparency about data provided to the LLMs. The latter is also necessary for creators to exercise their rights.

%% file: sections/2.4_desiderata.tex
\subsection{What Is Next for the LLM Data Ecosystem?}
\label{sec:desiderata}

As discussed in \cref{sec:stakeholders}, the LLM data ecosystem is characterized by a power asymmetry between the individuals (creators, users) and the intermediary organizations, especially those in a dominant market position. The LLM `disruption' left many individuals dissatisfied (\cref{sec:ecosystem-disruptions}), but unable to just bargain individually or `vote with their feet'. The LLM users currently carry all the copyright infringement risks (\cref{sec:ecosystem-disruptions}), in addition to epistemic risks~\cite{Barrett_2026_US_Invaded_Venezuela_and_Captured_Nicolas_Maduro_ChatGPT_Disagrees} and possibly worse terms overall.\footnote{Some popular services added various `AI' components that are hard or impossible to remove, and raised the prices of their services for everyone (e.g. Microsoft Office, Slack~\cite{Crider_2025_Microsoft_follows_Google_with_price_bump_forced_AI_365_bundles_PCWorld,Maruccia_2025_Salesforce_hikes_Slack_prices_adds_AI_tools_for_all_paid_users}). There is also a new tendency to simply revise terms of service to force consent to data use for training~\cite{Maruf_2024_changed_its_terms_of_service_to_let_its_AI_train_on_everyones_posts_Now_users_are_up_in_arms,OfficeofTechnologyandTheDivisionofPrivacyandIdentityProtection_2024_AI_and_other_Companies_Quietly_Changing_Your_Terms_of_Service_Could_Be_Unfair_or_Deceptive}. The very content of chatbot conversations and even documents processed by the chatbots, often highly confidential, may also be used for training~\cite{2025_How_your_data_is_used_to_improve_model_performance}. Finally, ads personalized on chat data are also coming~\cite{Bort_2025_Perplexity_CEO_says_its_browser_will_track_everything_users_do_online_to_sell_hyper_personalized_ads,Vanian_2025_Meta_greenlights_Facebook_Instagram_ads_based_on_your_AI_chats,Roth_2024_Googles_AI_search_summaries_officially_have_ads,Capoot_2026_OpenAI_ads_pilot_tops_100_million_in_annualized_revenue_in_under_2_months}, which seems as questionable for something marketed as an information access technology as for search engines~\cite[app. A]{BrinPage_1998_anatomy_of_large-scale_hypertextual_Web_search_engine}, and doesn't necessarily serve users best \cite{WuLiuEtAl_2026_Ads_in_AI_Chatbots_Analysis_of_How_Large_Language_Models_Navigate_Conflicts_of_Interest}.} The creators are watching publishers~\cite[e.g.][]{Palmer_Taylor_Francis_AI_Deal_Sets_Worrying_Precedent_for_Academic_Publishing,Salmons_2024_Routledge_Sells_Out_Authors_to_AI,AI_Licensing_for_Authors_Who_Owns_Rights_and_Whats_Fair_Split} and platforms~\cite[e.g.][]{Wiggers_2024_OpenAI_inks_deal_to_train_AI_on_Reddit_data,Edwards_2024_Stack_Overflow_users_sabotage_their_posts_after_OpenAI_deal} make licensing deals without their consent (at least 26 deals so far ~\cite{AIwatchdog_2025_Content_Licensing_Deals}). But even the publishers may not have much bargaining power,\footnote{Google currently does not allow publishers to opt out of `AI overviews' without opting out of Google search~\cite{Alba_2025_Google_Can_Train_Search_AI_With_Web_Content_Even_After_Opt-Out}. Some of the reported OpenAI deals provide OpenAI access to publisher past and future content for compensation that is relatively small compared to other big tech negotiations~\cite{David_2024_OpenAIs_news_publisher_deals_reportedly_top_out_at_5_million_year}.
} if it is possible for the industry to just forego any licensing (see \autoref{footnote:abuse}).
This would be true even if there were reliable data valuation methods: they could facilitate bargaining by making data contributions more legible and comparable, but would not directly address underlying power asymmetries.

Political economy suggests that this situation calls for countervailing power \cite{Galbraith_1954_Countervailing_Power}, which could be provided via such mechanisms as antitrust enforcement \cite[e.g.][]{Colangelo_2022_Enforcing_copyright_through_antitrust_strange_case_of_news_publishers_against_digital_platforms}, regulation \cite[e.g.][]{CBC_2024_Australia_wants_to_make_digital_platforms_pay_for_news_even_if_they_block_it_like_Meta_did_here} and collective action \cite{vincentCollectiveBargainingInformation2025}. 
This process has already started: e.g. there are already class action lawsuits \cite[e.g.][]{2023_Authors_Guild_John_Grisham_Jodi_Picoult_David_Baldacci_George_RR_Martin_and_13_Other_Authors_File_Class-Action_Suit_Against_OpenAI}, collective action for protection against job displacement~\cite{WGA_Agreement_Introduces_Key_Protections_for_TV_and_Film_Writers_Against_AI}, and the provision for copyright protection in the EU AI Act~\cite{EuropeanCommission_2025_Explanatory_Notice_and_Template_for_Public_Summary_of_Training_Content_for_general-purpose_AI_models_Shaping_Europes_digital_future}. In response to the creator backlash, there already are some developments in the AI industry: the above-mentioned licensing deals with publishers/platforms~\cite{AIwatchdog_2025_Content_Licensing_Deals}, promises of revenue sharing at some point~\cite{Sauer_2024_OpenAI_CEO_Sam_Altman_You_could_get_paid_one_day_for_AI_training_data_we_use}, at least one experiment with actual revenue sharing~\cite{Campbell_2025_Perplexity_has_cooked_up_new_way_to_pay_publishers_for_their_content}, and at least one `answer engine' service developed with a revenue-sharing business model~\cite{ProRataAI_2025_ProRata_Partners_with_Danish_Publishers_Group_DPCMO_to_Launch_First_Decentralized_Sovereign_AI_Answer_Engine}. There are also technical proposals for the web infrastructure, including controlling bot traffic at the DNS level~\cite{LeeGuptaEtAl_2025_Control_content_use_for_AI_training_with_Cloudflares_managed_robotstxt_and_blocking_for_monetized_content} or via new services ~\cite{Fischer_2024_AI_startup_TollBit_raises_24M_series, dpcmo_tollbit_2026}, and machine-readable license protocols~\cite{2025_IETF_working_group_will_further_develop_our_proposal_for_opt-out_vocabulary,Brandom_2025_RSS_co-creator_launches_new_protocol_for_AI_data_licensinga}. 

We anticipate that the current litigation and AI industry lobbying efforts, whatever their outcome,\footnote{The legal situation is still unclear even in individual jurisdictions, much less internationally. See \cref{sec:fair-use} for the current state of the `fair use' dispute in the U.S. In the EU, there are German court rulings against OpenAI for training on song lyrics~\cite{Ulea_2025_OpenAI_cannot_use_song_lyrics_without_paying_German_court_rules}, but in favor of LAION in a web scraping case \cite{24_Germany_Hamburg_District_Court_310_O22723_LAION_Robert_Kneschke}. The EU Court of Justice is currently considering the landmark case of Like Company v Google Ireland \cite{2025_Case_C-250_25_Like_Company_Request_for_preliminary_ruling_from_Budapest_Kornyeki_Torvenyszek_Hungary_lodged_on_3_April_2025_Like_Company_Google_Ireland_Limited,Like_Company_Google_CJEU_Holds_First-Ever_Hearing_on_Generative_AI_and_Copyright_on_10_March_2026_Bird_Bird}.} will not be the final say. 
Labor rights offer a relevant precedent: what we have now was shaped through decades of collective action, which was triggered by the abuses of the industrial age~\cite{Editorial_2024_Evolution_of_Labor_Law_Comprehensive_Historical_Overview,Rivas_2025_From_Sweatshops_to_Standards_History_of_US_Labor_Laws}. If this parallel holds, we now are in the beginning of a society-wide re-negotiation of data rights, which will take different forms and arrive at different results on different timescales in different parts of the world (depending on the local laws, cultural and industry norms, among other factors). Like with the labor rights, this will require iterative coalition-building,\footnote{The organizations representing users could include consumer protection and watchdog organizations (e.g. US Federal Trade Commission, the Midas Project). On the creator side, collective bargaining could be performed by guilds and unions (e.g. the Writers' Union of Canada) and publisher associations (e.g. Danish Press Publications' Collective Management Organisation). On the creator side, we also see new structures that centralize the collection of opt-out preferences (e.g. Do Not Train Registry) and data licensing preferences including attribution and compensation (e.g. Mozilla Data Collective). Such lightweight structures ease the organizational burden on the creators, but would likely still require collective action to enforce industry compliance.} electing representatives, formulating policy goals, and taking collective action in various sectors until either the current platforms offer fairer terms, or alternatives emerge in response to the new demands, and we arrive at a relatively stable equilibrium. Like with labor rights, collective action will not be easy to organize.\footnote{Improved control over data rights is a collective good, and collective action towards a collective good faces the fundamental difficulty in recruiting the individuals who would bear the costs of organizing: it is in everyone's interest to have someone else do this work \cite{Olson_1971_logic_of_collective_action_public_goods_and_theory_of_groups}. Still, the clear threat posed by commercial use of copyrighted materials by AI industry should help to counter that obstacle.} Like with labor rights, this process will never be quite `finished' (e.g. the union negotiations are still ongoing even in the Nordics). As soon as the structures representing the individuals weaken, the industry would have an incentive to take advantage of that fact.

From that perspective, we argue that data attribution research can play an important role in supporting the collective action process towards re-balancing the data economy, as we will outline next.

%% file: sections/4_attribution-worthiness.tex
\section{Human-centric Data Attribution}
\label{sec:framework}

If, as argued in \cref{sec:ecosystem-disruptions}, we are now in the beginning of AI data rights movement, data attribution is needed as a credit assignment mechanism, to provide the relevant economic, social and/or epistemic traces. But to develop attribution methods, we have to have an accurate and specific definition of the problem. This is where NLP research would benefit from (a) positioning the problem in the evolving socioeconomic context, (b) informing the stakeholders of their current technical options, and being informed of the real-world needs for attribution. 
To achieve this, we propose a three-step framework for operationalizing attribution for a specific LLM use case (\S\ref{sec:stakeholders-to-attribution}). In \S\ref{sec:framework-methods} we discuss the technical attribution criteria that can currently be used to realize these goals in practice.

\subsection{From Stakeholder Goals to Attribution Requirements}
\label{sec:stakeholders-to-attribution}

To reiterate, we consider data attribution as the broad problem of establishing if a model used a certain work to produce a certain output. Depending on the implementation, insofar as attribution provides a trace of what data was used, it could support the economic incentives, social signals, and/or epistemic traces relevant for the stakeholders. A key problem is that, as discussed in \cref{sec:stakeholders}, their needs vary substantially. For example, a fiction writer may be willing for LLMs to be trained on their work, but only if they are compensated. At the same time, academics may be willing for their texts to be used for RAG for free, but their main `currency' is citations and social credit, and that should be supported \cite{AtmakuriSinghEtAl_2025_Making_AI_citations_count_with_Asta}. This variety entails that it is neither possible nor desirable to have a single solution that would cover all the cases. 

\begin{figure}
    \centering
    \includegraphics[width=0.85\linewidth]{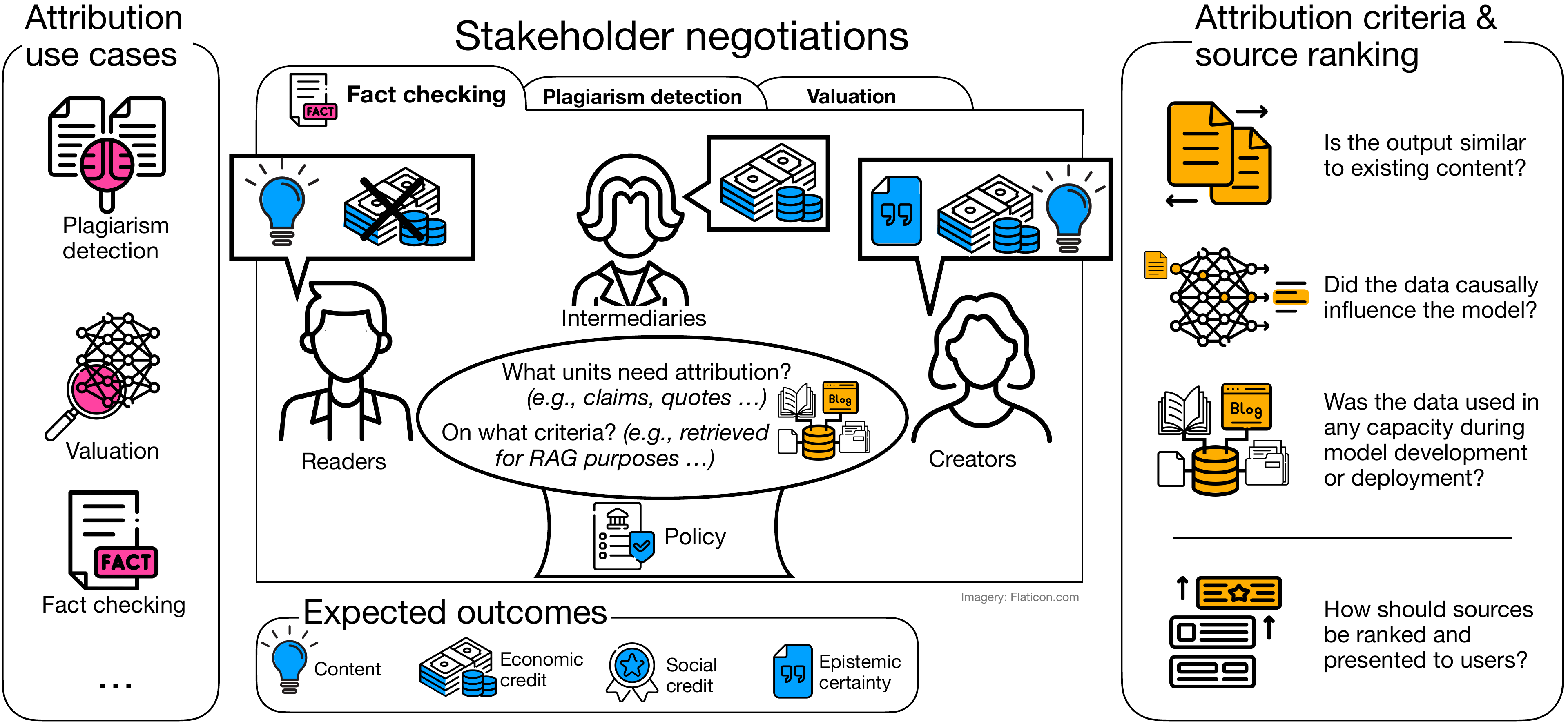}
    \caption{The human-centric attribution framework is grounded in case-specific stakeholder negotiations, which explicate the expected outcomes and agree on how attribution would be implemented. The example shows possible conflicting objectives for the fact checking use case. All stakeholders have conflicting economic objectives: the LLM service prefers to maximize its revenue, the users prefer to get the information for free, and the creators prefer the LLM service to share revenue. Epistemically, the creators and the public informedness policy may also prefer the LLM service to show cited quotes to avoid misrepresentation, while the users prefer simple generated answers, and the LLM service prefers to follow user preferences.} 
    \label{fig:attribution-worthiness-framework-overview}
    \Description[The human-centric attribution framework]{The human-centric attribution framework is grounded in case-specific stakeholder negotiations, which explicate and balance their expected outcomes and agree on the general criteria for implementing attribution.      
    }
\end{figure}

Our solution is the human-centric attribution framework shown in \autoref{fig:attribution-worthiness-framework-overview}. It supports operationalizing attribution for a given case based on stakeholder negotiations, either explicit (e.g. specific talks leading to an agreement) or implicit (e.g. a series of court cases with similar plaintiffs and defendants). 
Each attribution scenario is specified as follows. 

\paragraph{Step 1: Identify the Relevant Stakeholders and the LLM Use Case.} This, as well as Step 2, is in the sphere of policy and collective action (\cref{sec:desiderata}). The current discussion and litigation effort seems to be focused on the creators and intermediaries (e.g. news content accessed via RAG systems), but there could also be more parties, representing e.g. public-interest considerations and consumer groups.\footnote{Of all groups, consumers are the most numerous and the least connected, and hence likely to face the greatest challenges in collective action. Ideally, technology would be designed based on both participatory principles and research findings on known exploits of human psychology, while being regulated by consumer-oriented public policy. Long-term social priorities such as education, informedness and skill maintenance of the population may conflict with short-term commercial interests of the AI industry.}

\paragraph{Step 2: Negotiate the Intended Outcomes of Attribution.} 
Since there may be conflicting objectives for data attribution, the relevant stakeholders must negotiate to find a sustainable balance between their needs. As one possible example of conflicting negotiation objectives, let us consider information seeking queries to LLMs, where data attribution is needed for fact-checking. The competing stakeholder objectives are shown in Figure~\ref{fig:attribution-worthiness-framework-overview}. 

\paragraph{Step 3: Specify What Is Attributed and How Attribution Is Implemented.}
Step 3 is the most directly relevant to NLP research, which serves as a source of the current implementation options, and also receives motivation for what new methods are needed in practice. The stakeholders need to make an informed decision about the relevant \textit{attribution units}\footnote{See \citet{dengSurveyDataAttribution2025} for a discussion of various levels of attribution discussed in NLP literature (tokens, samples, groups of examples, domains). In stakeholder negotiations, we may find that different domains have different human-interpretable criteria for `substantial similarity',  possibly very granular and grounded in specific practices, with categories not readily implement-able in NLP. For example, fiction writers may find similarity in narrative structure important.} (i.e. what needs to be attributed -- claims, character profiles, certain stylistic features?), \textit{attribution criteria} (i.e. how the attribution would be determined), and \textit{ranking principles} (i.e. how the order of the displayed sources would be determined). 
At present, there are three broad groups of attribution criteria, discussed in detail in \cref{sec:framework-methods}. The core technical options need to be communicated to stakeholders in a clear and accessible way (ideally by independent experts), together with the options for ranking the sources for display (\cref{sec:ranking}). Any transparency measures for monitoring the outcomes and stakeholder feedback,\footnote{In the ideal world, all stakeholders would continually monitor the outcomes of the agreement via pre-agreed transparency measures (e.g. via analytics on the web traffic and revenue), as well as any relevant developments (e.g. court case outcomes and better technical methods). Terms could be subsequently re-negotiated if the outcomes do not match the expectations, or the negotiating position is strengthened due to external developments. A group may also reconsider its position because of a successful lobbying by a subgroup whose interests have not been taken into account initially, or as a result of a top-down initiative to collect feedback across its membership (e.g. via surveys or interviews). The resources and results of this process could be shared and re-used by similar stakeholder groups.} as well as conditions for re-negotiation, should ideally also be part of this discussion.

With further development, the proposed framework could be used directly for participatory design of attribution methods: to elicit the preferences of specific groups of stakeholders (e.g. via interviews or focus groups), to facilitate discussions about balancing the conflicting interests, and use the results to inform design of attribution methods. However, the practical outcomes of such work would likely be limited to academic research. The real world will in the meanwhile continue the implicit multi-step negotiations between the creators, the industry and the users via coalition building, collective action, and legislative efforts. In those negotiations, in any specific round, not all relevant parties may be at the table\footnote{E.g. Google currently justifies its practice to provide news content in AI overviews, saying that it is the `preferred user experience'~\cite{Dickey_2025_Penske_Media_Sues_Google_for_AI_Overview_News_Story_Summaries_Without_Publishers_Consent}. The publishers object: ``it's not Google's place to make this business decision on behalf of the people who are producing the goods being sold''~\cite{Germain_2025_Is_Google_about_to_destroy_web}.}, and the outcomes may reflect the balance of power and resources of stakeholders rather than only their interests. The weaker groups may then build a stronger coalition and force a further re-negotiation. The proposed framework can help to position the attribution problem in that broader process, and surface unaddressed attribution needs motivating future research: e.g. the current lawsuits signal a strong creator preference for compensation, which, irrespective of legal outcomes, motivates research on fair data pricing. In the long run, this process tracks the necessary updates to copyright laws, the goal of which is to ``strike the right balance between the interests of content creators, developers and investors and the public interest in being able to access and use creative content'' \cite[p.22]{2020_What_is_intellectual_property}. 

%\input{sections/4.5_practice}

%%%%%%%%%%%%%%%%%%%%%%%%%
\subsection{Technical Criteria for Data Attribution}
\label{sec:framework-methods}

At present, there are three broad groups of attribution criteria: similarity to existing content (\cref{criterion:similarity}), causal influence on the model (\cref{criterion:causality}), 
and whether the data was used for model development or deployment
(\cref{criterion:use}). 
They rely on different technical principles, and would likely result in fundamentally different outcomes for the stakeholders. It is possible that one negotiated case involves more than one type of attribution (e.g. both dataset-level for training, and for specific instances when they are used in generation).
\subsubsection{Is the Output Similar to Existing Content?}\label{criterion:similarity} A generated text could be conceptually and/or structurally similar to a text that had been provided to the model, and on this basis the attribution link would be established. This attribution may be the easiest to grasp for humans, but such similarity does \textit{not} indicate that the model relied on a given instance in the generation process~\cite{HuangYangEtAl_2024_Demystifying_Verbatim_Memorization_in_Large_Language_Models}. \citet{worledge2024unifying} describe such attribution as `corroborative'.
 
The relevant similarity criteria will vary widely across use cases, and should be agreed on by the relevant stakeholders. 
For programmers, it may be important to recognize  algorithms rather than strings  -- the way a problem is solved rather than the exact way the code is written.  
In academia, where writing style and text structure are relatively rigid and formulaic, attributing scientific claims may be more important than attributing n-gram overlaps or argument structure. For fiction writers, attributing claims may be negligible, while stylistic features may be essential.

\textit{Current Technical Feasibility.}
Measuring similarity between two text units is a core NLP task that has been studied for a long time and has many efficient solutions, including string-based \citep{Kusner-et-al-2015_wmd}, statistical~\cite{papineni-etal-2002-bleu}, based on linguistic structure \citep{oya-2020-syntactic}, and neural~\cite{mikolov2013efficientestimationwordrepresentations}. There are also newer, more computationally expensive methods including contextual embeddings~\cite{reimers-gurevych-2019-sentence}, LLM-based methods~\cite{vera2025embeddinggemmapowerfullightweighttext,behnamghader2024llmvec,su-etal-2023-one} and hybrid methods~\cite{splade}. However, most of this research has focused on generic distributional similarity, rather than the aspect-specific similarity that would e.g. enable a writer to compare their story arc to a possible source. Instruction-following methods could specify fine-grained similarity criteria~\cite{WellerChangEtAl_2025_FollowIR_Evaluating_and_Teaching_Information_Retrieval_Models_to_Follow_Instructions}, but this would require running an extra LLM just for attribution. To the best of our knowledge, there is only one proposed solution that supports the model generation process with similar snippets from the model training data, and it only supports verbatim string matches~\cite{LiuBlantonEtAl_2025_OLMoTrace_Tracing_Language_Model_Outputs_Back_to_Trillions_of_Training_Tokens}.

\textit{Suggested directions for future NLP research.} To support various attribution use cases, the NLP community would need efficient methods for identifying specific types of textual similarity, 
as well as efficient semantic search solutions for LLM-scale corpora. 
Any such solutions would also need to be evaluated for precision, recall and also potential bias for any specific attribution scenario.

\subsubsection{Did the Data Causally Influence the Model?} \label{criterion:causality}
Causality-based attribution (`contributive attribution' in \citet{worledge2024unifying}) aims to quantify the influence of training data points on model behavior. This is done by estimating the change in model outputs or parameters under counterfactual training scenarios, e.g. removing or reweighting data points.
In contrast to the similarity-based approach, text identified by causality-based methods may seem unconnected to the generated text. For attribution to training data, causality-based methods are functions that estimates the importance of a training instance given a specific model, its training data, a given input and the output generated in response to that input. Another relevant line of research is data curation, where data ablations or quality assessments are used to establish how much training on a given dataset contributed to model improvement on a given metric~\cite{wettig-et-al-2025_organizewebconstructingdomains, olmo2025olmo3}.  
\citet{shi2025flexolmoopenlanguagemodels} propose an architectural paradigm of federated training different experts of a Mixture-of-Experts model \citep{6797059, shazeerOutrageouslyLargeNeural2017} on isolated datasets from known owners. The trained router can then be used for attributing the contribution of different datasets to a models' output.

\textit{Current technical feasibility}
Causality-based methods are challenging to implement on LLM scale, as their costs scale with dataset size and parameter counts, making them prohibitively expensive. E.g.
\citet{wang2024economicsolutioncopyrightchallenges} propose using Shapley values, which estimate the utility of training data subsets to value data and compensate contributors, but for this the model has to be retrained multiple times to establish the utility of different data subsets. Model-centric approaches such as influence functions measure a data point's counterfactual effects on the model parameters~\cite{worledge2024unifying}. Given a generated output, the influence of each training sample can be calculated.
At present, there are several proposals for influence functions-based data attribution solutions at the smaller LLM scale~\cite{park2023trakattributingmodelbehavior,changScalableInfluenceFact2024, grosse2023studyinglargelanguagemodel}, but there are concerns about their performance~\cite{li2024influencefunctionsworklarge} and what they actually reflect~\cite{bae2022influencefunctionsanswerquestion}. 

\textit{Suggested directions for future NLP research.} Further work is needed on computationally efficient and model-faithful data attribution methods. 
Furthermore, since these methods reflect the model process, but the result may not be human-interpretable, it remains to be seen whether they could be effectively used by non-experts in practical data attribution scenarios. 
To use causal attribution methods for data valuation, future research also needs to consider how training goals and the interaction between different training data impact valuation. We need to understand for example to which extent coding data may be more valuable than fiction data for a code assistant model, or if the value of an additional fiction dataset may be decreased if one fiction dataset has already been included in training. 
Finally, the RAG setting poses faithfulness challenges that require research: should the credit of the citation change based on whether the retrieved information was `new' to the model, or was already available in parametric knowledge?

\subsubsection{Was the Data Used in Any Capacity During Model Development or Deployment?}\label{criterion:use}
For this criterion, any usage of a creator's data 
during model development (including any intermediary models) or deployment
elicits attribution. This holds even if the data is not similar to a particular output, does not causally influence it, and training on a given dataset did not improve the model performance on a metric of interest. 
This criterion is relevant in cases where attribution serves as acknowledgment rather than as a measure of causal influence, and any remuneration does not depend on a disclosure about how or to what extent that data shaped specific outputs or functionality.
For example, if a dataset is overall free to use with attribution (e.g. Wiki data), and the model is used for language assistance rather than information, a data partner may be satisfied if the use of their data is simply publicly acknowledged.

\textit{Current technical feasibility.}
With cooperation of all parties, this would be the easiest and computationally cheapest method to implement. 
It would require open training and development corpora, source use transparency in RAG setups and the creators' informed consent if their data is to be included. This could follow the recommendations described by~\cite{kyiGovernanceGenerativeAI2025}, standardized in a license for AI training that content owners can grant to model developers to establish ``collective or representative consent mechanisms''~\citep{kyiGovernanceGenerativeAI2025}. 

In the absence of such a cooperation, relatively little can be done as a comprehensive solution. Inference attack methods may be used by specific stakeholders to inspect  models that they suspect of violating their rights. For example, a model could be simply prompted to complete a snippet of proprietary text, and its success would indicate that the content was seen during training and memorized~\cite{carlini2022quantifying}.
Methods designed to detect data contamination, meaning overlaps between training and evaluation data of a language model, could also be extended to attribution. \citet{deng-etal-2024-unveiling} provide an overview, including detection methods that investigate if models can correctly reproduce key information (such as named entities), and sensitivity tests, which measure if a model has a preference for the canonical order of input data.

\textit{Suggested directions for future NLP research.} With inference attack methods, the onus is on the creator to find out about the violation of their rights, prove it, and take legal action. One could imagine solutions to support a signed-up list of authors, who would like to systematically monitor new models for rights violations, and coordinate for class action. Then a further challenge would be proving the exact source and extent of exposure (e.g. maybe some elements from copyrighted works were learned from their mentions in non-copyrighted materials).
Beyond inference attacks, NLP research could further advance model transparency, documentation and reporting practices (e.g. datasheets~\cite{GebruMorgensternEtAl_2020_Datasheets_for_Datasets}, model cards~\cite{MitchellWuEtAl_2019_Model_Cards_for_Model_Reporting}, see \cref{sec:related}), that could serve as conventions for such acknowledgment-based attribution.

\subsubsection{Selection and ranking} \label{sec:ranking}

Since a given generation may be supported by hundreds of data points, the attribution problem is accompanied by an orthogonal problem of selection and ranking, to keep the references manageable for the user. Some attribution methods may provide a natural ranking (e.g. determined by cosine similarity scores), but it is also possible that there are many results with nearly identical scores. There should be a fair strategy for selecting what the user would see first, where the selection criteria are understood and accepted by the stakeholders. We propose that the following implementation principles are considered:

\begin{enumerate}
    \item \textit{Uniqueness}: units should be considered for attribution when they are relatively rare (e.g. there is no need for a reference for ``the sky is blue'' or ``time is money'').
    \item \emph{Chronological precedence}: assuming there are multiple potential references, i.e., high similarity matches, the most attribution-worthy is the one that is the oldest. Without such mechanism, there would be an incentive to create multiple copies of content for the attribution credit~\cite{Rogers_2024_Google_Search_Ranks_AI_Spam_Above_Original_Reporting_in_News_Results}.
\item \emph{Extent of shared context} with the training document(s) the attributable unit appeared in. This characterizes how competitive the model output is with the model source. For example, if a given trope is reused in the context of another fiction story with the similar story arc, it increases its  attribution-worthiness. 
\item \emph{Random sampling as the last resort}. If after the application of the above criteria there are still too many candidates, the top spots should be allocated randomly. This is the part where attribution should avoid the known pitfalls of the current  platforms: 
search compromised with ads~\cite{AhmedHaskell-Dowland_2021_Is_Google_getting_worse_Increased_advertising_and_algorithm_changes_may_make_it_harder_to_find_what_youre_looking_for}, proprietary recommendation algorithms that unilaterally change the rules of the game~\cite{Gil_2024_Stop_Chasing_Algorithms_Heres_How_Creators_Can_Take_Control_of_Their_Content_and_Monetize_on_Their_Own_Terms}, and absence of due process~\cite{Walsh_2025_YouTube_error_that_couldve_cost_thousands,2025_Meta_Wrongfully_Disabling_Accounts_with_No_Human_Customer_Support,Thompson_2025_They_Criticized_Musk_on_Then_Their_Reach_Collapsed}.

\end{enumerate}

\input{sections/outlook}

%% file: sections/outlook.tex
\subsection{Outlook for Human-centric Data Attribution}
\label{sec:outlook}

%\paragraph{Data Attribution Negotiations in Practice}
%\label{sec:practice} 
As preliminary evidence that the proposed framework tracks the emerging negotiation practices in the LLM data economy, let us consider the recent deal between ProRata and the Danish Press Publications' Collective Management Organisation (DPCMO) \cite{ProRataAI_2025_ProRata_Partners_with_Danish_Publishers_Group_DPCMO_to_Launch_First_Decentralized_Sovereign_AI_Answer_Engine}. Its details identify all three elements of the proposed framework. For step 1 (stakeholders and use case), we know that the deal focuses on an `answer engine' for news articles, provided by ProRata to DPCMO (collectively representing 99\% of Danish news industry organizations). For step 2 (intended outcomes), the financial details are not public, but ProRata's website currently stipulates sharing 50\% of revenue with content partners \cite{ProRataAI_We_build_AI-powered_search_advertising_and_attribution_solutions_grounded_in_respect_for_content}. The existence of the deal implies that a mutually acceptable split was identified. For step 3 (implementation), the announcement states how attribution would be implemented (RAG), and that publishers remain in control what content can be used in this way. While this deal is specific to Danish news publishers, and may not represent the interests of the individual journalists, it signals that overall a strong creator group can find industry partners willing to offer better terms, and protect themselves against those who do not.\footnote{Subsequently, DPCMO sued OpenAI \cite{Ronde_2026_united_Danish_media_industry_takes_OpenAI_to_court} and partnered with TollBit to protect their content from unauthorized use \cite{Ronde_2026_DPCMO_Partners_with_TollBit_to_Offer_Solutions_to_Support_its_Members_DPCMO}).} The technical options under consideration were probably limited to what ProRata had implemented, but this deal signals an overall acceptability of RAG by this group of stakeholders, and motivates further work on overcoming the limitations of RAG \cite{LiuZhangEtAl_2023_Evaluating_Verifiability_in_Generative_Search_Engines,JazwinskaChandrasekar_AI_Search_Has_Citation_Problem}.
%define stakeholders, negotiate outcomes, specify attribution criteria

Would data attribution be enough? %\paragraph{Could synthetic data upend attribution?} %Strengthen the discussion of boundary conditions for the framework's effectiveness. Explicitly address when attribution and compensation might not suffice to sustain creator participation. Incorporate the rebuttal's arguments (historical creator resilience, non-economic motivations) into the paper with greater precision about how current conditions differ from historical comparisons in scale, speed, and synthetic data capability. Engage with the synthetic data complication (Iskander et al. 2024) that Reviewer VHEb raises.
According to \cite{Hooker_2025_On_Slow_Death_of_Scaling}, ``the cost of generating synthetic data is now low enough that we can treat the data space as malleable and something which can be optimized''. Given that numerous studies showed benefits of synthetic data in various NLP scenarios \cite[inter alia]{iskander-etal-2024-quality,viswanathan-etal-2025-synthetic,shimabucoro-etal-2024-llm}, it is fair to ask whether the efforts towards sustaining creators via data attribution are overall doomed.  
We cannot exclude such a scenario, but we note that cultural markets are not dependent only on prices -- otherwise we would never have any new writers, who necessarily compete against an endless supply of high-quality public domain books. One factor is social: we may prefer to pay for a recent book or a newspaper because we know and trust the source. This explains why many publishing authors actively build their audiences via social media, newsletters etc. \cite{Channels_of_Communication_Part_1_Authors_Guild_Launchpad_Webinar_3}, and why there already are scammy generated `books' that try to pass for a high-reputation human source~\cite{Knibbs_2024_Scammy_AI-Generated_Books_Are_Flooding_Amazon,Sommer_2025_AI-Generated_Books_on_Amazon_Are_Hurting_Authors_and_Publishing_Industry}.
Another factor is relevance to the current moment: facts, preferences and even linguistic patterns change all the time, and have to be continually sourced from humans \cite{rogers2026the}. Last but not least, the synthetic data still does not come out of nowhere. It is enabled by the the original training data of the generating model, which is at minimum covered by the criterion in \cref{criterion:use}. Synthetic data is still subject to attribution and compensation, otherwise there would be a strong incentive to simply launder any new data through intermediate models.

%% file: sections/5_attribution-in-practice.tex
\section{Moonshot: What Attribution Could Look Like in Practice}
\label{sec:moonshot}

As described in \cref{sec:framework}, the proposed framework does not by itself prescribe any specific solutions for attribution, which need to emerge out of iterative collective action process.  

As an illustration, we present a moonshot for how attribution could look like in practice, inspired by the `data as labor' philosophy~\cite{Lanier_2014_Who_owns_future,ArrietaIbarraGoffEtAl_2017_Should_We_Treat_Data_as_Labor_Moving_Beyond_Free}. In our hypothetical scenario, a LLM service provides creative writing assistance. The creators, the publishers, the users, and LLM service providers agreed on two kinds of attribution: general acknowledgment $+$ license fee for all training data, and instance-level attribution, triggered when the model makes suggestions that are substantially similar to rare creative elements in training data. This instance-level attribution should confer social credit to the author of the original text and support micro-payments. 

This arrangement would raise the cost of the LLM service for the users, possibly decreasing its user base and the overall number of queries. But it also has advantages: the users do not have to worry about plagiarism~\cite[e.g.][]{Gault_2025_Authors_Are_Accidentally_Leaving_AI_Prompts_In_their_Novels}, which could result in reputation damage or even litigation, and they know (and could even prove) the exact extent of their own contributions. The LLM service explicitly assumes the broker role and receives a share of micro-payments for the attribution-worthy outputs (some uses, such as grammar checks, would not qualify for that). It enjoys an improved reputation and possibility of competing based on its `creator library' (e.g. `this model is a better creative writing assistant than a competitor X, thanks to licensed content from <list of writers>'). 

The dataset-level attribution (e.g. simply based on use of specific works in training) is technically the simplest to implement. The parties could agree e.g. that the writers are acknowledged in a public list of contributors to LLM training data (which the LLM service could also use for marketing purposes). They would further receive some negotiated share of the one-time licensing fee, due whenever the data in question is used for training.

\begin{figure}[t]
    \centering
    \includegraphics[width=0.95\linewidth]{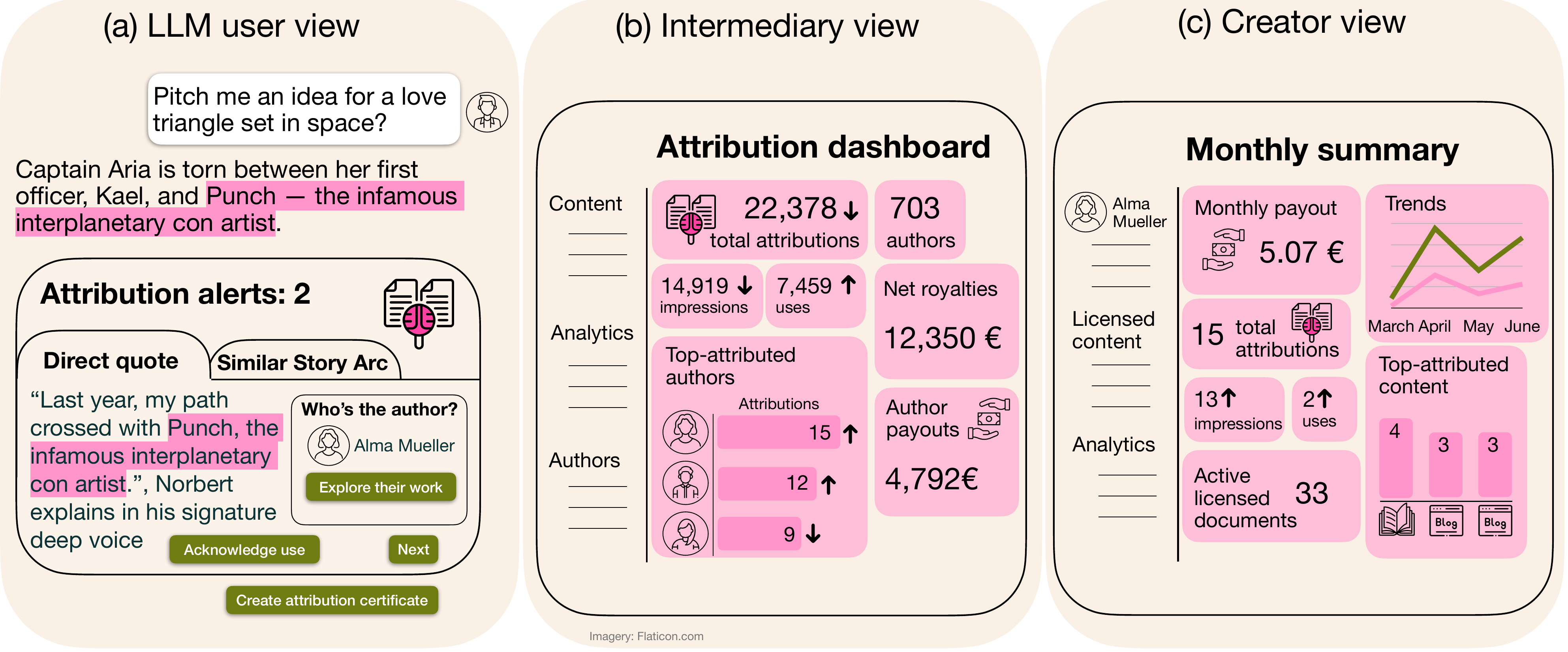}
    \caption{Moonshot: human-centric data attribution for LLM-assisted creative writing. We show how this process could look like in a user interface for three stakeholders: the LLM user, the creator of the attributed text, and the intermediary party.}
    \label{fig:moonshot}
     \Description[Moonshot: human-centric data attribution for LLM-assisted creative writing.]{Moonshot: human-centric data attribution for LLM-assisted creative writing. We show how this process could look like in a user interface for three stakeholders: the LLM user, the creator of the attributed text, and the intermediary party.}
\end{figure}

The instance-level attribution would be implemented in several negotiated ways, e.g. optimizing for the search of similar tropes, narrative arcs, and character profiles. Assuming that sufficiently precise and efficient methods are developed for this kind of attribution, \autoref{fig:moonshot} presents a moonshot for what the resulting workflow could look like. The implementation would follow the principles outlined in \cref{criterion:similarity}, so as to avoid attribution alerts for non-original elements (e.g. common metaphors or plot lines). However, relatively unique elements,\footnote{This process should take into account user-provided input and whether new content is being suggested. E.g. if the model only checks grammar and does not suggest content not already present in the user input, there is nothing to attribute.} especially when combined in a similar way to an existing published source), should trigger the attribution alert. 

As shown in \autoref{fig:moonshot}, the LLM user would then see an alert for the likely source\footnote{The example in \autoref{fig:moonshot} shows that the LLM response triggered two attribution alerts: a verbatim match and a match for a similar story arc. We visualize the verbatim match and show one attributed source for clarity, but there could be numerous relevant candidates (see \cref{sec:ranking}).} of the LLM suggestion, with the link to the profile of the original author. Let us consider this an `impression' of the attributed content. The shown alert will be added to a log file, and a micro-payment to the creator (including some negotiated nano-share of that for the intermediaries), would be triggered. The user would then make a choice to rely on this idea in some way, or go for some other idea. Should they choose to use the attributed content, they could signal this by pressing a button or simply copying some part of the displayed text, triggering a somewhat larger micro-payment for `use with acknowledgment'. 

The logs of what pieces of content were displayed and used could be stored and used for various analytics, providing highly useful information to all parties. \autoref{fig:moonshot} shows that the author of the attributed work could see which elements of their work were used, and what earnings that brought. The publisher, having access to the same information, could make more informed decisions about which authors to sign up a deal with. 

Based on these logs, a standard for the attribution certificates could be adopted by LLM services. They could then provide a new feature: transparency in professional use of LLMs without documentation burden. For example, the citizens could see exactly how much input an LLM had in drafting the policy for their municipality~\cite{Sanford_2025_As_WA_government_officials_embrace_AI_policies_are_still_catching_up}. In education, it would be easy to `show your work' to the teacher. Of course, there will always be bad actors, but a system embracing attribution would make it intuitive to the users what `responsible use of LLMs' means. 

Appropriately scoped attribution could turn assisted writing from a questionable practice into a legitimate interaction with a cultural technology~\cite{FarrellGopnikEtAl_2025_Large_AI_models_are_cultural_and_social_technologies}, with a clear creative space for new work, and an economic position for both the past and new contributors. Without attribution, the proponents of generative AI are themselves subject to the paradox of \textit{`fair use for me but not for thee'}: e.g. OpenAI terms of services prohibit other developers from training on the outputs of their models~\cite{OpenAI_2023_Terms_of_use}, and some of its `power users' object to reuse of their prompts~\cite{Wilkins_2026_Furious_AI_Users_Say_Their_Prompts_Are_Being_Plagiarized}.

%% file: sections/6_conclusion.tex
\section{Conclusion}

So far, LLM services have been in an adversarial relation with the creators whose work makes them possible. But it does not have to be this way. We are only in the beginning of a long process of legislative and normative re-negotiation, where data attribution research can be informed by and support the collective action processes. The proposed human-centric data attribution framework (\Cref{sec:framework}) bridges socioeconomic research and policy work with development of NLP methods, which could incentivize high-quality contributions to the data economy. To the former we provide a blueprint for how case-specific attribution requirements could be specified for technical implementation, and what is currently feasible. NLP researchers benefit from a clear grounding of their work in the real-world demands, and an external source of construct validity and evaluation criteria.

For the creators, the current state of data economy leaves much to be desired, but it is also `a crisis we shouldn't let go to waste' \cite{Germain_2025_Is_Google_about_to_destroy_web}. Pre-LLMs, much of the web had to rely on ad-based monetization, which incentivized its own questionable practices such as clickbait. LLM services backed by data attribution could provide new incentives to create content in direct response to demand for its substance. And they could do so in a clear, legitimate dialogue with other creators, crediting their influences and receiving their own acknowledgments.

\section*{Limitations and Future Work}

This work is an interdisciplinary endeavor, with origins and primary audience in NLP research on data attribution. We aim to position the task of data attribution in the current socioeconomic reality, and explicate what role it could play in shaping a more socially sustainable future. We expect that economists, governance experts, political scientists or legal scholars would describe the processes in \cref{sec:ecosystem} differently, using the rich theoretical apparatus of their disciplines. We hope that they would not judge this work too harshly (as long as the analysis is generally correct and actionable), and that they would engage with the community developing attribution methods to explain any missing pieces of this puzzle.

As described in \cref{sec:stakeholders}, the LLM ecosystem contains a multitude of stakeholders that perform similar roles with widely varying motivations and hence would respond to different incentives. Our overview of these incentives focuses on three main types of motivation (social, financial and intrinsic factors). A more fine-grained classification of possible objectives could be developed to assist stakeholder negotiations. It is possible that the process of such negotiations would identify necessary revisions to the framework (though the overall multi-stakeholder perspective is essential). In particular, more focus could be given to supporting the creators' moral rights (including the right to object to misrepresentation).

A more detailed blueprint for stakeholder negotiations and preference elicitation could be developed by experts in mediation and participatory design, and more implementation options for possible use cases could be proposed (e.g. based on the analysis of arguments of both parties in ongoing litigation). 

\section*{Generative AI Usage Statement}
Generative AI tools (Claude Sonnet 4.5, GPT-5.2) were used for language assistance in some parts of the paper (grammar checking and stylistic suggestions).

\section*{Ethical Considerations}
In this paper we do not describe any experiments with human subjects, users and/or deployed systems. Further, we do not rely on any sensitive user data.
This work focuses on mitigating negative impacts of LLM-based technology that is currently deployed.

\section*{Acknowledgments}
This work was supported by a research grant ([VIL60860]) from Villum Fonden. We are grateful to the anonymous reviewers and our colleagues for feedback and conversations, including Maria Antoniak, Toine Bogers, Christian Hardmeier, Bertram Højer, Luis Jaburi and Arturo Valdivia.

%% file: sections/2.2_concepts.tex
\section{Background: Relevant Legal and Professional-Conduct-Related Concepts}
\label{sec:concepts}

\subsection{Copyright}\label{sec:copyright} 
Copyright is a subtype of intellectual property that applies to ``creative expression of ideas in many different forms'' \cite[p.21-22][]{2020_What_is_intellectual_property}, and includes both economic and moral rights (see \cref{sec:ecosystem-concepts-moral-rights}). It may be implemented differently in various national legal systems. 
In the U.S. copyright law, legal action requires establishing substantial similarity, i.e. that two works are ``so alike in protected elements that one work infringes upon the other''~\cite{Hickey_2015_Reraming_similarity_analysis_in_copyright}. However, determining substantial similarity in legal practice is highly non-trivial (ibid). 
A relevant part of copyright law is subsidiary rights: they enable e.g. publishing houses to further license content translation or TV adaption.

\subsection{The U.S. `Fair use' Doctrine} \label{sec:fair-use} In the current LLM ecosystem, companies that build commercial large language models have mostly been non-compliant with copyright and licensing agreements. Most of the major developers are currently based in the U.S., and they argue that their application falls under the U.S. `fair use' doctrine~\cite{Belanger_2023_Grisham_Martin_join_authors_suing_OpenAI_There_is_nothing_fair_about_this_Updated}. That doctrine does not apply internationally, and even in the U.S. it is not intended as a blanket copyright exemption: its application depends on case-specific considerations of transformativeness, nature of copyrighted work, amount and substantiality, and effect on market for the original work~\cite[][p.5]{HendersonLiEtAl_2023_Foundation_Models_and_Fair_Use}. 
The interaction between these factors is complex and case-specific, and the U.S. legal battle for commercial LLMs is not yet settled. In particular, a contested area is whether the market impact has to be the market for a specific work from specific highly similar outputs, or the overall market dilution due to mass creation of competing texts in the same market (e.g. if a model was trained on specific kind of novel, and then used to flood the market for that kind of novels, impacting the original creators).

The current situation is that there is a ruling in favor of Anthropic that accepted the `fair use' argument for `transformative' nature of model training, but rejected the use of pirated books for that purpose \cite{Anthropic_Lawsuit_Update_Settlement_Reached_with_Pay-Out_to_Authors}. Another ruling granted `fair use' to Meta  \cite{MartinezBorovik_2025_No_harm_no_win_cautionary_tale_of_Kadrey_Meta_Platforms_Inc}. However, both of these cases are at the level of the district court, and neither case examined the market impact from the generated text. The latter ruling even explicitly referenced the fact that the plaintiffs did not make that `potentially winning' argument \cite[p.4]{Chhabria_2025_ORDER_Denying_Plaintiffs_482_Motion_for_Partial_Summary_Judgment_and_Granting_Metas_501_Cross-Motion_for_Partial_Summary_Judgment}. In January 2026, a group of music publishers sued Anthropic \cite{Concord_Music_Group_Inc_Anthropic_PBC} explicitly with that claim: ``\textit{Anthropic’s extensive copying of Publishers’ and other music publishers’ works also allows Anthropic to offer AI models that users can and do use to create vast quantities of new musical compositions and lyrics that compete with and dilute the market for Publishers’ legitimate product,
and divert royalties and income from legitimate composers and publishers.}'' \cite[p.37]{ConcordMusicGroupInc.CapitolCMGInc._2026_COMPLAINT_and_Demand_for_Jury_Trial_against_All_Defendants}. That case is not yet decided, and overall the  U.S. legal landscape continues to develop. 

One argument for the development of generative models in line with the U.S. `fair use' doctrine is an analogy with human learning: since humans are allowed to learn from existing sources, the same applies for models. However, the U.S. Copyright office disagrees: fair use for human learning also has limits, and it differs `in ways that are material to copyright analysis', namely precision, speed and scale \cite[p.48]{2025_Report_on_Copyright_and_Artificial_Intelligence_Part_3_Generative_AI_Training_pre-publication_version}. The ruling by Justice Chabria also dismissed this argument made by Meta \cite[p.17]{Chhabria_2025_ORDER_Denying_Plaintiffs_482_Motion_for_Partial_Summary_Judgment_and_Granting_Metas_501_Cross-Motion_for_Partial_Summary_Judgment}: ``\textit{by creating a tool that anyone can use, Meta’s copying has the  potential to exponentially multiply creative expression in a way that teaching individual people  does not.}''

\subsection{Moral Rights of the Authors} 
\label{sec:ecosystem-concepts-moral-rights}
While most countries have some version of copyright laws, which can result in criminal charges, the moral rights may be covered by civil law or even only be part of social conventions. Many countries are signatories to the Berne Convention for the Protection of Literary and Artistic Works, but it is not by itself legally binding. Article 6 asserts the moral right  ``to claim authorship; to object to certain modifications and other derogatory actions''~\cite{1979_Berne_Convention_for_Protection_of_Literary_and_Artistic_Works}, while leaving the means of redress to the legislation of individual countries. For example, the U.S. is a signatory to this convention, but has no federal law to comprehensively define such means of redress, relying instead on ``a patchwork of federal and state laws, as well as industry customs and other forms of private ordering''~\cite{Temple_2019_Authors_Attribution_and_Integrity_Examining_Moral_Rights_in_United_States_Report_of_Register_of_Copyrights}. In France, on the other hand, both the moral and pecuniary rights are separate categories recognized by law~\cite{Sutherland-Smith_2016_Authorship_Ownership_and_Plagiarism_in_Digital_Age}, and the moral rights include not only the right of integrity of the work, but also the right to recall it if the author changes their point of view.  

\subsection{Authorship} \label{sec:authorship} There are multiple traditions and views on authorship, in legal, literary, and academic spheres among others. According to~\cite{Sutherland-Smith_2016_Authorship_Ownership_and_Plagiarism_in_Digital_Age}, the legal origins of intellectual property in English common law are intertwined with the Romantic notion of an author as ``the sole creator or originator of the text''. Important aspects of authorship in this sense are the \textit{individual labor of the author} and the \textit{originality} of the resulting text. In literary theory, that view has been challenged for ignoring the external influences on the author, the intertextual relations, and the role of the reader in constructing the meaning out of written words~\cite{Barthes_1988_death_of_author}. However, that discussion is about the role of original author's intent in interpretation of literary texts, rather than the economic rights of writers.

The world of non-fiction has at least one more important criterion for the notion of authorship: \textit{responsibility}. The first principle listed in the code of ethics of the Society of Professional Journalists~\cite{2014_SPJs_Code_of_Ethics} is to ``take responsibility for the accuracy of their work''. In the scientific world, the definition of authorship of the Association for Computation Machinery emphasizes not only intellectual contributions to content and drafting of the academic work, but also that all authors are aware that the work is being published, and take equal responsibility for it~\cite{ACMPublicationsBoard_2023_ACM_Policy_on_Plagiarism_Misrepresentation_and_Falsification}. 

\subsection{Plagiarism} \label{sec:plagiarism} We consider plagiarism as ``{using someone else’s intellectual product (such as texts, ideas, or results), thereby implying that it is their own''~\cite{HelgessonEriksson_2015_Plagiarism_in_research}. The notion of plagiarism relies more on the social norms and the moral rights of authors than the copyright legislation: e.g. if a student copies a part of a copyrighted textbook in their essay, this would typically be viewed in the lens of academic misconduct rather than copyright infringement. The legal profession has its own norms over copying the text of other lawyers, also viewed through the lens of professional conduct~\cite{Rendleman_2020_Copy_That_What_is_plagiarism_in_practice_of_law}. 
The Vancouver Convention lists plagiarism as a form of scientific misconduct, together with fabrication or falsification of data and failure to disclose relationships and activities \cite[][p.8]{vancouver-convension-authorship}.

As other concepts explored in this section, plagiarism is complicated due to the variety of specific norms and expectations for attribution in various spheres of human activity, which also change over time. For example, the perception of what does not need citations because everybody in a given academic field already knows this changes over time, as that field develops conceptually. \citeauthor{HavilandMullin_2009_Who_owns_this_text_plagiarism_authorship_and_disciplinary_cultures} argue that academics engage in the practice of their disciplines to ``participate in knowledge-building communities, to know and to be known in the professional arenas that matter to them'' \citet[p.157-158]{HavilandMullin_2009_Who_owns_this_text_plagiarism_authorship_and_disciplinary_cultures}. Over time they learn how the knowledge-building works in their scientific community, and they are typically careful to observe those norms, even though they are not universal or easy to formalize. 

Likewise, in journalism, it is not always clear when attribution is required~\cite{lewis2013root}, and it is not uncommon to ``pick up on ideas that have previously been published elsewhere in a similar form''. Still, the blurriness and fluidity of human attribution norms does not mean that plagiarism is not a real phenomenon. According to~\cite{meier2024plagiat}, journalistic plagiarism happens ``when inventive ideas or creative wording are copied from others systematically and deliberately and, in particular, the reader is led to believe in a creative uniqueness that simply does not exist''. 

Different spheres of activity may have different attribution expectations for copying the form or content of a certain work. In the academic realm the focus is on the content. For example, the Association for Computational Machinery specifically addresses as forms of plagiarism ``intentionally paraphrasing  portions of another's work'', and ``using automated tools that rephrase existing work as one's text without proper attribution''~\cite{ACMPublicationsBoard_2023_ACM_Policy_on_Plagiarism_Misrepresentation_and_Falsificationa}. On this definition, if LLM developers were to deal with copyright problem by detecting verbatim matches to known sources, and rephrasing to avoid detection, this would still constitute plagiarism.

%% file: main.bbl
%%% -*-BibTeX-*-
%%% Do NOT edit. File created by BibTeX with style
%%% ACM-Reference-Format-Journals [18-Jan-2012].

\begin{thebibliography}{236}

%%% ====================================================================
%%% NOTE TO THE USER: you can override these defaults by providing
%%% customized versions of any of these macros before the \bibliography
%%% command.  Each of them MUST provide its own final punctuation,
%%% except for \shownote{} and \showURL{}.  The latter two
%%% do not use final punctuation, in order to avoid confusing it with
%%% the Web address.
%%%
%%% To suppress output of a particular field, define its macro to expand
%%% to an empty string, or better, \unskip, like this:
%%%
%%% \newcommand{\showURL}[1]{\unskip}   % LaTeX syntax
%%%
%%% \def \showURL #1{\unskip}           % plain TeX syntax
%%%
%%% ====================================================================

\ifx \showCODEN    \undefined \def \showCODEN     #1{\unskip}     \fi
\ifx \showISBNx    \undefined \def \showISBNx     #1{\unskip}     \fi
\ifx \showISBNxiii \undefined \def \showISBNxiii  #1{\unskip}     \fi
\ifx \showISSN     \undefined \def \showISSN      #1{\unskip}     \fi
\ifx \showLCCN     \undefined \def \showLCCN      #1{\unskip}     \fi
\ifx \shownote     \undefined \def \shownote      #1{#1}          \fi
\ifx \showarticletitle \undefined \def \showarticletitle #1{#1}   \fi
\ifx \showURL      \undefined \def \showURL       {\relax}        \fi
% The following commands are used for tagged output and should be
% invisible to TeX
\providecommand\bibfield[2]{#2}
\providecommand\bibinfo[2]{#2}
\providecommand\natexlab[1]{#1}
\providecommand\showeprint[2][]{arXiv:#2}

\bibitem[100({[n.\,d.]})]%
        {1000_Authors_for_Libraries}
 \bibinfo{year}{[n.\,d.]}\natexlab{}.
\newblock \bibinfo{title}{1000+ {{Authors}} for {{Libraries}}}.
\newblock
\urldef\tempurl%
\url{https://www.fightforthefuture.org/authors-for-libraries}
\showURL{%
\tempurl}


\bibitem[AI_({[n.\,d.]})]%
        {AI_Licensing_for_Authors_Who_Owns_Rights_and_Whats_Fair_Split}
 \bibinfo{year}{[n.\,d.]}\natexlab{}.
\newblock \bibinfo{title}{{{AI Licensing}} for {{Authors}}: {{Who Owns}} the {{Rights}} and {{What}}'s a {{Fair Split}}?}
\newblock
\urldef\tempurl%
\url{https://authorsguild.org/news/ai-licensing-for-authors-who-owns-the-rights-and-whats-a-fair-split/}
\showURL{%
\tempurl}


\bibitem[Ant({[n.\,d.]})]%
        {Anthropic_Lawsuit_Update_Settlement_Reached_with_Pay-Out_to_Authors}
The Authors Guild \bibinfo{year}{[n.\,d.]}\natexlab{}.
\newblock \bibinfo{booktitle}{\emph{Anthropic {{Lawsuit Update}}: {{Settlement Reached}} with {{Pay-Out}} to {{Authors}}}}.
\newblock The Authors Guild.
\newblock
\urldef\tempurl%
\url{https://authorsguild.org/news/anthropic-lawsuit-update-settlement-reached-with-pay-out-to-authors/}
\showURL{%
\tempurl}


\bibitem[Cas(2025)]%
        {CaseC250_25_LikeCompanyvGoogle}
Court of Justice of the European Union \bibinfo{year}{2025}\natexlab{}.
\newblock \bibinfo{booktitle}{\emph{Case C-250/25, \emph{Like Company v. Google Ireland Limited}}}.
\newblock Court of Justice of the European Union.
\newblock
\urldef\tempurl%
\url{https://curia.europa.eu/juris/showPdf.jsf?text=&docid=300681&pageIndex=0&doclang=EN&mode=req&dir=&occ=first&part=1&cid=5661670}
\showURL{%
\tempurl}
\newblock
\shownote{Request lodged 3 April 2025; referring court: Budapest Környéki Törvényszék (Hungary); decision to refer dated 10 March 2025}.


\bibitem[202(2025)]%
        {2025_Case_C-250_25_Like_Company_Request_for_preliminary_ruling_from_Budapest_Kornyeki_Torvenyszek_Hungary_lodged_on_3_April_2025_Like_Company_Google_Ireland_Limited}
 \bibinfo{year}{2025}\natexlab{}.
\newblock \bibinfo{title}{Case {{C-250}}/25, {{Like Company}}: {{Request}} for a Preliminary Ruling from the {{Budapest Környéki Törvényszék}} ({{Hungary}}) Lodged on 3~{{April}} 2025~– {{Like Company}} v {{Google Ireland Limited}}}.
\newblock
\urldef\tempurl%
\url{http://data.europa.eu/eli/C/2025/3039/oj}
\showURL{%
\tempurl}


\bibitem[Cha(2023)]%
        {Channels_of_Communication_Part_1_Authors_Guild_Launchpad_Webinar_3}
The Authors Guild \bibinfo{year}{2023}\natexlab{}.
\newblock \bibinfo{booktitle}{\emph{Channels of {{Communication}}, {{Part}} 1 ({{Authors Guild Launchpad Webinar}} \#3)}}.
\newblock The Authors Guild.
\newblock
\urldef\tempurl%
\url{https://authorsguild.org/resource/ag-launchpad-channels-of-communication-part-1/}
\showURL{%
\tempurl}


\bibitem[Con({[n.\,d.]})]%
        {Concord_Music_Group_Inc_Anthropic_PBC}
 \bibinfo{year}{[n.\,d.]}\natexlab{}.
\newblock \bibinfo{title}{Concord {{Music Group}}, {{Inc}}. v. {{Anthropic PBC}}}.
\newblock
\urldef\tempurl%
\url{https://www.courtlistener.com/docket/72199828/concord-music-group-inc-v-anthropic-pbc/}
\showURL{%
\tempurl}


\bibitem[24_(0024)]%
        {24_Germany_Hamburg_District_Court_310_O22723_LAION_Robert_Kneschke}
EUIPO \bibinfo{year}{0024}\natexlab{}.
\newblock \bibinfo{booktitle}{\emph{Germany - {{Hamburg District Court}}, 310 {{O}}.22723, {{LAION}} v {{Robert Kneschke}}}}.
\newblock EUIPO.
\newblock
\urldef\tempurl%
\url{https://www.euipo.europa.eu/en/law/recent-case-law/germany-hamburg-district-court-310-o-22723-laion-v-robert-kneschke}
\showURL{%
\tempurl}


\bibitem[Pro({[n.\,d.]})]%
        {ProRataAI_We_build_AI-powered_search_advertising_and_attribution_solutions_grounded_in_respect_for_content}
ProRata.AI \bibinfo{year}{[n.\,d.]}\natexlab{}.
\newblock \bibinfo{booktitle}{\emph{{{ProRata}}.{{AI}} - {{We}} Build {{AI-powered}} Search, Advertising, and Attribution Solutions Grounded in Respect for Content.}}
\newblock ProRata.AI.
\newblock
\urldef\tempurl%
\url{https://prorata.ai/}
\showURL{%
\tempurl}


\bibitem[Sur({[n.\,d.]})]%
        {Survey_Reveals_90_Percent_of_Writers_Believe_Authors_Should_Be_Compensated_for_Use_of_Their_Books_in_Training_Generative_AI}
 \bibinfo{year}{[n.\,d.]}\natexlab{}.
\newblock \bibinfo{title}{Survey {{Reveals}} 90 {{Percent}} of {{Writers Believe Authors Should Be Compensated}} for the {{Use}} of {{Their Books}} in {{Training Generative AI}}}.
\newblock
\urldef\tempurl%
\url{https://authorsguild.org/news/ai-survey-90-percent-of-writers-believe-authors-should-be-compensated-for-ai-training-use/}
\showURL{%
\tempurl}


\bibitem[WGA({[n.\,d.]})]%
        {WGA_Agreement_Introduces_Key_Protections_for_TV_and_Film_Writers_Against_AI}
 \bibinfo{year}{[n.\,d.]}\natexlab{}.
\newblock \bibinfo{title}{{{WGA Agreement Introduces Key Protections}} for {{TV}} and {{Film Writers Against AI}}}.
\newblock
\urldef\tempurl%
\url{https://authorsguild.org/news/wga-agreement-introduces-key-protections-for-tv-and-film-writers-against-ai/}
\showURL{%
\tempurl}


\bibitem[197(1979)]%
        {1979_Berne_Convention_for_Protection_of_Literary_and_Artistic_Works}
 \bibinfo{year}{1979}\natexlab{}.
\newblock \showarticletitle{Berne {{Convention}} for the {{Protection}} of {{Literary}} and {{Artistic Works}}}.
\newblock  (\bibinfo{date}{sep} \bibinfo{year}{1979}).
\newblock
\urldef\tempurl%
\url{https://www.wipo.int/wipolex/en/text/283693}
\showURL{%
\tempurl}


\bibitem[201(2014)]%
        {2014_SPJs_Code_of_Ethics}
 \bibinfo{year}{2014}\natexlab{}.
\newblock \bibinfo{title}{{{SPJ}}'s {{Code}} of {{Ethics}}}.
\newblock
\urldef\tempurl%
\url{https://www.spj.org/spj-code-of-ethics/}
\showURL{%
\tempurl}


\bibitem[202(2020)]%
        {2020_What_is_intellectual_property}
 \bibinfo{year}{2020}\natexlab{}.
\newblock \bibinfo{title}{What Is Intellectual Property?}
\newblock


\bibitem[202(2022)]%
        {2022_Statement_From_Terrence_Hart_General_Counsel_Association_of_American_Publishers_on_Disinformation_in_Internet_Archive_Case_AAP}
 \bibinfo{year}{2022}\natexlab{}.
\newblock \bibinfo{title}{Statement {{From Terrence Hart}}, {{General Counsel}}, {{Association}} of {{American Publishers}} on {{Disinformation}} in {{The Internet Archive Case}} - {{AAP}}}.
\newblock
\urldef\tempurl%
\url{https://publishers.org/news/statement-from-terrence-hart-general-counsel-association-of-american-publishers-on-the-internet-archive-case/}
\showURL{%
\tempurl}


\bibitem[202(2023)]%
        {2023_Authors_Guild_John_Grisham_Jodi_Picoult_David_Baldacci_George_RR_Martin_and_13_Other_Authors_File_Class-Action_Suit_Against_OpenAI}
 \bibinfo{year}{2023}\natexlab{}.
\newblock \bibinfo{title}{The {{Authors Guild}}, {{John Grisham}}, {{Jodi Picoult}}, {{David Baldacci}}, {{George R}}.{{R}}. {{Martin}}, and 13 {{Other Authors File Class-Action Suit Against OpenAI}}}.
\newblock
\urldef\tempurl%
\url{https://authorsguild.org/news/ag-and-authors-file-class-action-suit-against-openai/}
\showURL{%
\tempurl}


\bibitem[202(2025a)]%
        {2025_How_your_data_is_used_to_improve_model_performance}
 \bibinfo{year}{2025}\natexlab{a}.
\newblock \bibinfo{title}{How Your Data Is Used to Improve Model Performance}.
\newblock
\urldef\tempurl%
\url{https://help.openai.com/en/articles/5722486-how-your-data-is-used-to-improve-model-performance}
\showURL{%
\tempurl}


\bibitem[202(2025b)]%
        {2025_IETF_working_group_will_further_develop_our_proposal_for_opt-out_vocabulary}
 \bibinfo{year}{2025}\natexlab{b}.
\newblock \bibinfo{title}{{{IETF}} Working Group Will Further Develop Our Proposal for an Opt-out Vocabulary}.
\newblock
\urldef\tempurl%
\url{https://openfuture.eu/blog/ietf-working-group-will-further-develop-our-proposal-for-an-opt-out-vocabulary}
\showURL{%
\tempurl}


\bibitem[202(2025c)]%
        {2025_Meta_Wrongfully_Disabling_Accounts_with_No_Human_Customer_Support}
 \bibinfo{year}{2025}\natexlab{c}.
\newblock \bibinfo{title}{Meta {{Wrongfully Disabling Accounts}} with {{No Human Customer Support}}}.
\newblock
\urldef\tempurl%
\url{https://www.change.org/p/meta-wrongfully-disabling-accounts-with-no-human-customer-support}
\showURL{%
\tempurl}


\bibitem[202(2025d)]%
        {2025_Microsoft_Copilot_Terms_of_Use}
 \bibinfo{year}{2025}\natexlab{d}.
\newblock \bibinfo{title}{Microsoft {{Copilot Terms}} of {{Use}}}.
\newblock
\urldef\tempurl%
\url{https://www.microsoft.com/en-gb/microsoft-copilot/for-individuals/termsofuse}
\showURL{%
\tempurl}


\bibitem[van(2025)]%
        {vancouver-convension-authorship}
 \bibinfo{year}{2025}\natexlab{}.
\newblock \bibinfo{title}{Recommendations for the {{Conduct}}, {{Reporting}}, {{Editing}}, and {{Publication}} of {{Scholarly Work}} in {{Medical Journals}}}.
\newblock
\urldef\tempurl%
\url{https://www.icmje.org/icmje-recommendations.pdf}
\showURL{%
\tempurl}


\bibitem[202(2025e)]%
        {2025_Report_on_Copyright_and_Artificial_Intelligence_Part_3_Generative_AI_Training_pre-publication_version}
 \bibinfo{year}{2025}\natexlab{e}.
\newblock \bibinfo{booktitle}{\emph{Report on {{Copyright}} and {{Artificial Intelligence}}. {{Part}} 3: {{Generative AI Training}} (Pre-Publication Version)}}.
\newblock \bibinfo{type}{{T}echnical {R}eport}. \bibinfo{institution}{U.S. Copyright Office}.
\newblock
\urldef\tempurl%
\url{https://www.copyright.gov/ai/Copyright-and-Artificial-Intelligence-Part-3-Generative-AI-Training-Report-Pre-Publication-Version.pdf}
\showURL{%
\tempurl}


\bibitem[Abdalla et~al\mbox{.}(2023)]%
        {AbdallaWahleEtAl_2023_Elephant_in_Room_Analyzing_Presence_of_Big_Tech_in_Natural_Language_Processing_Researcha}
\bibfield{author}{\bibinfo{person}{Mohamed Abdalla}, \bibinfo{person}{Jan~Philip Wahle}, \bibinfo{person}{Terry Lima~Ruas}, \bibinfo{person}{Aur\{{\textbackslash}'e\}lie N\{{\textbackslash}'e\}v\{{\textbackslash}'e\}ol}, \bibinfo{person}{Fanny Ducel}, \bibinfo{person}{Saif Mohammad}, {and} \bibinfo{person}{Karen Fort}.} \bibinfo{year}{2023}\natexlab{}.
\newblock \showarticletitle{The {{Elephant}} in the {{Room}}: {{Analyzing}} the {{Presence}} of {{Big Tech}} in {{Natural Language Processing Research}}}. In \bibinfo{booktitle}{\emph{Proceedings of the 61st {{Annual Meeting}} of the {{Association}} for {{Computational Linguistics}} ({{Volume}} 1: {{Long Papers}})}}. \bibinfo{publisher}{Association for Computational Linguistics}, \bibinfo{address}{Toronto, Canada}, \bibinfo{pages}{13141--13160}.
\newblock
\urldef\tempurl%
\url{https://aclanthology.org/2023.acl-long.734/}
\showURL{%
\tempurl}


\bibitem[Acovino(2023)]%
        {Acovino_2023_Sci-Fi_magazine_stops_submissions_after_flood_of_AI_generated_stories}
\bibfield{author}{\bibinfo{person}{Vincent Acovino}.} \bibinfo{year}{2023}\natexlab{}.
\newblock \showarticletitle{Sci-{{Fi}} Magazine Stops Submissions after Flood of {{AI}} Generated Stories}.
\newblock \bibinfo{journal}{\emph{NPR}} (\bibinfo{date}{feb} \bibinfo{year}{2023}).
\newblock
\urldef\tempurl%
\url{https://www.npr.org/2023/02/23/1159118948/sci-fi-magazine-stops-submissions-after-flood-of-ai-generated-stories}
\showURL{%
\tempurl}


\bibitem[Agarwal and Sen(2026)]%
        {AgarwalSen_2026_Google_AI_Overviews_and_Publisher_Traffic_Evidence_from_Field_Experiment}
\bibfield{author}{\bibinfo{person}{Saharsh Agarwal} {and} \bibinfo{person}{Ananya Sen}.} \bibinfo{year}{2026}\natexlab{}.
\newblock \bibinfo{booktitle}{\emph{Google {{AI Overviews}} and {{Publisher Traffic}}: {{Evidence}} from a {{Field Experiment}}}}.
\newblock
\showeprint[Social Science Research Network]{6513059}
\href{https://doi.org/10.2139/ssrn.6513059}{doi:\nolinkurl{10.2139/ssrn.6513059}}


\bibitem[Ahmed and {Haskell-Dowland}(2021)]%
        {AhmedHaskell-Dowland_2021_Is_Google_getting_worse_Increased_advertising_and_algorithm_changes_may_make_it_harder_to_find_what_youre_looking_for}
\bibfield{author}{\bibinfo{person}{Mohiuddin Ahmed} {and} \bibinfo{person}{Paul {Haskell-Dowland}}.} \bibinfo{year}{2021}\natexlab{}.
\newblock \bibinfo{title}{Is {{Google}} Getting Worse? {{Increased}} Advertising and Algorithm Changes May Make It Harder to Find What You're Looking for}.
\newblock
\href{https://doi.org/10.64628/AA.av5ws3c54}{doi:\nolinkurl{10.64628/AA.av5ws3c54}}


\bibitem[{AI watchdog}(2025)]%
        {AIwatchdog_2025_Content_Licensing_Deals}
\bibfield{author}{\bibinfo{person}{{AI watchdog}}.} \bibinfo{year}{2025}\natexlab{}.
\newblock \bibinfo{title}{Content {{Licensing Deals}}}.
\newblock
\urldef\tempurl%
\url{https://aiwatch.dog/licensing}
\showURL{%
\tempurl}


\bibitem[Akyurek et~al\mbox{.}(2022)]%
        {akyurek-etal-2022-towards}
\bibfield{author}{\bibinfo{person}{Ekin Akyurek}, \bibinfo{person}{Tolga Bolukbasi}, \bibinfo{person}{Frederick Liu}, \bibinfo{person}{Binbin Xiong}, \bibinfo{person}{Ian Tenney}, \bibinfo{person}{Jacob Andreas}, {and} \bibinfo{person}{Kelvin Guu}.} \bibinfo{year}{2022}\natexlab{}.
\newblock \showarticletitle{Towards Tracing Knowledge in Language Models Back to the Training Data}. In \bibinfo{booktitle}{\emph{Findings of the Association for Computational Linguistics: EMNLP 2022}}, \bibfield{editor}{\bibinfo{person}{Yoav Goldberg}, \bibinfo{person}{Zornitsa Kozareva}, {and} \bibinfo{person}{Yue Zhang}} (Eds.). \bibinfo{publisher}{Association for Computational Linguistics}, \bibinfo{address}{Abu Dhabi, United Arab Emirates}, \bibinfo{pages}{2429--2446}.
\newblock
\href{https://doi.org/10.18653/v1/2022.findings-emnlp.180}{doi:\nolinkurl{10.18653/v1/2022.findings-emnlp.180}}


\bibitem[Alba(2025)]%
        {Alba_2025_Google_Can_Train_Search_AI_With_Web_Content_Even_After_Opt-Out}
\bibfield{author}{\bibinfo{person}{Davey Alba}.} \bibinfo{year}{2025}\natexlab{}.
\newblock \showarticletitle{Google {{Can Train Search AI With Web Content Even After Opt-Out}}}.
\newblock \bibinfo{journal}{\emph{Bloomberg.com}} (\bibinfo{date}{may} \bibinfo{year}{2025}).
\newblock
\urldef\tempurl%
\url{https://www.bloomberg.com/news/articles/2025-05-03/google-can-train-search-ai-with-web-content-even-after-opt-out}
\showURL{%
\tempurl}


\bibitem[Alliance(2024)]%
        {DatasetProvidersAlliance_2024_Machine_Learning_AI_Data_Licensing}
\bibfield{author}{\bibinfo{person}{Dataset~Providers Alliance}.} \bibinfo{year}{2024}\natexlab{}.
\newblock \bibinfo{title}{Machine {{Learning AI Data Licensing}}}.
\newblock
\urldef\tempurl%
\url{https://www.thedpa.ai}
\showURL{%
\tempurl}


\bibitem[Altman(2026)]%
        {Altman_2026_Our_principles}
\bibfield{author}{\bibinfo{person}{Sam Altman}.} \bibinfo{year}{2026}\natexlab{}.
\newblock \bibinfo{booktitle}{\emph{Our Principles}}.
\newblock OpenAI.
\newblock
\urldef\tempurl%
\url{https://openai.com/index/our-principles/}
\showURL{%
\tempurl}


\bibitem[Anthony et~al\mbox{.}(2009)]%
        {AnthonySmithEtAl_2009_Reputation_and_Reliability_in_Collective_Goods_Case_of_Online_Encyclopedia_Wikipedia}
\bibfield{author}{\bibinfo{person}{Denise Anthony}, \bibinfo{person}{Sean~W. Smith}, {and} \bibinfo{person}{Timothy Williamson}.} \bibinfo{year}{2009}\natexlab{}.
\newblock \showarticletitle{Reputation and {{Reliability}} in {{Collective Goods}}: {{The Case}} of the {{Online Encyclopedia Wikipedia}}}.
\newblock \bibinfo{journal}{\emph{Rationality and Society}} \bibinfo{volume}{21}, \bibinfo{number}{3} (\bibinfo{date}{aug} \bibinfo{year}{2009}), \bibinfo{pages}{283--306}.
\newblock
\showISSN{1043-4631}
\href{https://doi.org/10.1177/1043463109336804}{doi:\nolinkurl{10.1177/1043463109336804}}


\bibitem[Arrieta~Ibarra et~al\mbox{.}(2017)]%
        {ArrietaIbarraGoffEtAl_2017_Should_We_Treat_Data_as_Labor_Moving_Beyond_Free}
\bibfield{author}{\bibinfo{person}{Imanol Arrieta~Ibarra}, \bibinfo{person}{Leonard Goff}, \bibinfo{person}{Diego Jim{\'e}nez~Hern{\'a}ndez}, \bibinfo{person}{Jaron Lanier}, {and} \bibinfo{person}{E.~Glen Weyl}.} \bibinfo{year}{2017}\natexlab{}.
\newblock \bibinfo{title}{Should {{We Treat Data}} as {{Labor}}? {{Moving Beyond}} '{{Free}}'}.
\newblock
\showeprint[social science research network]{3093683}
\urldef\tempurl%
\url{https://papers.ssrn.com/abstract=3093683}
\showURL{%
\tempurl}


\bibitem[Atmakuri et~al\mbox{.}(2025)]%
        {AtmakuriSinghEtAl_2025_Making_AI_citations_count_with_Asta}
\bibfield{author}{\bibinfo{person}{Shriya Atmakuri}, \bibinfo{person}{Amanpreet Singh}, {and} \bibinfo{person}{Doug Downey}.} \bibinfo{year}{2025}\natexlab{}.
\newblock \bibinfo{booktitle}{\emph{Making {{AI}} Citations Count with {{Asta}}}}.
\newblock
\urldef\tempurl%
\url{https://allenai.org/blog/asta-citations}
\showURL{%
\tempurl}


\bibitem[Azcoitia et~al\mbox{.}(2023)]%
        {AzcoitiaIordanouEtAl_2023_Understanding_Price_of_Data_in_Commercial_Data_Marketplaces}
\bibfield{author}{\bibinfo{person}{Santiago~Andr{\'e}s Azcoitia}, \bibinfo{person}{Costas Iordanou}, {and} \bibinfo{person}{Nikolaos Laoutaris}.} \bibinfo{year}{2023}\natexlab{}.
\newblock \showarticletitle{Understanding the {{Price}} of {{Data}} in {{Commercial Data Marketplaces}}}. In \bibinfo{booktitle}{\emph{2023 {{IEEE}} 39th {{International Conference}} on {{Data Engineering}} ({{ICDE}})}}. \bibinfo{pages}{3718--3728}.
\newblock
\showISSN{2375-026X}
\href{https://doi.org/10.1109/ICDE55515.2023.00300}{doi:\nolinkurl{10.1109/ICDE55515.2023.00300}}


\bibitem[Bae et~al\mbox{.}(2022)]%
        {bae2022influencefunctionsanswerquestion}
\bibfield{author}{\bibinfo{person}{Juhan Bae}, \bibinfo{person}{Nathan Ng}, \bibinfo{person}{Alston Lo}, \bibinfo{person}{Marzyeh Ghassemi}, {and} \bibinfo{person}{Roger Grosse}.} \bibinfo{year}{2022}\natexlab{}.
\newblock \bibinfo{title}{If Influence Functions are the Answer, Then What is the Question?}
\newblock
\showeprint[arxiv]{2209.05364}~[cs.LG]
\urldef\tempurl%
\url{https://arxiv.org/abs/2209.05364}
\showURL{%
\tempurl}


\bibitem[Baio(2022)]%
        {Baio_2022_AI_Data_Laundering_How_Academic_and_Nonprofit_Researchers_Shield_Tech_Companies_from_Accountability}
\bibfield{author}{\bibinfo{person}{Andy Baio}.} \bibinfo{year}{2022}\natexlab{}.
\newblock \bibinfo{title}{{{AI Data Laundering}}: {{How Academic}} and {{Nonprofit Researchers Shield Tech Companies}} from {{Accountability}}}.
\newblock
\urldef\tempurl%
\url{https://waxy.org/2022/09/ai-data-laundering-how-academic-and-nonprofit-researchers-shield-tech-companies-from-accountability/}
\showURL{%
\tempurl}


\bibitem[Bandy and Vincent(2021)]%
        {BandyVincent_2021_Addressing_Documentation_Debt_in_Machine_Learning_Research_Retrospective_Datasheet_for_BookCorpus}
\bibfield{author}{\bibinfo{person}{Jack Bandy} {and} \bibinfo{person}{Nicholas Vincent}.} \bibinfo{year}{2021}\natexlab{}.
\newblock \showarticletitle{Addressing "{{Documentation Debt}}" in {{Machine Learning Research}}: {{A Retrospective Datasheet}} for {{BookCorpus}}}.
\newblock \bibinfo{journal}{\emph{arXiv:2105.05241 [cs]}} (\bibinfo{date}{may} \bibinfo{year}{2021}).
\newblock
\showeprint[arxiv]{2105.05241}~[cs]
\urldef\tempurl%
\url{http://arxiv.org/abs/2105.05241}
\showURL{%
\tempurl}


\bibitem[Barrett(2026)]%
        {Barrett_2026_US_Invaded_Venezuela_and_Captured_Nicolas_Maduro_ChatGPT_Disagrees}
\bibfield{author}{\bibinfo{person}{Brian Barrett}.} \bibinfo{year}{2026}\natexlab{}.
\newblock \showarticletitle{The {{US Invaded Venezuela}} and {{Captured Nicol\'as Maduro}}. {{ChatGPT Disagrees}}}.
\newblock \bibinfo{journal}{\emph{Wired}} (\bibinfo{date}{Jan.} \bibinfo{year}{2026}).
\newblock
\showISSN{1059-1028}
\urldef\tempurl%
\url{https://www.wired.com/story/us-invaded-venezuela-and-captured-nicolas-maduro-chatgpt-disagrees/}
\showURL{%
\tempurl}


\bibitem[Barthes(1988)]%
        {Barthes_1988_death_of_author}
\bibfield{author}{\bibinfo{person}{Roland Barthes}.} \bibinfo{year}{1988}\natexlab{}.
\newblock \showarticletitle{The Death of the Author}.
\newblock In \bibinfo{booktitle}{\emph{Image, Music, Text}}, \bibfield{editor}{\bibinfo{person}{Stephen Heath}} (Ed.). \bibinfo{publisher}{Noonday Press}, \bibinfo{pages}{142--148}.
\newblock
\urldef\tempurl%
\url{https://archive.org/details/imagemusictext0000bart\_e3d9/page/n7/mode/2up}
\showURL{%
\tempurl}


\bibitem[BehnamGhader et~al\mbox{.}(2024)]%
        {behnamghader2024llmvec}
\bibfield{author}{\bibinfo{person}{Parishad BehnamGhader}, \bibinfo{person}{Vaibhav Adlakha}, \bibinfo{person}{Marius Mosbach}, \bibinfo{person}{Dzmitry Bahdanau}, \bibinfo{person}{Nicolas Chapados}, {and} \bibinfo{person}{Siva Reddy}.} \bibinfo{year}{2024}\natexlab{}.
\newblock \showarticletitle{{LLM}2Vec: Large Language Models Are Secretly Powerful Text Encoders}. In \bibinfo{booktitle}{\emph{First Conference on Language Modeling}}.
\newblock
\urldef\tempurl%
\url{https://openreview.net/forum?id=IW1PR7vEBf}
\showURL{%
\tempurl}


\bibitem[Belanger(2023)]%
        {Belanger_2023_Grisham_Martin_join_authors_suing_OpenAI_There_is_nothing_fair_about_this_Updated}
\bibfield{author}{\bibinfo{person}{Ashley Belanger}.} \bibinfo{year}{2023}\natexlab{}.
\newblock \bibinfo{title}{Grisham, {{Martin}} Join Authors Suing {{OpenAI}}: ``{{There}} Is Nothing Fair about This'' [{{Updated}}]}.
\newblock
\urldef\tempurl%
\url{https://arstechnica.com/tech-policy/2023/09/george-r-r-martin-joins-authors-suing-openai-over-copyright-infringement/}
\showURL{%
\tempurl}


\bibitem[Belanger(2025a)]%
        {Belanger_2025_Lawsuit_Reddit_caught_Perplexity_red-handed_stealing_data_from_Google_results}
\bibfield{author}{\bibinfo{person}{Ashley Belanger}.} \bibinfo{year}{2025}\natexlab{a}.
\newblock \bibinfo{title}{Lawsuit: {{Reddit}} Caught {{Perplexity}} ``Red-Handed'' Stealing Data from {{Google}} Results}.
\newblock
\urldef\tempurl%
\url{https://arstechnica.com/tech-policy/2025/10/reddit-sues-to-block-perplexity-from-scraping-google-search-results/}
\showURL{%
\tempurl}


\bibitem[Belanger(2025b)]%
        {Belanger_2025_OpenAI_declares_AI_race_over_if_training_on_copyrighted_works_isnt_fair_use}
\bibfield{author}{\bibinfo{person}{Ashley Belanger}.} \bibinfo{year}{2025}\natexlab{b}.
\newblock \bibinfo{title}{{{OpenAI}} Declares {{AI}} Race ``over'' If Training on Copyrighted Works Isn't Fair Use}.
\newblock
\urldef\tempurl%
\url{https://arstechnica.com/tech-policy/2025/03/openai-urges-trump-either-settle-ai-copyright-debate-or-lose-ai-race-to-china/}
\showURL{%
\tempurl}


\bibitem[Bender and Friedman(2018)]%
        {BenderFriedman_2018_Data_Statements_for_Natural_Language_Processing_Toward_Mitigating_System_Bias_and_Enabling_Better_Science}
\bibfield{author}{\bibinfo{person}{Emily~M. Bender} {and} \bibinfo{person}{Batya Friedman}.} \bibinfo{year}{2018}\natexlab{}.
\newblock \showarticletitle{Data {{Statements}} for {{Natural Language Processing}}: {{Toward Mitigating System Bias}} and {{Enabling Better Science}}}.
\newblock \bibinfo{journal}{\emph{Transactions of the Association for Computational Linguistics}}  \bibinfo{volume}{6} (\bibinfo{year}{2018}), \bibinfo{pages}{587--604}.
\newblock
\href{https://doi.org/10.1162/tacl\_a\_00041}{doi:\nolinkurl{10.1162/tacl\_a\_00041}}


\bibitem[Biderman et~al\mbox{.}(2023)]%
        {BidermanSchoelkopfEtAl_2023_Pythia_Suite_for_Analyzing_Large_Language_Models_Across_Training_and_Scaling}
\bibfield{author}{\bibinfo{person}{Stella Biderman}, \bibinfo{person}{Hailey Schoelkopf}, \bibinfo{person}{Quentin~Gregory Anthony}, \bibinfo{person}{Herbie Bradley}, \bibinfo{person}{Kyle O'Brien}, \bibinfo{person}{Eric Hallahan}, \bibinfo{person}{Mohammad~Aflah Khan}, \bibinfo{person}{Shivanshu Purohit}, \bibinfo{person}{Usvsn~Sai Prashanth}, \bibinfo{person}{Edward Raff}, \bibinfo{person}{Aviya Skowron}, \bibinfo{person}{Lintang Sutawika}, {and} \bibinfo{person}{Oskar Van~Der Wal}.} \bibinfo{year}{2023}\natexlab{}.
\newblock \showarticletitle{Pythia: {{A Suite}} for {{Analyzing Large Language Models Across Training}} and {{Scaling}}}. In \bibinfo{booktitle}{\emph{Proceedings of the 40th {{International Conference}} on {{Machine Learning}}}}. \bibinfo{publisher}{PMLR}, \bibinfo{pages}{2397--2430}.
\newblock
\showISSN{2640-3498}
\urldef\tempurl%
\url{https://proceedings.mlr.press/v202/biderman23a.html}
\showURL{%
\tempurl}


\bibitem[Board(2023a)]%
        {ACMPublicationsBoard_2023_ACM_Policy_on_Plagiarism_Misrepresentation_and_Falsification}
\bibfield{author}{\bibinfo{person}{ACM~Publications Board}.} \bibinfo{year}{2023}\natexlab{a}.
\newblock \bibinfo{title}{{{ACM Policy}} on {{Plagiarism}}, {{Misrepresentation}}, and {{Falsification}}}.
\newblock
\urldef\tempurl%
\url{https://www.acm.org/publications/policies/plagiarism-overview}
\showURL{%
\tempurl}


\bibitem[Board(2023b)]%
        {ACMPublicationsBoard_2023_ACM_Policy_on_Plagiarism_Misrepresentation_and_Falsificationa}
\bibfield{author}{\bibinfo{person}{ACM~Publications Board}.} \bibinfo{year}{2023}\natexlab{b}.
\newblock \bibinfo{title}{{{ACM Policy}} on {{Plagiarism}}, {{Misrepresentation}}, and {{Falsification}}}.
\newblock
\urldef\tempurl%
\url{https://www.acm.org/publications/policies/plagiarism-overview}
\showURL{%
\tempurl}


\bibitem[Bort(2025)]%
        {Bort_2025_Perplexity_CEO_says_its_browser_will_track_everything_users_do_online_to_sell_hyper_personalized_ads}
\bibfield{author}{\bibinfo{person}{Julie Bort}.} \bibinfo{year}{2025}\natexlab{}.
\newblock \bibinfo{title}{Perplexity {{CEO}} Says Its Browser Will Track Everything Users Do Online to Sell 'hyper Personalized' Ads}.
\newblock
\urldef\tempurl%
\url{https://techcrunch.com/2025/04/24/perplexity-ceo-says-its-browser-will-track-everything-users-do-online-to-sell-hyper-personalized-ads/}
\showURL{%
\tempurl}


\bibitem[Brandom(2025)]%
        {Brandom_2025_RSS_co-creator_launches_new_protocol_for_AI_data_licensinga}
\bibfield{author}{\bibinfo{person}{Russell Brandom}.} \bibinfo{year}{2025}\natexlab{}.
\newblock \bibinfo{title}{{{RSS}} Co-Creator Launches New Protocol for {{AI}} Data Licensing}.
\newblock
\urldef\tempurl%
\url{https://techcrunch.com/2025/09/10/rss-co-creator-launches-new-protocol-for-ai-data-licensing/}
\showURL{%
\tempurl}


\bibitem[Brin and Page(1998)]%
        {BrinPage_1998_anatomy_of_large-scale_hypertextual_Web_search_engine}
\bibfield{author}{\bibinfo{person}{Sergey Brin} {and} \bibinfo{person}{Lawrence Page}.} \bibinfo{year}{1998}\natexlab{}.
\newblock \showarticletitle{The Anatomy of a Large-Scale Hypertextual {{Web}} Search Engine}.
\newblock  \bibinfo{volume}{30}, \bibinfo{number}{1} (\bibinfo{year}{1998}), \bibinfo{pages}{107--117}.
\newblock
\showISSN{0169-7552}
\href{https://doi.org/10.1016/S0169-7552(98)00110-X}{doi:\nolinkurl{10.1016/S0169-7552(98)00110-X}}


\bibitem[Brooks(2020)]%
        {Brooks_2020_Dilemma_of_Free_Facebooks_Monopsony_Power_and_Need_For_Antitrust_Renaissance}
\bibfield{author}{\bibinfo{person}{John Brooks}.} \bibinfo{year}{2020}\natexlab{}.
\newblock \bibinfo{title}{The {{Dilemma}} of '{{Free}}': {{Facebook}}'s {{Monopsony Power}} and the {{Need For}} an {{Antitrust Renaissance}}}.
\newblock
\showeprint[social science research network]{3531172}
\href{https://doi.org/10.2139/ssrn.3531172}{doi:\nolinkurl{10.2139/ssrn.3531172}}


\bibitem[Brown(2020)]%
        {Brown_2020_Should_Stay_or_Should_Leave_Exploring_Discontinued_Facebook_Use_After_Cambridge_Analytica_Scandal}
\bibfield{author}{\bibinfo{person}{Allison~J. Brown}.} \bibinfo{year}{2020}\natexlab{}.
\newblock \showarticletitle{``{{Should I Stay}} or {{Should I Leave}}?'': {{Exploring}} ({{Dis}})Continued {{Facebook Use After}} the {{Cambridge Analytica Scandal}}}.
\newblock \bibinfo{journal}{\emph{Social Media + Society}} \bibinfo{volume}{6}, \bibinfo{number}{1} (\bibinfo{date}{jan} \bibinfo{year}{2020}), \bibinfo{pages}{2056305120913884}.
\newblock
\showISSN{2056-3051}
\href{https://doi.org/10.1177/2056305120913884}{doi:\nolinkurl{10.1177/2056305120913884}}


\bibitem[Bruckman(2002)]%
        {Bruckman_2002_Studying_amateur_artist_perspective_on_disguising_data_collected_in_human_subjects_research_on_Internet}
\bibfield{author}{\bibinfo{person}{Amy Bruckman}.} \bibinfo{year}{2002}\natexlab{}.
\newblock \showarticletitle{Studying the Amateur Artist: {{A}} Perspective on Disguising Data Collected in Human Subjects Research on the {{Internet}}}.
\newblock \bibinfo{journal}{\emph{Ethics and Information Technology}} \bibinfo{volume}{4}, \bibinfo{number}{3} (\bibinfo{date}{Sept.} \bibinfo{year}{2002}), \bibinfo{pages}{217--231}.
\newblock
\showISSN{1572-8439}
\href{https://doi.org/10.1023/A:1021316409277}{doi:\nolinkurl{10.1023/A:1021316409277}}


\bibitem[Burkhardt and Rieder(2024)]%
        {BurkhardtRieder_2024_Foundation_models_are_platform_models_Prompting_and_political_economy_of_AI}
\bibfield{author}{\bibinfo{person}{Sarah Burkhardt} {and} \bibinfo{person}{Bernhard Rieder}.} \bibinfo{year}{2024}\natexlab{}.
\newblock \showarticletitle{Foundation Models Are Platform Models: {{Prompting}} and the Political Economy of {{AI}}}.
\newblock  \bibinfo{volume}{11}, \bibinfo{number}{2} (\bibinfo{year}{2024}), \bibinfo{pages}{20539517241247839}.
\newblock
\showISSN{2053-9517}
\href{https://doi.org/10.1177/20539517241247839}{doi:\nolinkurl{10.1177/20539517241247839}}


\bibitem[Campbell(2025)]%
        {Campbell_2025_Perplexity_has_cooked_up_new_way_to_pay_publishers_for_their_content}
\bibfield{author}{\bibinfo{person}{Ian~Carlos Campbell}.} \bibinfo{year}{2025}\natexlab{}.
\newblock \bibinfo{title}{Perplexity Has Cooked up a New Way to Pay Publishers for Their Content}.
\newblock
\urldef\tempurl%
\url{https://www.engadget.com/ai/perplexity-has-cooked-up-a-new-way-to-pay-publishers-for-their-content-204255019.html}
\showURL{%
\tempurl}


\bibitem[Capoot(2026)]%
        {Capoot_2026_OpenAI_ads_pilot_tops_100_million_in_annualized_revenue_in_under_2_months}
\bibfield{author}{\bibinfo{person}{Ashley Capoot}.} \bibinfo{year}{2026}\natexlab{}.
\newblock \bibinfo{booktitle}{\emph{{{OpenAI}} Ads Pilot Tops \$100 Million in Annualized Revenue in under 2 Months}}.
\newblock CNBC.
\newblock
\urldef\tempurl%
\url{https://www.cnbc.com/2026/03/26/openai-ads-pilot-tops-100-million-in-arr-in-under-2-months.html}
\showURL{%
\tempurl}


\bibitem[Carlini et~al\mbox{.}(2022)]%
        {carlini2022quantifying}
\bibfield{author}{\bibinfo{person}{Nicholas Carlini}, \bibinfo{person}{Daphne Ippolito}, \bibinfo{person}{Matthew Jagielski}, \bibinfo{person}{Katherine Lee}, \bibinfo{person}{Florian Tramer}, {and} \bibinfo{person}{Chiyuan Zhang}.} \bibinfo{year}{2022}\natexlab{}.
\newblock \showarticletitle{Quantifying memorization across neural language models}. In \bibinfo{booktitle}{\emph{The Eleventh International Conference on Learning Representations}}.
\newblock


\bibitem[CBC(2024)]%
        {CBC_2024_Australia_wants_to_make_digital_platforms_pay_for_news_even_if_they_block_it_like_Meta_did_here}
\bibfield{author}{\bibinfo{person}{CBC}.} \bibinfo{year}{2024}\natexlab{}.
\newblock \showarticletitle{Australia Wants to Make Digital Platforms Pay for News — Even If They Block It, like {{Meta}} Did Here}.
\newblock  (\bibinfo{year}{2024}).
\newblock
\urldef\tempurl%
\url{https://www.cbc.ca/news/world/australia-social-media-ban-1.7408426}
\showURL{%
\tempurl}


\bibitem[Chang et~al\mbox{.}(2024)]%
        {changScalableInfluenceFact2024}
\bibfield{author}{\bibinfo{person}{Tyler~A. Chang}, \bibinfo{person}{Dheeraj Rajagopal}, \bibinfo{person}{Tolga Bolukbasi}, \bibinfo{person}{Lucas Dixon}, {and} \bibinfo{person}{Ian Tenney}.} \bibinfo{year}{2024}\natexlab{}.
\newblock \showarticletitle{Scalable {{Influence}} and {{Fact Tracing}} for {{Large Language Model Pretraining}}}. In \bibinfo{booktitle}{\emph{The {{Thirteenth International Conference}} on {{Learning Representations}}}}.
\newblock


\bibitem[Chatterji et~al\mbox{.}(2025)]%
        {ChatterjiCunninghamEtAl_2025_How_People_Use_ChatGPT}
\bibfield{author}{\bibinfo{person}{Aaron Chatterji}, \bibinfo{person}{Thomas Cunningham}, \bibinfo{person}{David~J Deming}, \bibinfo{person}{Zoe Hitzig}, \bibinfo{person}{Christopher Ong}, \bibinfo{person}{Carl~Yan Shan}, {and} \bibinfo{person}{Kevin Wadman}.} \bibinfo{year}{2025}\natexlab{}.
\newblock \showarticletitle{How {{People Use ChatGPT}}}.
\newblock \bibinfo{journal}{\emph{National Bureau of Economic Research}}  \bibinfo{volume}{34255} (\bibinfo{date}{Sept.} \bibinfo{year}{2025}).
\newblock
\urldef\tempurl%
\url{http://www.nber.org/papers/w34255}
\showURL{%
\tempurl}


\bibitem[Chhabria(2025)]%
        {Chhabria_2025_ORDER_Denying_Plaintiffs_482_Motion_for_Partial_Summary_Judgment_and_Granting_Metas_501_Cross-Motion_for_Partial_Summary_Judgment}
\bibfield{author}{\bibinfo{person}{Vince Chhabria}.} \bibinfo{year}{2025}\natexlab{}.
\newblock \bibinfo{title}{{{ORDER Denying}} the {{Plaintiffs}}' 482 {{Motion}} for {{Partial Summary Judgment}} and {{Granting Meta}}'s 501 {{Cross-Motion}} for {{Partial Summary Judgment}}.}
\newblock
\urldef\tempurl%
\url{https://storage.courtlistener.com/recap/gov.uscourts.cand.415175/gov.uscourts.cand.415175.598.0_2.pdf}
\showURL{%
\tempurl}


\bibitem[Choe et~al\mbox{.}(2024)]%
        {ChoeAhnEtAl_2024_What_is_Your_Data_Worth_to_GPT_LLM-Scale_Data_Valuation_with_Influence_Functions}
\bibfield{author}{\bibinfo{person}{Sang~Keun Choe}, \bibinfo{person}{Hwijeen Ahn}, \bibinfo{person}{Juhan Bae}, \bibinfo{person}{Kewen Zhao}, \bibinfo{person}{Minsoo Kang}, \bibinfo{person}{Youngseog Chung}, \bibinfo{person}{Adithya Pratapa}, \bibinfo{person}{Willie Neiswanger}, \bibinfo{person}{Emma Strubell}, \bibinfo{person}{Teruko Mitamura}, \bibinfo{person}{Jeff Schneider}, \bibinfo{person}{Eduard Hovy}, \bibinfo{person}{Roger Grosse}, {and} \bibinfo{person}{Eric Xing}.} \bibinfo{year}{2024}\natexlab{}.
\newblock \bibinfo{title}{What Is {{Your Data Worth}} to {{GPT}}? {{LLM-Scale Data Valuation}} with {{Influence Functions}}}.
\newblock
\showeprint[arxiv]{2405.13954}~[cs]
\href{https://doi.org/10.48550/arXiv.2405.13954}{doi:\nolinkurl{10.48550/arXiv.2405.13954}}


\bibitem[Clark et~al\mbox{.}(2025)]%
        {clark2025epistemicalignmentmediatingframework}
\bibfield{author}{\bibinfo{person}{Nicholas Clark}, \bibinfo{person}{Hua Shen}, \bibinfo{person}{Bill Howe}, {and} \bibinfo{person}{Tanushree Mitra}.} \bibinfo{year}{2025}\natexlab{}.
\newblock \bibinfo{title}{Epistemic Alignment: A Mediating Framework for User-LLM Knowledge Delivery}.
\newblock
\showeprint[arxiv]{2504.01205}~[cs.HC]
\urldef\tempurl%
\url{https://arxiv.org/abs/2504.01205}
\showURL{%
\tempurl}


\bibitem[Colangelo(2022)]%
        {Colangelo_2022_Enforcing_copyright_through_antitrust_strange_case_of_news_publishers_against_digital_platforms}
\bibfield{author}{\bibinfo{person}{Giuseppe Colangelo}.} \bibinfo{year}{2022}\natexlab{}.
\newblock \showarticletitle{Enforcing Copyright through Antitrust? {{The}} Strange Case of News Publishers against Digital Platforms}.
\newblock \bibinfo{journal}{\emph{Journal of Antitrust Enforcement}} \bibinfo{volume}{10}, \bibinfo{number}{1} (\bibinfo{date}{mar} \bibinfo{year}{2022}), \bibinfo{pages}{133--161}.
\newblock
\showISSN{2050-0696}
\href{https://doi.org/10.1093/jaenfo/jnab009}{doi:\nolinkurl{10.1093/jaenfo/jnab009}}


\bibitem[Commission(2025)]%
        {EuropeanCommission_2025_Explanatory_Notice_and_Template_for_Public_Summary_of_Training_Content_for_general-purpose_AI_models_Shaping_Europes_digital_future}
\bibfield{author}{\bibinfo{person}{European Commission}.} \bibinfo{year}{2025}\natexlab{}.
\newblock \bibinfo{title}{Explanatory {{Notice}} and {{Template}} for the {{Public Summary}} of {{Training Content}} for General-Purpose {{AI}} Models \textbar{} {{Shaping Europe}}'s Digital Future}.
\newblock
\urldef\tempurl%
\url{https://digital-strategy.ec.europa.eu/en/library/explanatory-notice-and-template-public-summary-training-content-general-purpose-ai-models}
\showURL{%
\tempurl}


\bibitem[Concord Music~Group(2026)]%
        {ConcordMusicGroupInc.CapitolCMGInc._2026_COMPLAINT_and_Demand_for_Jury_Trial_against_All_Defendants}
\bibfield{author}{\bibinfo{person}{Capitol CMG~Inc. Concord Music~Group, Inc.}} \bibinfo{year}{2026}\natexlab{}.
\newblock \bibinfo{title}{{{COMPLAINT}} and {{Demand}} for {{Jury Trial}} against {{All Defendants}}}.
\newblock
\urldef\tempurl%
\url{https://storage.courtlistener.com/recap/gov.uscourts.cand.463362/gov.uscourts.cand.463362.1.0.pdf}
\showURL{%
\tempurl}


\bibitem[Cooper et~al\mbox{.}(2025)]%
        {CooperGokaslanEtAl_2025_Extracting_memorized_pieces_of_copyrighted_books_from_open-weight_language_models}
\bibfield{author}{\bibinfo{person}{A.~Feder Cooper}, \bibinfo{person}{Aaron Gokaslan}, \bibinfo{person}{Amy~B. Cyphert}, \bibinfo{person}{Christopher~De Sa}, \bibinfo{person}{Mark~A. Lemley}, \bibinfo{person}{Daniel~E. Ho}, {and} \bibinfo{person}{Percy Liang}.} \bibinfo{year}{2025}\natexlab{}.
\newblock \bibinfo{title}{Extracting Memorized Pieces of (Copyrighted) Books from Open-Weight Language Models}.
\newblock
\showeprint[arxiv]{2505.12546}~[cs]
\href{https://doi.org/10.48550/arXiv.2505.12546}{doi:\nolinkurl{10.48550/arXiv.2505.12546}}


\bibitem[Crider(2025)]%
        {Crider_2025_Microsoft_follows_Google_with_price_bump_forced_AI_365_bundles_PCWorld}
\bibfield{author}{\bibinfo{person}{Michael Crider}.} \bibinfo{year}{2025}\natexlab{}.
\newblock \showarticletitle{Microsoft Follows {{Google}} with Price Bump, Forced {{AI}} 365 Bundles \textbar{} {{PCWorld}}}.
\newblock \bibinfo{journal}{\emph{PCWorld}} (\bibinfo{date}{jan} \bibinfo{year}{2025}).
\newblock
\urldef\tempurl%
\url{https://www.pcworld.com/article/2581179/microsoft-follows-google-with-price-bump-forced-ai-365-bundles.html}
\showURL{%
\tempurl}


\bibitem[{Danish Press Council and Media Organizations (DPCMO)}(2026)]%
        {dpcmo_tollbit_2026}
\bibfield{author}{\bibinfo{person}{{Danish Press Council and Media Organizations (DPCMO)}}.} \bibinfo{year}{2026}\natexlab{}.
\newblock \bibinfo{title}{DPCMO Partners with TollBit to Offer Solutions to Support its Members}.
\newblock \bibinfo{howpublished}{\url{https://dpcmo.dk/dpcmo-partners-with-tollbit-to-offer-solutions-to-support-its-members/}}.
\newblock


\bibitem[David(2024)]%
        {David_2024_OpenAIs_news_publisher_deals_reportedly_top_out_at_5_million_year}
\bibfield{author}{\bibinfo{person}{Emilia David}.} \bibinfo{year}{2024}\natexlab{}.
\newblock \bibinfo{title}{{{OpenAI}}'s News Publisher Deals Reportedly Top out at \$5 Million a Year}.
\newblock
\urldef\tempurl%
\url{https://www.theverge.com/2024/1/4/24025409/openai-training-data-lowball-nyt-ai-copyright}
\showURL{%
\tempurl}


\bibitem[de~la Merced and Kaye(2025)]%
        {MercedKaye_2025_Exclusive_OpenAI_Secures_Another_Giant_Funding_Deal}
\bibfield{author}{\bibinfo{person}{Andrew Ross SorkinBernhard WarnerSarah KesslerMichael~J. de~la Merced} {and} \bibinfo{person}{Danielle Kaye}.} \bibinfo{year}{2025}\natexlab{}.
\newblock \showarticletitle{Exclusive: {{OpenAI Secures Another Giant Funding Deal}}}.
\newblock \bibinfo{journal}{\emph{The New York Times}} (\bibinfo{date}{aug} \bibinfo{year}{2025}).
\newblock
\showISSN{0362-4331}
\urldef\tempurl%
\url{https://www.nytimes.com/2025/08/01/business/dealbook/openai-ai-mega-funding-deal.html}
\showURL{%
\tempurl}


\bibitem[Deng et~al\mbox{.}(2024)]%
        {deng-etal-2024-unveiling}
\bibfield{author}{\bibinfo{person}{Chunyuan Deng}, \bibinfo{person}{Yilun Zhao}, \bibinfo{person}{Yuzhao Heng}, \bibinfo{person}{Yitong Li}, \bibinfo{person}{Jiannan Cao}, \bibinfo{person}{Xiangru Tang}, {and} \bibinfo{person}{Arman Cohan}.} \bibinfo{year}{2024}\natexlab{}.
\newblock \showarticletitle{Unveiling the Spectrum of Data Contamination in Language Model: A Survey from Detection to Remediation}. In \bibinfo{booktitle}{\emph{Findings of the Association for Computational Linguistics: ACL 2024}}, \bibfield{editor}{\bibinfo{person}{Lun-Wei Ku}, \bibinfo{person}{Andre Martins}, {and} \bibinfo{person}{Vivek Srikumar}} (Eds.). \bibinfo{publisher}{Association for Computational Linguistics}, \bibinfo{address}{Bangkok, Thailand}, \bibinfo{pages}{16078--16092}.
\newblock
\href{https://doi.org/10.18653/v1/2024.findings-acl.951}{doi:\nolinkurl{10.18653/v1/2024.findings-acl.951}}


\bibitem[Deng et~al\mbox{.}(2025)]%
        {dengSurveyDataAttribution2025}
\bibfield{author}{\bibinfo{person}{Junwei Deng}, \bibinfo{person}{Yuzheng Hu}, \bibinfo{person}{Pingbang Hu}, \bibinfo{person}{Ting-wei Li}, \bibinfo{person}{Shixuan Liu}, \bibinfo{person}{Jiachen~T. Wang}, \bibinfo{person}{Dan Ley}, \bibinfo{person}{Qirun Dai}, \bibinfo{person}{Benhao Huang}, \bibinfo{person}{Jin Huang}, \bibinfo{person}{Cathy Jiao}, \bibinfo{person}{Hoang~Anh Just}, \bibinfo{person}{Yijun Pan}, \bibinfo{person}{Jingyan Shen}, \bibinfo{person}{Yiwen Tu}, \bibinfo{person}{Weiyi Wang}, \bibinfo{person}{Xinhe Wang}, \bibinfo{person}{Shichang Zhang}, \bibinfo{person}{Shiyuan Zhang}, \bibinfo{person}{Ruoxi Jia}, \bibinfo{person}{Himabindu Lakkaraju}, \bibinfo{person}{Hao Peng}, \bibinfo{person}{Weijing Tang}, \bibinfo{person}{Chenyan Xiong}, \bibinfo{person}{Jieyu Zhao}, \bibinfo{person}{Hanghang Tong}, \bibinfo{person}{Han Zhao}, {and} \bibinfo{person}{Jiaqi~W. Ma}.} \bibinfo{year}{2025}\natexlab{}.
\newblock \bibinfo{title}{A {{Survey}} of {{Data Attribution}}: {{Methods}}, {{Applications}}, and {{Evaluation}} in the {{Era}} of {{Generative AI}}}.
\newblock
\showeprint[social science research network]{5451054}
\href{https://doi.org/10.2139/ssrn.5451054}{doi:\nolinkurl{10.2139/ssrn.5451054}}


\bibitem[Devlin et~al\mbox{.}(2019)]%
        {DevlinChangEtAl_2019_BERT_Pre-training_of_Deep_Bidirectional_Transformers_for_Language_Understanding}
\bibfield{author}{\bibinfo{person}{Jacob Devlin}, \bibinfo{person}{Ming-Wei Chang}, \bibinfo{person}{Kenton Lee}, {and} \bibinfo{person}{Kristina Toutanova}.} \bibinfo{year}{2019}\natexlab{}.
\newblock \showarticletitle{{{BERT}}: {{Pre-training}} of {{Deep Bidirectional Transformers}} for {{Language Understanding}}}. In \bibinfo{booktitle}{\emph{Proceedings of the 2019 {{Conference}} of the {{North American Chapter}} of the {{Association}} for {{Computational Linguistics}}: {{Human Language Technologies}}, {{Volume}} 1 ({{Long}} and {{Short Papers}})}}. \bibinfo{pages}{4171--4186}.
\newblock
\urldef\tempurl%
\url{https://aclweb.org/anthology/papers/N/N19/N19-1423/}
\showURL{%
\tempurl}


\bibitem[Dickey(2025)]%
        {Dickey_2025_Penske_Media_Sues_Google_for_AI_Overview_News_Story_Summaries_Without_Publishers_Consent}
\bibfield{author}{\bibinfo{person}{Josh Dickey}.} \bibinfo{year}{2025}\natexlab{}.
\newblock \bibinfo{title}{Penske {{Media Sues Google}} for {{AI}} '{{Overview}}' {{News Story Summaries Without Publishers}}' {{Consent}}}.
\newblock
\urldef\tempurl%
\url{https://www.thewrap.com/penske-media-sues-google-ai-overview-news-story-summaries/}
\showURL{%
\tempurl}


\bibitem[Doctorow(2025)]%
        {Doctorow_2025_Enshittification}
\bibfield{author}{\bibinfo{person}{Cory Doctorow}.} \bibinfo{year}{2025}\natexlab{}.
\newblock \bibinfo{booktitle}{\emph{Enshittification}}.
\newblock \bibinfo{publisher}{Verso Books}, \bibinfo{address}{London}.
\newblock
\showISBNx{978-1-83674-222-7}
\urldef\tempurl%
\url{https://guardianbookshop.com/enshittification-9781836742227/}
\showURL{%
\tempurl}


\bibitem[Dodge et~al\mbox{.}(2021)]%
        {DodgeSapEtAl_2021_Documenting_Large_Webtext_Corpora_Case_Study_on_Colossal_Clean_Crawled_Corpus}
\bibfield{author}{\bibinfo{person}{Jesse Dodge}, \bibinfo{person}{Maarten Sap}, \bibinfo{person}{Ana Marasovi{\'c}}, \bibinfo{person}{William Agnew}, \bibinfo{person}{Gabriel Ilharco}, \bibinfo{person}{Dirk Groeneveld}, \bibinfo{person}{Margaret Mitchell}, {and} \bibinfo{person}{Matt Gardner}.} \bibinfo{year}{2021}\natexlab{}.
\newblock \showarticletitle{Documenting {{Large Webtext Corpora}}: {{A Case Study}} on the {{Colossal Clean Crawled Corpus}}}. In \bibinfo{booktitle}{\emph{Proceedings of the 2021 {{Conference}} on {{Empirical Methods}} in {{Natural Language Processing}}}}. \bibinfo{publisher}{Association for Computational Linguistics}, \bibinfo{address}{Online and Punta Cana, Dominican Republic}, \bibinfo{pages}{1286--1305}.
\newblock
\href{https://doi.org/10.18653/v1/2021.emnlp-main.98}{doi:\nolinkurl{10.18653/v1/2021.emnlp-main.98}}


\bibitem[{Editorial}(2024)]%
        {Editorial_2024_Evolution_of_Labor_Law_Comprehensive_Historical_Overview}
\bibfield{author}{\bibinfo{person}{{Editorial}}.} \bibinfo{year}{2024}\natexlab{}.
\newblock \bibinfo{title}{The {{Evolution}} of {{Labor Law}}: {{A Comprehensive Historical Overview}}}.
\newblock
\urldef\tempurl%
\url{https://lawslearned.com/history-of-labor-law/}
\showURL{%
\tempurl}


\bibitem[Edwards(2024)]%
        {Edwards_2024_Stack_Overflow_users_sabotage_their_posts_after_OpenAI_deal}
\bibfield{author}{\bibinfo{person}{Benj Edwards}.} \bibinfo{year}{2024}\natexlab{}.
\newblock \bibinfo{title}{Stack {{Overflow}} Users Sabotage Their Posts after {{OpenAI}} Deal}.
\newblock
\urldef\tempurl%
\url{https://arstechnica.com/information-technology/2024/05/stack-overflow-users-sabotage-their-posts-after-openai-deal/}
\showURL{%
\tempurl}


\bibitem[Eiko(2022)]%
        {Eiko_2022_Welcome_to_Hotel_Elsevier_you_can_check-out_any_time_you_like_not_Eiko_Fried}
\bibfield{author}{\bibinfo{person}{Eiko}.} \bibinfo{year}{2022}\natexlab{}.
\newblock \bibinfo{title}{Welcome to {{Hotel Elsevier}}: You Can Check-out Any Time You like \dots{} Not >> {{Eiko Fried}}}.
\newblock
\urldef\tempurl%
\url{https://eiko-fried.com/welcome-to-hotel-elsevier-you-can-check-out-any-time-you-like-not/}
\showURL{%
\tempurl}


\bibitem[Elazar et~al\mbox{.}(2023)]%
        {ElazarBhagiaEtAl_2023_Whats_In_My_Big_Data}
\bibfield{author}{\bibinfo{person}{Yanai Elazar}, \bibinfo{person}{Akshita Bhagia}, \bibinfo{person}{Ian~Helgi Magnusson}, \bibinfo{person}{Abhilasha Ravichander}, \bibinfo{person}{Dustin Schwenk}, \bibinfo{person}{Alane Suhr}, \bibinfo{person}{Evan~Pete Walsh}, \bibinfo{person}{Dirk Groeneveld}, \bibinfo{person}{Luca Soldaini}, \bibinfo{person}{Sameer Singh}, \bibinfo{person}{Hannaneh Hajishirzi}, \bibinfo{person}{Noah~A. Smith}, {and} \bibinfo{person}{Jesse Dodge}.} \bibinfo{year}{2023}\natexlab{}.
\newblock \showarticletitle{What's {{In My Big Data}}?}. In \bibinfo{booktitle}{\emph{The {{Twelfth International Conference}} on {{Learning Representations}}}}.
\newblock
\urldef\tempurl%
\url{https://openreview.net/forum?id=RvfPnOkPV4}
\showURL{%
\tempurl}


\bibitem[Fallah et~al\mbox{.}(2024)]%
        {FallahJordanEtAl_2024_On_Three-Layer_Data_Markets}
\bibfield{author}{\bibinfo{person}{Alireza Fallah}, \bibinfo{person}{Michael~I. Jordan}, \bibinfo{person}{Ali Makhdoumi}, {and} \bibinfo{person}{Azarakhsh Malekian}.} \bibinfo{year}{2024}\natexlab{}.
\newblock \bibinfo{title}{On {{Three-Layer Data Markets}}}.
\newblock
\urldef\tempurl%
\url{https://arxiv.org/abs/2402.09697v4}
\showURL{%
\tempurl}


\bibitem[family=Tan and Thelen(2026)]%
        {TanThelen_2026_Cloud_Capitalism_and_AI_Transition}
\bibfield{author}{\bibinfo{person}{given-i=JS family=Tan, given=JS} {and} \bibinfo{person}{Kathleen Thelen}.} \bibinfo{year}{2026}\natexlab{}.
\newblock \showarticletitle{Cloud {{Capitalism}} and the {{AI Transition}}}.
\newblock  \bibinfo{volume}{54}, \bibinfo{number}{2} (\bibinfo{year}{2026}), \bibinfo{pages}{184--214}.
\newblock
\showISSN{0032-3292}
\href{https://doi.org/10.1177/00323292251396395}{doi:\nolinkurl{10.1177/00323292251396395}}


\bibitem[Farrell et~al\mbox{.}(2025)]%
        {FarrellGopnikEtAl_2025_Large_AI_models_are_cultural_and_social_technologies}
\bibfield{author}{\bibinfo{person}{Henry Farrell}, \bibinfo{person}{Alison Gopnik}, \bibinfo{person}{Cosma Shalizi}, {and} \bibinfo{person}{James Evans}.} \bibinfo{year}{2025}\natexlab{}.
\newblock \showarticletitle{Large {{AI}} Models Are Cultural and Social Technologies}.
\newblock \bibinfo{journal}{\emph{Science}} \bibinfo{volume}{387}, \bibinfo{number}{6739} (\bibinfo{date}{mar} \bibinfo{year}{2025}), \bibinfo{pages}{1153--1156}.
\newblock
\href{https://doi.org/10.1126/science.adt9819}{doi:\nolinkurl{10.1126/science.adt9819}}


\bibitem[Fischer(2024)]%
        {Fischer_2024_AI_startup_TollBit_raises_24M_series}
\bibfield{author}{\bibinfo{person}{Sara Fischer}.} \bibinfo{year}{2024}\natexlab{}.
\newblock \bibinfo{title}{{{AI}} Startup {{TollBit}} Raises \${{24M}} Series {{A}}}.
\newblock
\urldef\tempurl%
\url{https://www.axios.com/2024/10/22/ai-startup-tollbit-media-publishers}
\showURL{%
\tempurl}


\bibitem[Florida(2022)]%
        {Florida_2022_The_Rise_of_the_Creator_Economy}
\bibfield{author}{\bibinfo{person}{Richard Florida}.} \bibinfo{year}{2022}\natexlab{}.
\newblock \bibinfo{title}{The {{Rise}} of the {{Creator}} {{Economy}}}.
\newblock
\urldef\tempurl%
\url{https://creativeclass.com/reports/The\_Rise\_of\_the\_Creator\_Economy.pdf}
\showURL{%
\tempurl}


\bibitem[Formal et~al\mbox{.}(2021)]%
        {splade}
\bibfield{author}{\bibinfo{person}{Thibault Formal}, \bibinfo{person}{Benjamin Piwowarski}, {and} \bibinfo{person}{St\'{e}phane Clinchant}.} \bibinfo{year}{2021}\natexlab{}.
\newblock \showarticletitle{SPLADE: Sparse Lexical and Expansion Model for First Stage Ranking}. In \bibinfo{booktitle}{\emph{Proceedings of the 44th International ACM SIGIR Conference on Research and Development in Information Retrieval}} (Virtual Event, Canada) \emph{(\bibinfo{series}{SIGIR '21})}. \bibinfo{publisher}{Association for Computing Machinery}, \bibinfo{address}{New York, NY, USA}, \bibinfo{pages}{2288–2292}.
\newblock
\showISBNx{9781450380379}
\href{https://doi.org/10.1145/3404835.3463098}{doi:\nolinkurl{10.1145/3404835.3463098}}


\bibitem[Forte and Bruckman(2005)]%
        {ForteBruckman_2005_Why_Do_People_Write_for_Wikipedia_Incentives_to_Contribute_to_Open-Content_Publishing}
\bibfield{author}{\bibinfo{person}{Andrea Forte} {and} \bibinfo{person}{Amy Bruckman}.} \bibinfo{year}{2005}\natexlab{}.
\newblock \showarticletitle{Why {{Do People Write}} for {{Wikipedia}}? {{Incentives}} to {{Contribute}} to {{Open-Content Publishing}}}.
\newblock  (\bibinfo{date}{Nov.} \bibinfo{year}{2005}).
\newblock


\bibitem[{Foxglove Legal}(2025)]%
        {Foxglove2025_GoogleNewsChallenge}
\bibfield{author}{\bibinfo{person}{{Foxglove Legal}}.} \bibinfo{year}{2025}\natexlab{}.
\newblock \bibinfo{booktitle}{\emph{NEW CASE: Foxglove launches international legal challenge to Google’s worldwide theft of news!}}
\newblock Foxglove.
\newblock
\urldef\tempurl%
\url{https://foxglove.org.uk}
\showURL{%
\tempurl}


\bibitem[Franceschi-Bicchierai(2025)]%
        {Franceschi-Bicchierai2025_PerplexityScraping}
\bibfield{author}{\bibinfo{person}{Lorenzo Franceschi-Bicchierai}.} \bibinfo{year}{2025}\natexlab{}.
\newblock \bibinfo{booktitle}{\emph{Perplexity accused of scraping websites that explicitly blocked AI scraping}}.
\newblock TechCrunch.
\newblock
\urldef\tempurl%
\url{https://techcrunch.com/2025/08/04/perplexity-accused-of-scraping-websites-that-explicitly-blocked-ai-scraping/}
\showURL{%
\tempurl}


\bibitem[Galbraith(1954)]%
        {Galbraith_1954_Countervailing_Power}
\bibfield{author}{\bibinfo{person}{John~Kenneth Galbraith}.} \bibinfo{year}{1954}\natexlab{}.
\newblock \showarticletitle{Countervailing {{Power}}}.
\newblock  \bibinfo{volume}{44}, \bibinfo{number}{2} (\bibinfo{year}{1954}), \bibinfo{pages}{1--6}.
\newblock
\showISSN{0002-8282}
\urldef\tempurl%
\url{https://www.jstor.org/stable/1818317}
\showURL{%
\tempurl}


\bibitem[Gault(2025)]%
        {Gault_2025_Authors_Are_Accidentally_Leaving_AI_Prompts_In_their_Novels}
\bibfield{author}{\bibinfo{person}{Matthew Gault}.} \bibinfo{year}{2025}\natexlab{}.
\newblock \bibinfo{booktitle}{\emph{Authors {{Are Accidentally Leaving AI Prompts In}} Their {{Novels}}}}.
\newblock 404 Media.
\newblock
\urldef\tempurl%
\url{https://www.404media.co/authors-are-accidentally-leaving-ai-prompts-in-their-novels/}
\showURL{%
\tempurl}


\bibitem[Gebru et~al\mbox{.}(2020)]%
        {GebruMorgensternEtAl_2020_Datasheets_for_Datasets}
\bibfield{author}{\bibinfo{person}{Timnit Gebru}, \bibinfo{person}{Jamie Morgenstern}, \bibinfo{person}{Briana Vecchione}, \bibinfo{person}{Jennifer~Wortman Vaughan}, \bibinfo{person}{Hanna Wallach}, \bibinfo{person}{Hal Daum{\'e}~III}, {and} \bibinfo{person}{Kate Crawford}.} \bibinfo{year}{2020}\natexlab{}.
\newblock \showarticletitle{Datasheets for {{Datasets}}}.
\newblock \bibinfo{journal}{\emph{arXiv:1803.09010 [cs]}} (\bibinfo{date}{mar} \bibinfo{year}{2020}).
\newblock
\showeprint[arxiv]{1803.09010}~[cs]
\urldef\tempurl%
\url{http://arxiv.org/abs/1803.09010}
\showURL{%
\tempurl}


\bibitem[Germain(2025)]%
        {Germain_2025_Is_Google_about_to_destroy_web}
\bibfield{author}{\bibinfo{person}{Thomas Germain}.} \bibinfo{year}{2025}\natexlab{}.
\newblock \showarticletitle{Is {{Google}} about to Destroy the Web?}
\newblock \bibinfo{journal}{\emph{BBC}} (\bibinfo{date}{jun} \bibinfo{year}{2025}).
\newblock
\urldef\tempurl%
\url{https://www.bbc.com/future/article/20250611-ai-mode-is-google-about-to-change-the-internet-forever}
\showURL{%
\tempurl}


\bibitem[Gil(2024)]%
        {Gil_2024_Stop_Chasing_Algorithms_Heres_How_Creators_Can_Take_Control_of_Their_Content_and_Monetize_on_Their_Own_Terms}
\bibfield{author}{\bibinfo{person}{Carlos Gil}.} \bibinfo{year}{2024}\natexlab{}.
\newblock \bibinfo{title}{Stop {{Chasing Algorithms}} --- {{Here}}'s {{How Creators Can Take Control}} of {{Their Content}} and {{Monetize}} on {{Their Own Terms}}}.
\newblock
\urldef\tempurl%
\url{https://www.entrepreneur.com/science-technology/why-relying-on-social-media-for-income-is-a-losing-game-for/481348}
\showURL{%
\tempurl}


\bibitem[Gnewuch et~al\mbox{.}(2026)]%
        {GnewuchWahleEtAl_2026_Big_tech-funded_AI_papers_have_higher_citation_impact_greater_insularity_and_larger_recency_bias}
\bibfield{author}{\bibinfo{person}{Max~Martin Gnewuch}, \bibinfo{person}{Jan~Philip Wahle}, \bibinfo{person}{Terry Ruas}, {and} \bibinfo{person}{Bela Gipp}.} \bibinfo{year}{2026}\natexlab{}.
\newblock \showarticletitle{Big Tech-Funded {{AI}} Papers Have Higher Citation Impact, Greater Insularity, and Larger Recency Bias}. In \bibinfo{booktitle}{\emph{2026 International Conference on Artificial Intelligence, Computer, Data Sciences and Applications ({{ACDSA}})}} (2026). \bibinfo{pages}{1--7}.
\newblock
\href{https://doi.org/10.1109/ACDSA67686.2026.11468172}{doi:\nolinkurl{10.1109/ACDSA67686.2026.11468172}}


\bibitem[Griffiths(2026)]%
        {Griffiths_2026_Microsoft_says_Copilot_isnt_just_for_entertainment_purposes_after_its_terms_of_service_language_goes_viral}
\bibfield{author}{\bibinfo{person}{Brent~D. Griffiths}.} \bibinfo{year}{2026}\natexlab{}.
\newblock \bibinfo{booktitle}{\emph{Microsoft Says {{Copilot}} Isn't Just 'for Entertainment Purposes' after Its Terms of Service Language Goes Viral}}.
\newblock Business Insider.
\newblock
\urldef\tempurl%
\url{https://www.businessinsider.com/microsoft-copilot-entertainment-purposes-terms-of-service-agreement-2026-4}
\showURL{%
\tempurl}


\bibitem[Groeneveld et~al\mbox{.}(2024)]%
        {groeneveldOLMoAcceleratingScience2024}
\bibfield{author}{\bibinfo{person}{Dirk Groeneveld}, \bibinfo{person}{Iz Beltagy}, \bibinfo{person}{Pete Walsh}, \bibinfo{person}{Akshita Bhagia}, \bibinfo{person}{Rodney Kinney}, \bibinfo{person}{Oyvind Tafjord}, \bibinfo{person}{Ananya~Harsh Jha}, \bibinfo{person}{Hamish Ivison}, \bibinfo{person}{Ian Magnusson}, \bibinfo{person}{Yizhong Wang}, \bibinfo{person}{Shane Arora}, \bibinfo{person}{David Atkinson}, \bibinfo{person}{Russell Authur}, \bibinfo{person}{Khyathi~Raghavi Chandu}, \bibinfo{person}{Arman Cohan}, \bibinfo{person}{Jennifer Dumas}, \bibinfo{person}{Yanai Elazar}, \bibinfo{person}{Yuling Gu}, \bibinfo{person}{Jack Hessel}, \bibinfo{person}{Tushar Khot}, \bibinfo{person}{William Merrill}, \bibinfo{person}{Jacob Morrison}, \bibinfo{person}{Niklas Muennighoff}, \bibinfo{person}{Aakanksha Naik}, \bibinfo{person}{Crystal Nam}, \bibinfo{person}{Matthew~E. Peters}, \bibinfo{person}{Valentina Pyatkin}, \bibinfo{person}{Abhilasha Ravichander}, \bibinfo{person}{Dustin Schwenk}, \bibinfo{person}{Saurabh
  Shah}, \bibinfo{person}{Will Smith}, \bibinfo{person}{Emma Strubell}, \bibinfo{person}{Nishant Subramani}, \bibinfo{person}{Mitchell Wortsman}, \bibinfo{person}{Pradeep Dasigi}, \bibinfo{person}{Nathan Lambert}, \bibinfo{person}{Kyle Richardson}, \bibinfo{person}{Luke Zettlemoyer}, \bibinfo{person}{Jesse Dodge}, \bibinfo{person}{Kyle Lo}, \bibinfo{person}{Luca Soldaini}, \bibinfo{person}{Noah~A. Smith}, {and} \bibinfo{person}{Hannaneh Hajishirzi}.} \bibinfo{year}{2024}\natexlab{}.
\newblock \bibinfo{title}{{{OLMo}}: {{Accelerating}} the {{Science}} of {{Language Models}}}.
\newblock
\showeprint[arxiv]{2402.00838}~[cs]
\href{https://doi.org/10.48550/arXiv.2402.00838}{doi:\nolinkurl{10.48550/arXiv.2402.00838}}


\bibitem[Grosse et~al\mbox{.}(2023)]%
        {grosse2023studyinglargelanguagemodel}
\bibfield{author}{\bibinfo{person}{Roger Grosse}, \bibinfo{person}{Juhan Bae}, \bibinfo{person}{Cem Anil}, \bibinfo{person}{Nelson Elhage}, \bibinfo{person}{Alex Tamkin}, \bibinfo{person}{Amirhossein Tajdini}, \bibinfo{person}{Benoit Steiner}, \bibinfo{person}{Dustin Li}, \bibinfo{person}{Esin Durmus}, \bibinfo{person}{Ethan Perez}, \bibinfo{person}{Evan Hubinger}, \bibinfo{person}{Kamilė Lukošiūtė}, \bibinfo{person}{Karina Nguyen}, \bibinfo{person}{Nicholas Joseph}, \bibinfo{person}{Sam McCandlish}, \bibinfo{person}{Jared Kaplan}, {and} \bibinfo{person}{Samuel~R. Bowman}.} \bibinfo{year}{2023}\natexlab{}.
\newblock \bibinfo{title}{Studying Large Language Model Generalization with Influence Functions}.
\newblock
\showeprint[arxiv]{2308.03296}~[cs.LG]
\urldef\tempurl%
\url{https://arxiv.org/abs/2308.03296}
\showURL{%
\tempurl}


\bibitem[Gupta and Pruthi(2025)]%
        {GuptaPruthi_2025_All_That_Glitters_is_Not_Novel_Plagiarism_in_AI_Generated_Researcha}
\bibfield{author}{\bibinfo{person}{Tarun Gupta} {and} \bibinfo{person}{Danish Pruthi}.} \bibinfo{year}{2025}\natexlab{}.
\newblock \showarticletitle{All {{That Glitters}} Is {{Not Novel}}: {{Plagiarism}} in {{AI Generated Research}}}. In \bibinfo{booktitle}{\emph{Proceedings of the 63rd {{Annual Meeting}} of the {{Association}} for {{Computational Linguistics}} ({{Volume}} 1: {{Long Papers}})}} (Vienna, Austria, 2025-07), \bibfield{editor}{\bibinfo{person}{Wanxiang Che}, \bibinfo{person}{Joyce Nabende}, \bibinfo{person}{Ekaterina Shutova}, {and} \bibinfo{person}{Mohammad~Taher Pilehvar}} (Eds.). \bibinfo{publisher}{Association for Computational Linguistics}, \bibinfo{pages}{25721--25738}.
\newblock
\showISBNx{979-8-89176-251-0}
\href{https://doi.org/10.18653/v1/2025.acl-long.1249}{doi:\nolinkurl{10.18653/v1/2025.acl-long.1249}}


\bibitem[Gupta and Pruthi(2025)]%
        {gupta-pruthi-2025-glitters}
\bibfield{author}{\bibinfo{person}{Tarun Gupta} {and} \bibinfo{person}{Danish Pruthi}.} \bibinfo{year}{2025}\natexlab{}.
\newblock \showarticletitle{All That Glitters is Not Novel: Plagiarism in {AI} Generated Research}. In \bibinfo{booktitle}{\emph{Proceedings of the 63rd Annual Meeting of the Association for Computational Linguistics (Volume 1: Long Papers)}}, \bibfield{editor}{\bibinfo{person}{Wanxiang Che}, \bibinfo{person}{Joyce Nabende}, \bibinfo{person}{Ekaterina Shutova}, {and} \bibinfo{person}{Mohammad~Taher Pilehvar}} (Eds.). \bibinfo{publisher}{Association for Computational Linguistics}, \bibinfo{address}{Vienna, Austria}, \bibinfo{pages}{25721--25738}.
\newblock
\showISBNx{979-8-89176-251-0}
\href{https://doi.org/10.18653/v1/2025.acl-long.1249}{doi:\nolinkurl{10.18653/v1/2025.acl-long.1249}}


\bibitem[Haim et~al\mbox{.}(2022)]%
        {HaimVardiEtAl_2022_Reconstructing_Training_Data_From_Trained_Neural_Networks}
\bibfield{author}{\bibinfo{person}{Niv Haim}, \bibinfo{person}{Gal Vardi}, \bibinfo{person}{Gilad Yehudai}, \bibinfo{person}{Ohad Shamir}, {and} \bibinfo{person}{Michal Irani}.} \bibinfo{year}{2022}\natexlab{}.
\newblock \showarticletitle{Reconstructing {{Training Data From Trained Neural Networks}}}. In \bibinfo{booktitle}{\emph{Advances in {{Neural Information Processing Systems}}}}.
\newblock
\urldef\tempurl%
\url{https://openreview.net/forum?id=Sxk8Bse3RKO}
\showURL{%
\tempurl}


\bibitem[Hansen(2024)]%
        {Hansen_2024_Text_Data_Mining_Research_DMCA_Exemption_Renewed_and_Expanded}
\bibfield{author}{\bibinfo{person}{Dave Hansen}.} \bibinfo{year}{2024}\natexlab{}.
\newblock \bibinfo{title}{Text {{Data Mining Research DMCA Exemption Renewed}} and {{Expanded}}}.
\newblock
\urldef\tempurl%
\url{https://www.authorsalliance.org/2024/10/25/text-data-mining-research-dmca-exemption-renewed-and-expanded/}
\showURL{%
\tempurl}


\bibitem[Hashim et~al\mbox{.}(2018)]%
        {HashimKannanEtAl_2018_Central_Role_of_Moral_Obligations_in_Determining_Intentions_to_Engage_in_Digital_Piracy}
\bibfield{author}{\bibinfo{person}{Matthew~J. Hashim}, \bibinfo{person}{Karthik~N. Kannan}, {and} \bibinfo{person}{Duane~T. Wegener}.} \bibinfo{year}{2018}\natexlab{}.
\newblock \showarticletitle{Central {{Role}} of {{Moral Obligations}} in {{Determining Intentions}} to {{Engage}} in {{Digital Piracy}}}.
\newblock \bibinfo{journal}{\emph{Journal of Management Information Systems}} \bibinfo{volume}{35}, \bibinfo{number}{3} (\bibinfo{date}{jul} \bibinfo{year}{2018}), \bibinfo{pages}{934--963}.
\newblock
\showISSN{0742-1222, 1557-928X}
\href{https://doi.org/10.1080/07421222.2018.1481670}{doi:\nolinkurl{10.1080/07421222.2018.1481670}}


\bibitem[Haviland and Mullin(2009)]%
        {HavilandMullin_2009_Who_owns_this_text_plagiarism_authorship_and_disciplinary_cultures}
\bibfield{editor}{\bibinfo{person}{Carol~Peterson Haviland} {and} \bibinfo{person}{Joan~A. Mullin}} (Eds.). \bibinfo{year}{2009}\natexlab{}.
\newblock \bibinfo{booktitle}{\emph{Who Owns This Text? Plagiarism, Authorship, and Disciplinary Cultures}}.
\newblock \bibinfo{publisher}{Utah State University Press}, \bibinfo{address}{Logan, Utah}.
\newblock
\showISBNx{978-0-87421-728-5 978-0-87421-729-2}
\showLCCN{PN167 .W46 2009}


\bibitem[Helgesson and Eriksson(2015)]%
        {HelgessonEriksson_2015_Plagiarism_in_research}
\bibfield{author}{\bibinfo{person}{Gert Helgesson} {and} \bibinfo{person}{Stefan Eriksson}.} \bibinfo{year}{2015}\natexlab{}.
\newblock \showarticletitle{Plagiarism in Research}.
\newblock \bibinfo{journal}{\emph{Medicine, Health Care and Philosophy}} \bibinfo{volume}{18}, \bibinfo{number}{1} (\bibinfo{date}{feb} \bibinfo{year}{2015}), \bibinfo{pages}{91--101}.
\newblock
\showISSN{1572-8633}
\href{https://doi.org/10.1007/s11019-014-9583-8}{doi:\nolinkurl{10.1007/s11019-014-9583-8}}


\bibitem[Henderson et~al\mbox{.}(2023)]%
        {HendersonLiEtAl_2023_Foundation_Models_and_Fair_Use}
\bibfield{author}{\bibinfo{person}{Peter Henderson}, \bibinfo{person}{Xuechen Li}, \bibinfo{person}{Dan Jurafsky}, \bibinfo{person}{Tatsunori Hashimoto}, \bibinfo{person}{Mark~A. Lemley}, {and} \bibinfo{person}{Percy Liang}.} \bibinfo{year}{2023}\natexlab{}.
\newblock \bibinfo{title}{Foundation {{Models}} and {{Fair Use}}}.
\newblock
\showeprint[arxiv]{2303.15715}~[cs]
\href{https://doi.org/10.48550/arXiv.2303.15715}{doi:\nolinkurl{10.48550/arXiv.2303.15715}}


\bibitem[Hickey(2015)]%
        {Hickey_2015_Reraming_similarity_analysis_in_copyright}
\bibfield{author}{\bibinfo{person}{Kevin~J Hickey}.} \bibinfo{year}{2015}\natexlab{}.
\newblock \showarticletitle{Reframing Similarity Analysis in Copyright}.
\newblock   \bibinfo{volume}{93} (\bibinfo{year}{2015}), \bibinfo{pages}{681--731}.
\newblock


\bibitem[Holder and Ghaffary(2024)]%
        {HolderGhaffary_2024_Sam_Altman-Backed_Group_Completes_Largest_US_Study_on_Basic_Income}
\bibfield{author}{\bibinfo{person}{Sarah Holder} {and} \bibinfo{person}{Shirin Ghaffary}.} \bibinfo{year}{2024}\natexlab{}.
\newblock \showarticletitle{Sam {{Altman-Backed Group Completes Largest US Study}} on {{Basic Income}}}.
\newblock  (\bibinfo{year}{2024}).
\newblock
\urldef\tempurl%
\url{https://www.bloomberg.com/news/articles/2024-07-22/ubi-study-backed-by-openai-s-sam-altman-bolsters-support-for-basic-income}
\showURL{%
\tempurl}


\bibitem[Hooker(2025)]%
        {Hooker_2025_On_Slow_Death_of_Scaling}
\bibfield{author}{\bibinfo{person}{Sara Hooker}.} \bibinfo{year}{2025}\natexlab{}.
\newblock \bibinfo{booktitle}{\emph{On the {{Slow Death}} of {{Scaling}}}}.
\newblock
\showeprint[Social Science Research Network]{5877662}
\href{https://doi.org/10.2139/ssrn.5877662}{doi:\nolinkurl{10.2139/ssrn.5877662}}


\bibitem[Huang et~al\mbox{.}(2024)]%
        {HuangYangEtAl_2024_Demystifying_Verbatim_Memorization_in_Large_Language_Models}
\bibfield{author}{\bibinfo{person}{Jing Huang}, \bibinfo{person}{Diyi Yang}, {and} \bibinfo{person}{Christopher Potts}.} \bibinfo{year}{2024}\natexlab{}.
\newblock \showarticletitle{Demystifying {{Verbatim Memorization}} in {{Large Language Models}}}. In \bibinfo{booktitle}{\emph{Proceedings of the 2024 {{Conference}} on {{Empirical Methods}} in {{Natural Language Processing}}}}, \bibfield{editor}{\bibinfo{person}{Yaser {Al-Onaizan}}, \bibinfo{person}{Mohit Bansal}, {and} \bibinfo{person}{Yun-Nung Chen}} (Eds.). \bibinfo{publisher}{Association for Computational Linguistics}, \bibinfo{address}{Miami, Florida, USA}, \bibinfo{pages}{10711--10732}.
\newblock
\href{https://doi.org/10.18653/v1/2024.emnlp-main.598}{doi:\nolinkurl{10.18653/v1/2024.emnlp-main.598}}


\bibitem[Hutchinson et~al\mbox{.}(2021)]%
        {HutchinsonSmartEtAl_2021_Towards_Accountability_for_Machine_Learning_Datasets_Practices_from_Software_Engineering_and_Infrastructure}
\bibfield{author}{\bibinfo{person}{Ben Hutchinson}, \bibinfo{person}{Andrew Smart}, \bibinfo{person}{Alex Hanna}, \bibinfo{person}{Emily Denton}, \bibinfo{person}{Christina Greer}, \bibinfo{person}{Oddur Kjartansson}, \bibinfo{person}{Parker Barnes}, {and} \bibinfo{person}{Margaret Mitchell}.} \bibinfo{year}{2021}\natexlab{}.
\newblock \showarticletitle{Towards {{Accountability}} for {{Machine Learning Datasets}}: {{Practices}} from {{Software Engineering}} and {{Infrastructure}}}.
\newblock \bibinfo{journal}{\emph{arXiv:2010.13561 [cs]}} (\bibinfo{date}{jan} \bibinfo{year}{2021}).
\newblock
\showeprint[arxiv]{2010.13561}~[cs]
\urldef\tempurl%
\url{http://arxiv.org/abs/2010.13561}
\showURL{%
\tempurl}


\bibitem[Iskander et~al\mbox{.}(2024)]%
        {iskander-etal-2024-quality}
\bibfield{author}{\bibinfo{person}{Shadi Iskander}, \bibinfo{person}{Sofia Tolmach}, \bibinfo{person}{Ori Shapira}, \bibinfo{person}{Nachshon Cohen}, {and} \bibinfo{person}{Zohar Karnin}.} \bibinfo{year}{2024}\natexlab{}.
\newblock \showarticletitle{Quality Matters: Evaluating Synthetic Data for Tool-Using {LLM}s}. In \bibinfo{booktitle}{\emph{Proceedings of the 2024 Conference on Empirical Methods in Natural Language Processing}}, \bibfield{editor}{\bibinfo{person}{Yaser Al-Onaizan}, \bibinfo{person}{Mohit Bansal}, {and} \bibinfo{person}{Yun-Nung Chen}} (Eds.). \bibinfo{publisher}{Association for Computational Linguistics}, \bibinfo{address}{Miami, Florida, USA}, \bibinfo{pages}{4958--4976}.
\newblock
\href{https://doi.org/10.18653/v1/2024.emnlp-main.285}{doi:\nolinkurl{10.18653/v1/2024.emnlp-main.285}}


\bibitem[Jacobs et~al\mbox{.}(1991)]%
        {6797059}
\bibfield{author}{\bibinfo{person}{Robert~A. Jacobs}, \bibinfo{person}{Michael~I. Jordan}, \bibinfo{person}{Steven~J. Nowlan}, {and} \bibinfo{person}{Geoffrey~E. Hinton}.} \bibinfo{year}{1991}\natexlab{}.
\newblock \showarticletitle{Adaptive Mixtures of Local Experts}.
\newblock \bibinfo{journal}{\emph{Neural Computation}} \bibinfo{volume}{3}, \bibinfo{number}{1} (\bibinfo{year}{1991}), \bibinfo{pages}{79--87}.
\newblock
\href{https://doi.org/10.1162/neco.1991.3.1.79}{doi:\nolinkurl{10.1162/neco.1991.3.1.79}}


\bibitem[Jayasundara(2022)]%
        {Jayasundara_2022_Study_on_Risk_of_Prosecution_and_Perceived_Proximity_on_State_University_Undergraduates_Behavioural_Intention_for_e-Book_Piracy}
\bibfield{author}{\bibinfo{person}{C.~C. Jayasundara}.} \bibinfo{year}{2022}\natexlab{}.
\newblock \showarticletitle{A {{Study}} on the {{Risk}} of {{Prosecution}} and {{Perceived Proximity}} on {{State University Undergraduates}}' {{Behavioural Intention}} for e-{{Book Piracy}}}.
\newblock \bibinfo{journal}{\emph{New Review of Academic Librarianship}} \bibinfo{volume}{28}, \bibinfo{number}{4} (\bibinfo{date}{oct} \bibinfo{year}{2022}), \bibinfo{pages}{406--434}.
\newblock
\showISSN{1361-4533, 1740-7834}
\href{https://doi.org/10.1080/13614533.2021.1976655}{doi:\nolinkurl{10.1080/13614533.2021.1976655}}


\bibitem[Ja{\'z}wi{\'n}ska and Chandrasekar({[n.\,d.]})]%
        {JazwinskaChandrasekar_AI_Search_Has_Citation_Problem}
\bibfield{author}{\bibinfo{person}{Klaudia Ja{\'z}wi{\'n}ska} {and} \bibinfo{person}{Aisvarya Chandrasekar}.} \bibinfo{year}{[n.\,d.]}\natexlab{}.
\newblock \bibinfo{title}{{{AI Search Has}} a {{Citation Problem}}}.
\newblock
\urldef\tempurl%
\url{https://www.cjr.org/tow\_center/we-compared-eight-ai-search-engines-theyre-all-bad-at-citing-news.php}
\showURL{%
\tempurl}


\bibitem[Jernite et~al\mbox{.}(2022)]%
        {JerniteNguyenEtAl_2022_Data_Governance_in_Age_of_Large-Scale_Data-Driven_Language_Technology}
\bibfield{author}{\bibinfo{person}{Yacine Jernite}, \bibinfo{person}{Huu Nguyen}, \bibinfo{person}{Stella Biderman}, \bibinfo{person}{Anna Rogers}, \bibinfo{person}{Maraim Masoud}, \bibinfo{person}{Valentin Danchev}, \bibinfo{person}{Samson Tan}, \bibinfo{person}{Alexandra~Sasha Luccioni}, \bibinfo{person}{Nishant Subramani}, \bibinfo{person}{Isaac Johnson}, \bibinfo{person}{Gerard Dupont}, \bibinfo{person}{Jesse Dodge}, \bibinfo{person}{Kyle Lo}, \bibinfo{person}{Zeerak Talat}, \bibinfo{person}{Dragomir Radev}, \bibinfo{person}{Aaron Gokaslan}, \bibinfo{person}{Somaieh Nikpoor}, \bibinfo{person}{Peter Henderson}, \bibinfo{person}{Rishi Bommasani}, {and} \bibinfo{person}{Margaret Mitchell}.} \bibinfo{year}{2022}\natexlab{}.
\newblock \showarticletitle{Data {{Governance}} in the {{Age}} of {{Large-Scale Data-Driven Language Technology}}}. In \bibinfo{booktitle}{\emph{2022 {{ACM Conference}} on {{Fairness}}, {{Accountability}}, and {{Transparency}}}} \emph{(\bibinfo{series}{{{FAccT}} '22})}. \bibinfo{publisher}{Association for Computing Machinery}, \bibinfo{address}{New York, NY, USA}, \bibinfo{pages}{2206--2222}.
\newblock
\showISBNx{978-1-4503-9352-2}
\href{https://doi.org/10.1145/3531146.3534637}{doi:\nolinkurl{10.1145/3531146.3534637}}


\bibitem[Jones et~al\mbox{.}(2025)]%
        {JonesNewmanEtAl_2025_AI-Generated_Slop_in_Online_Biomedical_Science_Educational_Videos_Mixed_Methods_Study_of_Prevalence_Characteristics_and_Hazards_to_Learners_and_Teachers}
\bibfield{author}{\bibinfo{person}{Eric~M. Jones}, \bibinfo{person}{Jane~D. Newman}, \bibinfo{person}{Boyun Kim}, {and} \bibinfo{person}{Emily~J. Fogle}.} \bibinfo{year}{2025}\natexlab{}.
\newblock \showarticletitle{{{AI-Generated}} “{{Slop}}” in {{Online Biomedical Science Educational Videos}}: {{Mixed Methods Study}} of {{Prevalence}}, {{Characteristics}}, and {{Hazards}} to {{Learners}} and {{Teachers}}}.
\newblock  \bibinfo{volume}{11}, \bibinfo{number}{1} (\bibinfo{year}{2025}), \bibinfo{pages}{e80084}.
\newblock
\href{https://doi.org/10.2196/80084}{doi:\nolinkurl{10.2196/80084}}


\bibitem[Jordan(2025)]%
        {Jordan_2025_Collectivist_Economic_Perspective_on_AI}
\bibfield{author}{\bibinfo{person}{Michael~I. Jordan}.} \bibinfo{year}{2025}\natexlab{}.
\newblock \bibinfo{title}{A {{Collectivist}}, {{Economic Perspective}} on {{AI}}}.
\newblock
\showeprint[arxiv]{2507.06268}~[cs]
\href{https://doi.org/10.48550/arXiv.2507.06268}{doi:\nolinkurl{10.48550/arXiv.2507.06268}}


\bibitem[Kairam et~al\mbox{.}(2024)]%
        {KairamBernsteinEtAl_2024_Community-Driven_Models_for_Research_on_Social_Platforms}
\bibfield{author}{\bibinfo{person}{Sanjay Kairam}, \bibinfo{person}{Michael~S. Bernstein}, \bibinfo{person}{Amy~S. Bruckman}, \bibinfo{person}{Stevie Chancellor}, \bibinfo{person}{Eshwar Chandrasekharan}, \bibinfo{person}{Munmun De~Choudhury}, \bibinfo{person}{Casey Fiesler}, \bibinfo{person}{Hanlin Li}, \bibinfo{person}{Nicholas Proferes}, \bibinfo{person}{Manoel Horta~Ribeiro}, \bibinfo{person}{C.~Estelle Smith}, {and} \bibinfo{person}{Galen~Cassebeer Weld}.} \bibinfo{year}{2024}\natexlab{}.
\newblock \showarticletitle{Community-{{Driven Models}} for {{Research}} on {{Social Platforms}}}. In \bibinfo{booktitle}{\emph{Companion {{Publication}} of the 2024 {{Conference}} on {{Computer-Supported Cooperative Work}} and {{Social Computing}}}} \emph{(\bibinfo{series}{{{CSCW Companion}} '24})}. \bibinfo{publisher}{Association for Computing Machinery}, \bibinfo{address}{New York, NY, USA}, \bibinfo{pages}{684--688}.
\newblock
\showISBNx{979-8-4007-1114-5}
\href{https://doi.org/10.1145/3678884.3687141}{doi:\nolinkurl{10.1145/3678884.3687141}}


\bibitem[Kandpal and Raffel(2025)]%
        {KandpalRaffel_2025_Position_Most_Expensive_Part_of_LLM_should_be_its_Training_Data}
\bibfield{author}{\bibinfo{person}{Nikhil Kandpal} {and} \bibinfo{person}{Colin Raffel}.} \bibinfo{year}{2025}\natexlab{}.
\newblock \showarticletitle{Position: {{The Most Expensive Part}} of an {{LLM}} *should* Be Its {{Training Data}}}.
\newblock
\urldef\tempurl%
\url{https://openreview.net/forum?id=L6RpQ1h4Nx}
\showURL{%
\tempurl}


\bibitem[Khosla(2024)]%
        {Khosla_2024_Roadmap_to_AI_Utopia}
\bibfield{author}{\bibinfo{person}{Vinod Khosla}.} \bibinfo{year}{2024}\natexlab{}.
\newblock \bibinfo{title}{A {{Roadmap}} to {{AI Utopia}}}.
\newblock
\urldef\tempurl%
\url{https://time.com/7174892/a-roadmap-to-ai-utopia/}
\showURL{%
\tempurl}


\bibitem[Kim(2024)]%
        {Kim_2024_Data_Scraping_for_Generative_AI_To_What_Extent}
\bibfield{author}{\bibinfo{person}{Jae Yeon~Cecilia Kim}.} \bibinfo{year}{2024}\natexlab{}.
\newblock \showarticletitle{Data {{Scraping}} for {{Generative AI}} - {{To What Extent}}?}
\newblock \bibinfo{journal}{\emph{Brooklyn Journal of Corporate, Financial \& Commercial Law}}  \bibinfo{volume}{19} (\bibinfo{year}{2024}), \bibinfo{pages}{179--200}.
\newblock
\urldef\tempurl%
\url{https://brooklynworks.brooklaw.edu/bjcfcl/vol19/iss1/8/}
\showURL{%
\tempurl}


\bibitem[Klassen and Fiesler(2022)]%
        {KlassenFiesler_2022_This_Isnt_Your_Data_Friend_Black_Twitter_as_Case_Study_on_Research_Ethics_for_Public_Data}
\bibfield{author}{\bibinfo{person}{Shamika Klassen} {and} \bibinfo{person}{Casey Fiesler}.} \bibinfo{year}{2022}\natexlab{}.
\newblock \showarticletitle{``{{This Isn}}'t {{Your Data}}, {{Friend}}'': {{Black Twitter}} as a {{Case Study}} on {{Research Ethics}} for {{Public Data}}}.
\newblock \bibinfo{journal}{\emph{Social Media + Society}} \bibinfo{volume}{8}, \bibinfo{number}{4} (\bibinfo{date}{Oct.} \bibinfo{year}{2022}), \bibinfo{pages}{20563051221144317}.
\newblock
\showISSN{2056-3051}
\href{https://doi.org/10.1177/20563051221144317}{doi:\nolinkurl{10.1177/20563051221144317}}


\bibitem[Knibbs(2024)]%
        {Knibbs_2024_Scammy_AI-Generated_Books_Are_Flooding_Amazon}
\bibfield{author}{\bibinfo{person}{Katie Knibbs}.} \bibinfo{year}{2024}\natexlab{}.
\newblock \showarticletitle{Scammy {{AI-Generated Books Are Flooding Amazon}}}.
\newblock  (\bibinfo{year}{2024}).
\newblock
\showISSN{1059-1028}
\urldef\tempurl%
\url{https://www.wired.com/story/scammy-ai-generated-books-flooding-amazon/}
\showURL{%
\tempurl}


\bibitem[Koh and Liang(2017)]%
        {10.5555/3305381.3305576}
\bibfield{author}{\bibinfo{person}{Pang~Wei Koh} {and} \bibinfo{person}{Percy Liang}.} \bibinfo{year}{2017}\natexlab{}.
\newblock \showarticletitle{Understanding black-box predictions via influence functions}. In \bibinfo{booktitle}{\emph{Proceedings of the 34th International Conference on Machine Learning - Volume 70}} (Sydney, NSW, Australia) \emph{(\bibinfo{series}{ICML'17})}. \bibinfo{publisher}{JMLR.org}, \bibinfo{pages}{1885–1894}.
\newblock


\bibitem[Kreutzer et~al\mbox{.}(2022)]%
        {KreutzerCaswellEtAl_2022_Quality_at_Glance_Audit_of_WebCrawled_Multilingual_Datasets}
\bibfield{author}{\bibinfo{person}{Julia Kreutzer}, \bibinfo{person}{Isaac Caswell}, \bibinfo{person}{Lisa Wang}, \bibinfo{person}{Ahsan Wahab}, \bibinfo{person}{Daan {van Esch}}, \bibinfo{person}{Nasanbayar {Ulzii-Orshikh}}, \bibinfo{person}{Allahsera Tapo}, \bibinfo{person}{Nishant Subramani}, \bibinfo{person}{Artem Sokolov}, \bibinfo{person}{Claytone Sikasote}, \bibinfo{person}{Monang Setyawan}, \bibinfo{person}{Supheakmungkol Sarin}, \bibinfo{person}{Sokhar Samb}, \bibinfo{person}{Beno{\^i}t Sagot}, \bibinfo{person}{Clara Rivera}, \bibinfo{person}{Annette Rios}, \bibinfo{person}{Isabel Papadimitriou}, \bibinfo{person}{Salomey Osei}, \bibinfo{person}{Pedro~Ortiz Suarez}, \bibinfo{person}{Iroro Orife}, \bibinfo{person}{Kelechi Ogueji}, \bibinfo{person}{Andre~Niyongabo Rubungo}, \bibinfo{person}{Toan~Q. Nguyen}, \bibinfo{person}{Mathias M{\"u}ller}, \bibinfo{person}{Andr{\'e} M{\"u}ller}, \bibinfo{person}{Shamsuddeen~Hassan Muhammad}, \bibinfo{person}{Nanda Muhammad}, \bibinfo{person}{Ayanda Mnyakeni},
  \bibinfo{person}{Jamshidbek Mirzakhalov}, \bibinfo{person}{Tapiwanashe Matangira}, \bibinfo{person}{Colin Leong}, \bibinfo{person}{Nze Lawson}, \bibinfo{person}{Sneha Kudugunta}, \bibinfo{person}{Yacine Jernite}, \bibinfo{person}{Mathias Jenny}, \bibinfo{person}{Orhan Firat}, \bibinfo{person}{Bonaventure F.~P. Dossou}, \bibinfo{person}{Sakhile Dlamini}, \bibinfo{person}{Nisansa {de Silva}}, \bibinfo{person}{Sakine {\c C}abuk~Ball{\i}}, \bibinfo{person}{Stella Biderman}, \bibinfo{person}{Alessia Battisti}, \bibinfo{person}{Ahmed Baruwa}, \bibinfo{person}{Ankur Bapna}, \bibinfo{person}{Pallavi Baljekar}, \bibinfo{person}{Israel~Abebe Azime}, \bibinfo{person}{Ayodele Awokoya}, \bibinfo{person}{Duygu Ataman}, \bibinfo{person}{Orevaoghene Ahia}, \bibinfo{person}{Oghenefego Ahia}, \bibinfo{person}{Sweta Agrawal}, {and} \bibinfo{person}{Mofetoluwa Adeyemi}.} \bibinfo{year}{2022}\natexlab{}.
\newblock \showarticletitle{Quality at a {{Glance}}: {{An Audit}} of {{Web-Crawled Multilingual Datasets}}}.
\newblock \bibinfo{journal}{\emph{Transactions of the Association for Computational Linguistics}}  \bibinfo{volume}{10} (\bibinfo{date}{jan} \bibinfo{year}{2022}), \bibinfo{pages}{50--72}.
\newblock
\showISSN{2307-387X}
\href{https://doi.org/10.1162/tacl\_a\_00447}{doi:\nolinkurl{10.1162/tacl\_a\_00447}}


\bibitem[Kusner et~al\mbox{.}(2015)]%
        {Kusner-et-al-2015_wmd}
\bibfield{author}{\bibinfo{person}{Matt Kusner}, \bibinfo{person}{Yu Sun}, \bibinfo{person}{Nicholas Kolkin}, {and} \bibinfo{person}{Kilian Weinberger}.} \bibinfo{year}{2015}\natexlab{}.
\newblock \showarticletitle{From Word Embeddings To Document Distances}. In \bibinfo{booktitle}{\emph{Proceedings of the 32nd International Conference on Machine Learning}} \emph{(\bibinfo{series}{Proceedings of Machine Learning Research}, Vol.~\bibinfo{volume}{37})}, \bibfield{editor}{\bibinfo{person}{Francis Bach} {and} \bibinfo{person}{David Blei}} (Eds.). \bibinfo{publisher}{PMLR}, \bibinfo{address}{Lille, France}, \bibinfo{pages}{957--966}.
\newblock
\urldef\tempurl%
\url{https://proceedings.mlr.press/v37/kusnerb15.html}
\showURL{%
\tempurl}


\bibitem[Kyi et~al\mbox{.}(2025)]%
        {kyiGovernanceGenerativeAI2025}
\bibfield{author}{\bibinfo{person}{Lin Kyi}, \bibinfo{person}{Amruta Mahuli}, \bibinfo{person}{M.~Six Silberman}, \bibinfo{person}{Reuben Binns}, \bibinfo{person}{Jun Zhao}, {and} \bibinfo{person}{Asia~J. Biega}.} \bibinfo{year}{2025}\natexlab{}.
\newblock \showarticletitle{Governance of {{Generative AI}} in {{Creative Work}}: {{Consent}}, {{Credit}}, {{Compensation}}, and {{Beyond}}}. In \bibinfo{booktitle}{\emph{Proceedings of the 2025 {{CHI Conference}} on {{Human Factors}} in {{Computing Systems}}}} \emph{(\bibinfo{series}{{{CHI}} '25})}. \bibinfo{publisher}{Association for Computing Machinery}, \bibinfo{address}{New York, NY, USA}, \bibinfo{pages}{1--16}.
\newblock
\showISBNx{979-8-4007-1394-1}
\href{https://doi.org/10.1145/3706598.3713799}{doi:\nolinkurl{10.1145/3706598.3713799}}


\bibitem[Lai and Chang(2011)]%
        {LaiChang_2011_User_attitudes_toward_dedicated_e-book_readers_for_reading_effects_of_convenience_compatibility_and_media_richness}
\bibfield{author}{\bibinfo{person}{Jung-Yu Lai} {and} \bibinfo{person}{Chih-Yen Chang}.} \bibinfo{year}{2011}\natexlab{}.
\newblock \showarticletitle{User Attitudes toward Dedicated E-book Readers for Reading: {{The}} Effects of Convenience, Compatibility and Media Richness}.
\newblock \bibinfo{journal}{\emph{Online Information Review}} \bibinfo{volume}{35}, \bibinfo{number}{4} (\bibinfo{date}{aug} \bibinfo{year}{2011}), \bibinfo{pages}{558--580}.
\newblock
\showISSN{1468-4527}
\href{https://doi.org/10.1108/14684521111161936}{doi:\nolinkurl{10.1108/14684521111161936}}


\bibitem[Landymore(2025)]%
        {Landymore_2025_OpenAI_Successfully_Sheds_Its_Roots_as_Ethical_Non-Profit}
\bibfield{author}{\bibinfo{person}{Frank Landymore}.} \bibinfo{year}{2025}\natexlab{}.
\newblock \bibinfo{title}{{{OpenAI Successfully Sheds Its Roots}} as an {{Ethical Non-Profit}}}.
\newblock
\urldef\tempurl%
\url{https://futurism.com/artificial-intelligence/openai-sheds-roots-ethical-non-profit}
\showURL{%
\tempurl}


\bibitem[Lanier(2014)]%
        {Lanier_2014_Who_owns_future}
\bibfield{author}{\bibinfo{person}{Jaron Lanier}.} \bibinfo{year}{2014}\natexlab{}.
\newblock \bibinfo{booktitle}{\emph{Who Owns the Future?}}
\newblock \bibinfo{publisher}{Penguin Books}, \bibinfo{address}{London}.
\newblock
\showISBNx{978-0-241-95721-9}
\showLCCN{303.483 3}


\bibitem[Lee et~al\mbox{.}(2025)]%
        {LeeGuptaEtAl_2025_Control_content_use_for_AI_training_with_Cloudflares_managed_robotstxt_and_blocking_for_monetized_content}
\bibfield{author}{\bibinfo{person}{Jin-Hee Lee}, \bibinfo{person}{Dipunj Gupta}, \bibinfo{person}{Brian Mitchell}, \bibinfo{person}{Reid Tatoris}, {and} \bibinfo{person}{Henry Clausen}.} \bibinfo{year}{2025}\natexlab{}.
\newblock \bibinfo{title}{Control Content Use for {{AI}} Training with {{Cloudflare}}'s Managed Robots.Txt and Blocking for Monetized Content}.
\newblock
\urldef\tempurl%
\url{https://blog.cloudflare.com/control-content-use-for-ai-training/}
\showURL{%
\tempurl}


\bibitem[Lewis and Zhong(2013)]%
        {lewis2013root}
\bibfield{author}{\bibinfo{person}{Norman~P Lewis} {and} \bibinfo{person}{Bu Zhong}.} \bibinfo{year}{2013}\natexlab{}.
\newblock \showarticletitle{The root of journalistic plagiarism: Contested attribution beliefs}.
\newblock \bibinfo{journal}{\emph{Journalism \& Mass Communication Quarterly}} \bibinfo{volume}{90}, \bibinfo{number}{1} (\bibinfo{year}{2013}), \bibinfo{pages}{148--166}.
\newblock


\bibitem[Lewis et~al\mbox{.}(2020)]%
        {lewis2020retrieval}
\bibfield{author}{\bibinfo{person}{Patrick Lewis}, \bibinfo{person}{Ethan Perez}, \bibinfo{person}{Aleksandra Piktus}, \bibinfo{person}{Fabio Petroni}, \bibinfo{person}{Vladimir Karpukhin}, \bibinfo{person}{Naman Goyal}, \bibinfo{person}{Heinrich K{\"u}ttler}, \bibinfo{person}{Mike Lewis}, \bibinfo{person}{Wen-tau Yih}, \bibinfo{person}{Tim Rockt{\"a}schel}, {et~al\mbox{.}}} \bibinfo{year}{2020}\natexlab{}.
\newblock \showarticletitle{Retrieval-augmented generation for knowledge-intensive nlp tasks}.
\newblock \bibinfo{journal}{\emph{Advances in neural information processing systems}}  \bibinfo{volume}{33} (\bibinfo{year}{2020}), \bibinfo{pages}{9459--9474}.
\newblock


\bibitem[Li et~al\mbox{.}(2024)]%
        {li2024influencefunctionsworklarge}
\bibfield{author}{\bibinfo{person}{Zhe Li}, \bibinfo{person}{Wei Zhao}, \bibinfo{person}{Yige Li}, {and} \bibinfo{person}{Jun Sun}.} \bibinfo{year}{2024}\natexlab{}.
\newblock \bibinfo{title}{Do Influence Functions Work on Large Language Models?}
\newblock
\showeprint[arxiv]{2409.19998}~[cs.CL]
\urldef\tempurl%
\url{https://arxiv.org/abs/2409.19998}
\showURL{%
\tempurl}


\bibitem[Lin et~al\mbox{.}(2025)]%
        {LinShanEtAl_2025_Stop_DDoS_Attacking_Research_Community_with_AI-Generated_Survey_Papers}
\bibfield{author}{\bibinfo{person}{Jianghao Lin}, \bibinfo{person}{Rong Shan}, \bibinfo{person}{Jiachen Zhu}, \bibinfo{person}{Yunjia Xi}, \bibinfo{person}{Yong Yu}, {and} \bibinfo{person}{Weinan Zhang}.} \bibinfo{year}{2025}\natexlab{}.
\newblock \bibinfo{booktitle}{\emph{Stop {{DDoS Attacking}} the {{Research Community}} with {{AI-Generated Survey Papers}}}}.
\newblock
\showeprint[arXiv]{2510.09686}~[cs]
\href{https://doi.org/10.48550/arXiv.2510.09686}{doi:\nolinkurl{10.48550/arXiv.2510.09686}}


\bibitem[Liu et~al\mbox{.}(2025)]%
        {LiuBlantonEtAl_2025_OLMoTrace_Tracing_Language_Model_Outputs_Back_to_Trillions_of_Training_Tokens}
\bibfield{author}{\bibinfo{person}{Jiacheng Liu}, \bibinfo{person}{Taylor Blanton}, \bibinfo{person}{Yanai Elazar}, \bibinfo{person}{Sewon Min}, \bibinfo{person}{Yen-Sung Chen}, \bibinfo{person}{Arnavi {Chheda-Kothary}}, \bibinfo{person}{Huy Tran}, \bibinfo{person}{Byron Bischoff}, \bibinfo{person}{Eric Marsh}, \bibinfo{person}{Michael Schmitz}, \bibinfo{person}{Cassidy Trier}, \bibinfo{person}{Aaron Sarnat}, \bibinfo{person}{Jenna James}, \bibinfo{person}{Jon Borchardt}, \bibinfo{person}{Bailey Kuehl}, \bibinfo{person}{Evie Yu-Yen Cheng}, \bibinfo{person}{Karen Farley}, \bibinfo{person}{Taira Anderson}, \bibinfo{person}{David Albright}, \bibinfo{person}{Carissa Schoenick}, \bibinfo{person}{Luca Soldaini}, \bibinfo{person}{Dirk Groeneveld}, \bibinfo{person}{Rock~Yuren Pang}, \bibinfo{person}{Pang~Wei Koh}, \bibinfo{person}{Noah~A. Smith}, \bibinfo{person}{Sophie Lebrecht}, \bibinfo{person}{Yejin Choi}, \bibinfo{person}{Hannaneh Hajishirzi}, \bibinfo{person}{Ali Farhadi}, {and} \bibinfo{person}{Jesse Dodge}.}
  \bibinfo{year}{2025}\natexlab{}.
\newblock \showarticletitle{{{OLMoTrace}}: {{Tracing Language Model Outputs Back}} to {{Trillions}} of {{Training Tokens}}}. In \bibinfo{booktitle}{\emph{Proceedings of the 63rd {{Annual Meeting}} of the {{Association}} for {{Computational Linguistics}} ({{Volume}} 3: {{System Demonstrations}})}}, \bibfield{editor}{\bibinfo{person}{Pushkar Mishra}, \bibinfo{person}{Smaranda Muresan}, {and} \bibinfo{person}{Tao Yu}} (Eds.). \bibinfo{publisher}{Association for Computational Linguistics}, \bibinfo{address}{Vienna, Austria}, \bibinfo{pages}{178--188}.
\newblock
\showISBNx{979-8-89176-253-4}
\href{https://doi.org/10.18653/v1/2025.acl-demo.18}{doi:\nolinkurl{10.18653/v1/2025.acl-demo.18}}


\bibitem[Liu et~al\mbox{.}(2021)]%
        {LiuJiangEtAl_2021_Why_Audiences_Donate_Money_to_Content_Creators_Uses_and_Gratifications_Perspective}
\bibfield{author}{\bibinfo{person}{Lili Liu}, \bibinfo{person}{Jiujiu Jiang}, \bibinfo{person}{Shanjiao Ren}, {and} \bibinfo{person}{Linwei Hu}.} \bibinfo{year}{2021}\natexlab{}.
\newblock \showarticletitle{Why {{Audiences Donate Money}} to {{Content Creators}}? {{A Uses}} and {{Gratifications Perspective}}}. In \bibinfo{booktitle}{\emph{{{HCI International}} 2021 - {{Late Breaking Posters}}}}, \bibfield{editor}{\bibinfo{person}{Constantine Stephanidis}, \bibinfo{person}{Margherita Antona}, {and} \bibinfo{person}{Stavroula Ntoa}} (Eds.). \bibinfo{publisher}{Springer International Publishing}, \bibinfo{address}{Cham}, \bibinfo{pages}{390--398}.
\newblock
\showISBNx{978-3-030-90179-0}
\href{https://doi.org/10.1007/978-3-030-90179-0\_50}{doi:\nolinkurl{10.1007/978-3-030-90179-0\_50}}


\bibitem[Liu et~al\mbox{.}(2023)]%
        {LiuZhangEtAl_2023_Evaluating_Verifiability_in_Generative_Search_Engines}
\bibfield{author}{\bibinfo{person}{Nelson Liu}, \bibinfo{person}{Tianyi Zhang}, {and} \bibinfo{person}{Percy Liang}.} \bibinfo{year}{2023}\natexlab{}.
\newblock \showarticletitle{Evaluating {{Verifiability}} in {{Generative Search Engines}}}. In \bibinfo{booktitle}{\emph{Findings of the {{Association}} for {{Computational Linguistics}}: {{EMNLP}} 2023}}, \bibfield{editor}{\bibinfo{person}{Houda Bouamor}, \bibinfo{person}{Juan Pino}, {and} \bibinfo{person}{Kalika Bali}} (Eds.). \bibinfo{publisher}{Association for Computational Linguistics}, \bibinfo{address}{Singapore}, \bibinfo{pages}{7001--7025}.
\newblock
\href{https://doi.org/10.18653/v1/2023.findings-emnlp.467}{doi:\nolinkurl{10.18653/v1/2023.findings-emnlp.467}}


\bibitem[Liu et~al\mbox{.}(2024)]%
        {liu-et-al-2024}
\bibfield{author}{\bibinfo{person}{Xiaoze Liu}, \bibinfo{person}{Ting Sun}, \bibinfo{person}{Tianyang Xu}, \bibinfo{person}{Feijie Wu}, \bibinfo{person}{Cunxiang Wang}, \bibinfo{person}{Xiaoqian Wang}, {and} \bibinfo{person}{Jing Gao}.} \bibinfo{year}{2024}\natexlab{}.
\newblock \showarticletitle{{SHIELD}: Evaluation and Defense Strategies for Copyright Compliance in {LLM} Text Generation}. In \bibinfo{booktitle}{\emph{Proceedings of the 2024 Conference on Empirical Methods in Natural Language Processing}}, \bibfield{editor}{\bibinfo{person}{Yaser Al-Onaizan}, \bibinfo{person}{Mohit Bansal}, {and} \bibinfo{person}{Yun-Nung Chen}} (Eds.). \bibinfo{publisher}{Association for Computational Linguistics}, \bibinfo{address}{Miami, Florida, USA}, \bibinfo{pages}{1640--1670}.
\newblock
\href{https://doi.org/10.18653/v1/2024.emnlp-main.98}{doi:\nolinkurl{10.18653/v1/2024.emnlp-main.98}}


\bibitem[Livni et~al\mbox{.}(2024)]%
        {LivniMoranEtAl_2024_Credit_attribution_and_stable_compression}
\bibfield{author}{\bibinfo{person}{Roi Livni}, \bibinfo{person}{Shay Moran}, \bibinfo{person}{Kobbi Nissim}, {and} \bibinfo{person}{Chirag Pabbaraju}.} \bibinfo{year}{2024}\natexlab{}.
\newblock \showarticletitle{Credit Attribution and Stable Compression}. In \bibinfo{booktitle}{\emph{Proceedings of the 38th {{International Conference}} on {{Neural Information Processing Systems}}}} \emph{(\bibinfo{series}{{{NIPS}} '24}, Vol.~\bibinfo{volume}{37})}. \bibinfo{publisher}{Curran Associates Inc.}, \bibinfo{address}{Red Hook, NY, USA}, \bibinfo{pages}{2663--2685}.
\newblock
\showISBNx{979-8-3313-1438-5}


\bibitem[Longpre et~al\mbox{.}(2024)]%
        {longpreConsentCrisisRapid2024}
\bibfield{author}{\bibinfo{person}{Shayne Longpre}, \bibinfo{person}{Robert Mahari}, \bibinfo{person}{Ariel Lee}, \bibinfo{person}{Campbell Lund}, \bibinfo{person}{Hamidah Oderinwale}, \bibinfo{person}{William Brannon}, \bibinfo{person}{Nayan Saxena}, \bibinfo{person}{Naana {Obeng-Marnu}}, \bibinfo{person}{Tobin South}, \bibinfo{person}{Cole Hunter}, \bibinfo{person}{Kevin Klyman}, \bibinfo{person}{Christopher Klamm}, \bibinfo{person}{Hailey Schoelkopf}, \bibinfo{person}{Nikhil Singh}, \bibinfo{person}{Manuel Cherep}, \bibinfo{person}{Ahmad Anis}, \bibinfo{person}{An Dinh}, \bibinfo{person}{Caroline Chitongo}, \bibinfo{person}{Da Yin}, \bibinfo{person}{Damien Sileo}, \bibinfo{person}{Deividas Mataciunas}, \bibinfo{person}{Diganta Misra}, \bibinfo{person}{Emad Alghamdi}, \bibinfo{person}{Enrico Shippole}, \bibinfo{person}{Jianguo Zhang}, \bibinfo{person}{Joanna Materzynska}, \bibinfo{person}{Kun Qian}, \bibinfo{person}{Kush Tiwary}, \bibinfo{person}{Lester Miranda}, \bibinfo{person}{Manan Dey},
  \bibinfo{person}{Minnie Liang}, \bibinfo{person}{Mohammed Hamdy}, \bibinfo{person}{Niklas Muennighoff}, \bibinfo{person}{Seonghyeon Ye}, \bibinfo{person}{Seungone Kim}, \bibinfo{person}{Shrestha Mohanty}, \bibinfo{person}{Vipul Gupta}, \bibinfo{person}{Vivek Sharma}, \bibinfo{person}{Vu~Minh Chien}, \bibinfo{person}{Xuhui Zhou}, \bibinfo{person}{Yizhi Li}, \bibinfo{person}{Caiming Xiong}, \bibinfo{person}{Luis Villa}, \bibinfo{person}{Stella Biderman}, \bibinfo{person}{Hanlin Li}, \bibinfo{person}{Daphne Ippolito}, \bibinfo{person}{Sara Hooker}, \bibinfo{person}{Jad Kabbara}, {and} \bibinfo{person}{Sandy Pentland}.} \bibinfo{year}{2024}\natexlab{}.
\newblock \bibinfo{title}{Consent in {{Crisis}}: {{The Rapid Decline}} of the {{AI Data Commons}}}.
\newblock
\showeprint[arxiv]{2407.14933}~[cs]
\href{https://doi.org/10.48550/arXiv.2407.14933}{doi:\nolinkurl{10.48550/arXiv.2407.14933}}


\bibitem[Lorenz(2026)]%
        {Lorenz_2026_Dark-Money_Campaign_Is_Paying_Influencers_to_Frame_Chinese_AI_as_Threat}
\bibfield{author}{\bibinfo{person}{Taylor Lorenz}.} \bibinfo{year}{2026}\natexlab{}.
\newblock \showarticletitle{A {{Dark-Money Campaign Is Paying Influencers}} to {{Frame Chinese AI}} as a {{Threat}}}.
\newblock  (\bibinfo{year}{2026}).
\newblock
\showISSN{1059-1028}
\urldef\tempurl%
\url{https://www.wired.com/story/super-pac-backed-by-openai-and-palantir-is-paying-tiktok-influencers-to-fear-monger-about-china/}
\showURL{%
\tempurl}


\bibitem[Love et~al\mbox{.}(2025)]%
        {LoveSolonEtAl_2025_Google_Removes_Language_on_Weapons_From_Public_AI_Principles}
\bibfield{author}{\bibinfo{person}{Julia Love}, \bibinfo{person}{Olivia Solon}, {and} \bibinfo{person}{Davey Alba}.} \bibinfo{year}{2025}\natexlab{}.
\newblock \showarticletitle{Google {{Removes Language}} on {{Weapons From Public AI Principles}}}.
\newblock \bibinfo{journal}{\emph{Bloomberg.com}} (\bibinfo{date}{feb} \bibinfo{year}{2025}).
\newblock
\urldef\tempurl%
\url{https://www.bloomberg.com/news/articles/2025-02-04/google-removes-language-on-weapons-from-public-ai-principles}
\showURL{%
\tempurl}


\bibitem[Mak(2026)]%
        {Mak_2026_OpenAIs_spin_on_universal_basic_income}
\bibfield{author}{\bibinfo{person}{Aaron Mak}.} \bibinfo{year}{2026}\natexlab{}.
\newblock \bibinfo{booktitle}{\emph{{{OpenAI}}’s Spin on Universal Basic Income}}.
\newblock POLITICO.
\newblock
\urldef\tempurl%
\url{https://www.politico.com/newsletters/digital-future-daily/2026/04/07/openais-spin-on-universal-basic-income-00861981}
\showURL{%
\tempurl}


\bibitem[Manning(2005)]%
        {Manning_2005_Monopsony_in_Motion_Imperfect_Competition_in_Labor_Markets}
\bibfield{author}{\bibinfo{person}{Alan Manning}.} \bibinfo{year}{2005}\natexlab{}.
\newblock \bibinfo{booktitle}{\emph{Monopsony in {{Motion}}: {{Imperfect Competition}} in {{Labor Markets}}}}.
\newblock \bibinfo{publisher}{Princeton University Press}, \bibinfo{address}{Princeton, N.J}.
\newblock
\showISBNx{978-0-691-12328-8 978-1-4008-5067-9}
\href{https://doi.org/10.1515/9781400850679}{doi:\nolinkurl{10.1515/9781400850679}}


\bibitem[Martinez and Borovik(2025)]%
        {MartinezBorovik_2025_No_harm_no_win_cautionary_tale_of_Kadrey_Meta_Platforms_Inc}
\bibfield{author}{\bibinfo{person}{David Martinez} {and} \bibinfo{person}{Belle Borovik}.} \bibinfo{year}{2025}\natexlab{}.
\newblock \bibinfo{booktitle}{\emph{No Harm, No Win: {{A}} Cautionary Tale of {{Kadrey}} v. {{Meta Platforms Inc}}.}}
\newblock Los Angeles \& San Francisco Daily Journal.
\newblock
\urldef\tempurl%
\url{https://www.robinskaplan.com/newsroom/insights/no-harm-no-win-a-cautionary-tale-of-kadrey-v-meta-platforms-inc}
\showURL{%
\tempurl}


\bibitem[Maruccia(2025)]%
        {Maruccia_2025_Salesforce_hikes_Slack_prices_adds_AI_tools_for_all_paid_users}
\bibfield{author}{\bibinfo{person}{Alfonso Maruccia}.} \bibinfo{year}{2025}\natexlab{}.
\newblock \bibinfo{title}{Salesforce Hikes {{Slack}} Prices, Adds {{AI}} Tools for All Paid Users}.
\newblock
\urldef\tempurl%
\url{https://www.techspot.com/news/108366-salesforce-latest-price-increase-comes-promise-more-ai.html}
\showURL{%
\tempurl}


\bibitem[Maruf(2024)]%
        {Maruf_2024_changed_its_terms_of_service_to_let_its_AI_train_on_everyones_posts_Now_users_are_up_in_arms}
\bibfield{author}{\bibinfo{person}{Ramishah Maruf}.} \bibinfo{year}{2024}\natexlab{}.
\newblock \bibinfo{title}{X Changed Its Terms of Service to Let Its {{AI}} Train on Everyone's Posts. {{Now}} Users Are up in Arms}.
\newblock
\urldef\tempurl%
\url{https://www.cnn.com/2024/10/21/tech/x-twitter-terms-of-service}
\showURL{%
\tempurl}


\bibitem[Mehrotra and Marchman(2025)]%
        {MEHROTRA-MARCHMAN-perplexity-bullshit-machine}
\bibfield{author}{\bibinfo{person}{Dhruv Mehrotra} {and} \bibinfo{person}{Tim Marchman}.} \bibinfo{year}{2025}\natexlab{}.
\newblock \bibinfo{booktitle}{\emph{Perplexity Is a Bullshit Machine}}.
\newblock Wired.
\newblock
\urldef\tempurl%
\url{https://www.wired.com/story/perplexity-is-a-bullshit-machine/}
\showURL{%
\tempurl}


\bibitem[Meier(2024)]%
        {meier2024plagiat}
\bibfield{author}{\bibinfo{person}{Klaus Meier}.} \bibinfo{year}{2024}\natexlab{}.
\newblock \showarticletitle{Was ist ein Plagiat im Journalismus?: Ma{\ss}st{\"a}be, nach denen sich Redaktionen richten k{\"o}nnen}.
\newblock \bibinfo{journal}{\emph{Journalistik: Zeitschrift f{\"u}r Journalismusforschung}} \bibinfo{volume}{7}, \bibinfo{number}{2} (\bibinfo{year}{2024}), \bibinfo{pages}{204--210}.
\newblock


\bibitem[Menick et~al\mbox{.}(2022)]%
        {menick2022teachinglanguagemodelssupport}
\bibfield{author}{\bibinfo{person}{Jacob Menick}, \bibinfo{person}{Maja Trebacz}, \bibinfo{person}{Vladimir Mikulik}, \bibinfo{person}{John Aslanides}, \bibinfo{person}{Francis Song}, \bibinfo{person}{Martin Chadwick}, \bibinfo{person}{Mia Glaese}, \bibinfo{person}{Susannah Young}, \bibinfo{person}{Lucy Campbell-Gillingham}, \bibinfo{person}{Geoffrey Irving}, {and} \bibinfo{person}{Nat McAleese}.} \bibinfo{year}{2022}\natexlab{}.
\newblock \bibinfo{title}{Teaching language models to support answers with verified quotes}.
\newblock
\showeprint[arxiv]{2203.11147}~[cs.CL]
\urldef\tempurl%
\url{https://arxiv.org/abs/2203.11147}
\showURL{%
\tempurl}


\bibitem[Mikolov et~al\mbox{.}(2013)]%
        {mikolov2013efficientestimationwordrepresentations}
\bibfield{author}{\bibinfo{person}{Tomas Mikolov}, \bibinfo{person}{Kai Chen}, \bibinfo{person}{Greg Corrado}, {and} \bibinfo{person}{Jeffrey Dean}.} \bibinfo{year}{2013}\natexlab{}.
\newblock \bibinfo{title}{Efficient Estimation of Word Representations in Vector Space}.
\newblock
\showeprint[arxiv]{1301.3781}~[cs.CL]
\urldef\tempurl%
\url{https://arxiv.org/abs/1301.3781}
\showURL{%
\tempurl}


\bibitem[Mitchell et~al\mbox{.}(2023)]%
        {iiMeasuringData2023}
\bibfield{author}{\bibinfo{person}{Margaret Mitchell}, \bibinfo{person}{Alexandra~Sasha Luccioni}, \bibinfo{person}{Nathan Lambert}, \bibinfo{person}{Marissa Gerchick}, \bibinfo{person}{Angelina {McMillan-Major}}, \bibinfo{person}{Ezinwanne Ozoani}, \bibinfo{person}{Nazneen Rajani}, \bibinfo{person}{Tristan Thrush}, \bibinfo{person}{Yacine Jernite}, {and} \bibinfo{person}{Douwe Kiela}.} \bibinfo{year}{2023}\natexlab{}.
\newblock \bibinfo{title}{Measuring {{Data}}}.
\newblock
\showeprint[arxiv]{2212.05129}~[cs]
\href{https://doi.org/10.48550/arXiv.2212.05129}{doi:\nolinkurl{10.48550/arXiv.2212.05129}}


\bibitem[Mitchell et~al\mbox{.}(2019)]%
        {MitchellWuEtAl_2019_Model_Cards_for_Model_Reporting}
\bibfield{author}{\bibinfo{person}{Margaret Mitchell}, \bibinfo{person}{Simone Wu}, \bibinfo{person}{Andrew Zaldivar}, \bibinfo{person}{Parker Barnes}, \bibinfo{person}{Lucy Vasserman}, \bibinfo{person}{Ben Hutchinson}, \bibinfo{person}{Elena Spitzer}, \bibinfo{person}{Inioluwa~Deborah Raji}, {and} \bibinfo{person}{Timnit Gebru}.} \bibinfo{year}{2019}\natexlab{}.
\newblock \showarticletitle{Model {{Cards}} for {{Model Reporting}}}. In \bibinfo{booktitle}{\emph{Proceedings of the {{Conference}} on {{Fairness}}, {{Accountability}}, and {{Transparency}}}} \emph{(\bibinfo{series}{{{FAT}}* '19})}. \bibinfo{publisher}{Association for Computing Machinery}, \bibinfo{address}{New York, NY, USA}, \bibinfo{pages}{220--229}.
\newblock
\showISBNx{978-1-4503-6125-5}
\href{https://doi.org/10.1145/3287560.3287596}{doi:\nolinkurl{10.1145/3287560.3287596}}


\bibitem[Morris et~al\mbox{.}(2022)]%
        {MorrisGrehlEtAl_2022_On_what_motivates_us_detailed_review_of_intrinsic_extrinsic_motivation}
\bibfield{author}{\bibinfo{person}{Laurel~S. Morris}, \bibinfo{person}{Mora~M. Grehl}, \bibinfo{person}{Sarah~B. Rutter}, \bibinfo{person}{Marishka Mehta}, {and} \bibinfo{person}{Margaret~L. Westwater}.} \bibinfo{year}{2022}\natexlab{}.
\newblock \showarticletitle{On What Motivates Us: A Detailed Review of Intrinsic v. Extrinsic Motivation}.
\newblock \bibinfo{journal}{\emph{Psychological Medicine}} \bibinfo{volume}{52}, \bibinfo{number}{10} (\bibinfo{date}{jul} \bibinfo{year}{2022}), \bibinfo{pages}{1801--1816}.
\newblock
\showISSN{0033-2917, 1469-8978}
\href{https://doi.org/10.1017/S0033291722001611}{doi:\nolinkurl{10.1017/S0033291722001611}}


\bibitem[Morrone(2024)]%
        {Morrone_2024_New_report_60_of_OpenAI_models_responses_contain_plagiarism}
\bibfield{author}{\bibinfo{person}{Megan Morrone}.} \bibinfo{year}{2024}\natexlab{}.
\newblock \bibinfo{title}{New Report: 60\% of {{OpenAI}} Model's Responses Contain Plagiarism}.
\newblock
\urldef\tempurl%
\url{https://www.axios.com/2024/02/22/copyleaks-openai-chatgpt-plagiarism}
\showURL{%
\tempurl}


\bibitem[Murphy(2021)]%
        {Murphy_2021_Facebook_confronts_growth_problems_as_number_of_young_users_in_US_declines}
\bibfield{author}{\bibinfo{person}{Hannah Murphy}.} \bibinfo{year}{2021}\natexlab{}.
\newblock \showarticletitle{Facebook Confronts Growth Problems as Number of Young Users in {{US}} Declines}.
\newblock \bibinfo{journal}{\emph{Financial Times}} (\bibinfo{date}{oct} \bibinfo{year}{2021}).
\newblock
\urldef\tempurl%
\url{https://www.ft.com/content/4304f14a-1b06-46d8-a066-42bb1b3c200c}
\showURL{%
\tempurl}


\bibitem[Nayak(2019)]%
        {Nayak_2019_Understanding_searches_better_than_ever_before}
\bibfield{author}{\bibinfo{person}{Pandu Nayak}.} \bibinfo{year}{2019}\natexlab{}.
\newblock \bibinfo{title}{Understanding Searches Better than Ever Before}.
\newblock
\urldef\tempurl%
\url{https://blog.google/products-and-platforms/products/search/search-language-understanding-bert/}
\showURL{%
\tempurl}


\bibitem[Nelson and Kim(2021)]%
        {NelsonKim_2021_Improve_Trust_Increase_Loyalty_Analyzing_Relationship_Between_News_Credibility_and_Consumption}
\bibfield{author}{\bibinfo{person}{Jacob~L. Nelson} {and} \bibinfo{person}{Su~Jung Kim}.} \bibinfo{year}{2021}\natexlab{}.
\newblock \showarticletitle{Improve {{Trust}}, {{Increase Loyalty}}? {{Analyzing}} the {{Relationship Between News Credibility}} and {{Consumption}}}.
\newblock \bibinfo{journal}{\emph{Journalism Practice}} \bibinfo{volume}{15}, \bibinfo{number}{3} (\bibinfo{date}{mar} \bibinfo{year}{2021}), \bibinfo{pages}{348--365}.
\newblock
\showISSN{1751-2786, 1751-2794}
\href{https://doi.org/10.1080/17512786.2020.1719874}{doi:\nolinkurl{10.1080/17512786.2020.1719874}}


\bibitem[Nelson(1999)]%
        {Nelson_1999_Xanalogical_structure_needed_now_more_than_ever_parallel_documents_deep_links_to_content_deep_versioning_and_deep_re-use}
\bibfield{author}{\bibinfo{person}{Theodor~Holm Nelson}.} \bibinfo{year}{1999}\natexlab{}.
\newblock \showarticletitle{Xanalogical Structure, Needed Now More than Ever: Parallel Documents, Deep Links to Content, Deep Versioning, and Deep Re-Use}.
\newblock \bibinfo{journal}{\emph{ACM Comput. Surv.}} \bibinfo{volume}{31}, \bibinfo{number}{4es} (\bibinfo{date}{dec} \bibinfo{year}{1999}), \bibinfo{pages}{33--es}.
\newblock
\showISSN{0360-0300}
\href{https://doi.org/10.1145/345966.346033}{doi:\nolinkurl{10.1145/345966.346033}}


\bibitem[Newman(2026)]%
        {Newman_2026_Journalism_media_and_technology_trends_and_predictions_2026_Reuters_Institute_for_Study_of_Journalism}
\bibfield{author}{\bibinfo{person}{Nic Newman}.} \bibinfo{year}{2026}\natexlab{}.
\newblock \bibinfo{title}{Journalism, Media, and Technology Trends and Predictions 2026 \textbar{} {{Reuters Institute}} for the {{Study}} of {{Journalism}}}.
\newblock
\urldef\tempurl%
\url{http://reutersinstitute.politics.ox.ac.uk/journalism-media-and-technology-trends-and-predictions-2026}
\showURL{%
\tempurl}


\bibitem[Nicholls(2024a)]%
        {Nicholls_2024_Facebook_wont_keep_paying_Australian_media_outlets_for_their_content_Are_we_about_to_get_another_news_ban}
\bibfield{author}{\bibinfo{person}{Rob Nicholls}.} \bibinfo{year}{2024}\natexlab{a}.
\newblock \bibinfo{title}{Facebook Won't Keep Paying {{Australian}} Media Outlets for Their Content. {{Are}} We about to Get Another News Ban?}
\newblock
\href{https://doi.org/10.64628/AA.4nmed99tc}{doi:\nolinkurl{10.64628/AA.4nmed99tc}}


\bibitem[Nicholls(2024b)]%
        {Nicholls_2024_Facebook_wont_keep_paying_Australian_media_outlets_for_their_content_Are_we_about_to_get_another_news_bana}
\bibfield{author}{\bibinfo{person}{Rob Nicholls}.} \bibinfo{year}{2024}\natexlab{b}.
\newblock \bibinfo{title}{Facebook Won't Keep Paying {{Australian}} Media Outlets for Their Content. {{Are}} We about to Get Another News Ban?}
\newblock
\href{https://doi.org/10.64628/AA.4nmed99tc}{doi:\nolinkurl{10.64628/AA.4nmed99tc}}


\bibitem[Nowbar(2023)]%
        {Nowbar_2023_Microsoft_announces_new_Copilot_Copyright_Commitment_for_customers}
\bibfield{author}{\bibinfo{person}{Brad~Smith Nowbar, Hossein}.} \bibinfo{year}{2023}\natexlab{}.
\newblock \bibinfo{title}{Microsoft Announces New {{Copilot Copyright Commitment}} for Customers}.
\newblock
\urldef\tempurl%
\url{https://blogs.microsoft.com/on-the-issues/2023/09/07/copilot-copyright-commitment-ai-legal-concerns/}
\showURL{%
\tempurl}


\bibitem[{Office of Technology {and} The Division of Privacy {and} Identity Protection}(2024)]%
        {OfficeofTechnologyandTheDivisionofPrivacyandIdentityProtection_2024_AI_and_other_Companies_Quietly_Changing_Your_Terms_of_Service_Could_Be_Unfair_or_Deceptive}
\bibfield{author}{\bibinfo{person}{{Office of Technology {and} The Division of Privacy {and} Identity Protection}}.} \bibinfo{year}{2024}\natexlab{}.
\newblock \bibinfo{title}{{{AI}} (and Other) {{Companies}}: {{Quietly Changing Your Terms}} of {{Service Could Be Unfair}} or {{Deceptive}}}.
\newblock
\urldef\tempurl%
\url{https://www.ftc.gov/policy/advocacy-research/tech-at-ftc/2024/02/ai-other-companies-quietly-changing-your-terms-service-could-be-unfair-or-deceptive}
\showURL{%
\tempurl}


\bibitem[Olmo et~al\mbox{.}(2025)]%
        {olmo2025olmo3}
\bibfield{author}{\bibinfo{person}{Team Olmo}, \bibinfo{person}{:}, \bibinfo{person}{Allyson Ettinger}, \bibinfo{person}{Amanda Bertsch}, \bibinfo{person}{Bailey Kuehl}, \bibinfo{person}{David Graham}, \bibinfo{person}{David Heineman}, \bibinfo{person}{Dirk Groeneveld}, \bibinfo{person}{Faeze Brahman}, \bibinfo{person}{Finbarr Timbers}, \bibinfo{person}{Hamish Ivison}, \bibinfo{person}{Jacob Morrison}, \bibinfo{person}{Jake Poznanski}, \bibinfo{person}{Kyle Lo}, \bibinfo{person}{Luca Soldaini}, \bibinfo{person}{Matt Jordan}, \bibinfo{person}{Mayee Chen}, \bibinfo{person}{Michael Noukhovitch}, \bibinfo{person}{Nathan Lambert}, \bibinfo{person}{Pete Walsh}, \bibinfo{person}{Pradeep Dasigi}, \bibinfo{person}{Robert Berry}, \bibinfo{person}{Saumya Malik}, \bibinfo{person}{Saurabh Shah}, \bibinfo{person}{Scott Geng}, \bibinfo{person}{Shane Arora}, \bibinfo{person}{Shashank Gupta}, \bibinfo{person}{Taira Anderson}, \bibinfo{person}{Teng Xiao}, \bibinfo{person}{Tyler Murray}, \bibinfo{person}{Tyler Romero},
  \bibinfo{person}{Victoria Graf}, \bibinfo{person}{Akari Asai}, \bibinfo{person}{Akshita Bhagia}, \bibinfo{person}{Alexander Wettig}, \bibinfo{person}{Alisa Liu}, \bibinfo{person}{Aman Rangapur}, \bibinfo{person}{Chloe Anastasiades}, \bibinfo{person}{Costa Huang}, \bibinfo{person}{Dustin Schwenk}, \bibinfo{person}{Harsh Trivedi}, \bibinfo{person}{Ian Magnusson}, \bibinfo{person}{Jaron Lochner}, \bibinfo{person}{Jiacheng Liu}, \bibinfo{person}{Lester James~V. Miranda}, \bibinfo{person}{Maarten Sap}, \bibinfo{person}{Malia Morgan}, \bibinfo{person}{Michael Schmitz}, \bibinfo{person}{Michal Guerquin}, \bibinfo{person}{Michael Wilson}, \bibinfo{person}{Regan Huff}, \bibinfo{person}{Ronan~Le Bras}, \bibinfo{person}{Rui Xin}, \bibinfo{person}{Rulin Shao}, \bibinfo{person}{Sam Skjonsberg}, \bibinfo{person}{Shannon~Zejiang Shen}, \bibinfo{person}{Shuyue~Stella Li}, \bibinfo{person}{Tucker Wilde}, \bibinfo{person}{Valentina Pyatkin}, \bibinfo{person}{Will Merrill}, \bibinfo{person}{Yapei Chang},
  \bibinfo{person}{Yuling Gu}, \bibinfo{person}{Zhiyuan Zeng}, \bibinfo{person}{Ashish Sabharwal}, \bibinfo{person}{Luke Zettlemoyer}, \bibinfo{person}{Pang~Wei Koh}, \bibinfo{person}{Ali Farhadi}, \bibinfo{person}{Noah~A. Smith}, {and} \bibinfo{person}{Hannaneh Hajishirzi}.} \bibinfo{year}{2025}\natexlab{}.
\newblock \bibinfo{title}{Olmo 3}.
\newblock
\showeprint[arxiv]{2512.13961}~[cs.CL]
\urldef\tempurl%
\url{https://arxiv.org/abs/2512.13961}
\showURL{%
\tempurl}


\bibitem[Olson(1971)]%
        {Olson_1971_logic_of_collective_action_public_goods_and_theory_of_groups}
\bibfield{author}{\bibinfo{person}{Mancur Olson}.} \bibinfo{year}{1971}\natexlab{}.
\newblock \bibinfo{booktitle}{\emph{The Logic of Collective Action: Public Goods and the Theory of Groups}}.
\newblock Number 124 in \bibinfo{series}{Harvard {{Economic Studies}}}. \bibinfo{publisher}{Harvard university press}, \bibinfo{address}{Cambridge (Mass.) London}.
\newblock
\showISBNx{978-0-674-53751-4}
\showLCCN{323.042}


\bibitem[OpenAI(2023)]%
        {OpenAI_2023_Terms_of_use}
\bibfield{author}{\bibinfo{person}{OpenAI}.} \bibinfo{year}{2023}\natexlab{}.
\newblock \bibinfo{title}{Terms of Use}.
\newblock
\urldef\tempurl%
\url{https://openai.com/policies/terms-of-use}
\showURL{%
\tempurl}


\bibitem[OpenAI(2025)]%
        {OpenAI_2025_OpenAI_Response_OSTP_NSF_RFI_Notice_Request_for_Information_on_Development_of_Artificial_Intelligence_AI_Action_Plan}
\bibfield{author}{\bibinfo{person}{OpenAI}.} \bibinfo{year}{2025}\natexlab{}.
\newblock \bibinfo{title}{[{{OpenAI Response}}] {{OSTP}}/{{NSF RFI}}: {{Notice Request}} for {{Information}} on the {{Development}} of an {{Artificial Intelligence}} ({{AI}}) {{Action Plan}}}.
\newblock
\urldef\tempurl%
\url{https://cdn.openai.com/global-affairs/ostp-rfi/ec680b75-d539-4653-b297-8bcf6e5f7686/openai-response-ostp-nsf-rfi-notice-request-for-information-on-the-development-of-an-artificial-intelligence-ai-action-plan.pdf}
\showURL{%
\tempurl}


\bibitem[Orosz(2025)]%
        {Orosz_2025_Stack_overflow_is_almost_dead}
\bibfield{author}{\bibinfo{person}{Gergely Orosz}.} \bibinfo{year}{2025}\natexlab{}.
\newblock \bibinfo{booktitle}{\emph{Stack Overflow Is Almost Dead}}.
\newblock The Pragmatic Engineer.
\newblock
\urldef\tempurl%
\url{https://blog.pragmaticengineer.com/stack-overflow-is-almost-dead/}
\showURL{%
\tempurl}


\bibitem[Oya(2020)]%
        {oya-2020-syntactic}
\bibfield{author}{\bibinfo{person}{Masanori Oya}.} \bibinfo{year}{2020}\natexlab{}.
\newblock \showarticletitle{Syntactic similarity of the sentences in a multi-lingual parallel corpus based on the {E}uclidean distance of their dependency trees}. In \bibinfo{booktitle}{\emph{Proceedings of the 34th Pacific Asia Conference on Language, Information and Computation}}, \bibfield{editor}{\bibinfo{person}{Minh~Le Nguyen}, \bibinfo{person}{Mai~Chi Luong}, {and} \bibinfo{person}{Sanghoun Song}} (Eds.). \bibinfo{publisher}{Association for Computational Linguistics}, \bibinfo{address}{Hanoi, Vietnam}, \bibinfo{pages}{225--233}.
\newblock
\urldef\tempurl%
\url{https://aclanthology.org/2020.paclic-1.26/}
\showURL{%
\tempurl}


\bibitem[Pahwa(2024)]%
        {Pahwa_2024_How_Quora_Died}
\bibfield{author}{\bibinfo{person}{Nitish Pahwa}.} \bibinfo{year}{2024}\natexlab{}.
\newblock \showarticletitle{How {{Quora Died}}}.
\newblock  (\bibinfo{year}{2024}).
\newblock
\showISSN{1091-2339}
\urldef\tempurl%
\url{https://slate.com/technology/2024/02/quora-what-happened-ai-decline.html}
\showURL{%
\tempurl}


\bibitem[Palmer({[n.\,d.]})]%
        {Palmer_Taylor_Francis_AI_Deal_Sets_Worrying_Precedent_for_Academic_Publishing}
\bibfield{author}{\bibinfo{person}{Kathryn Palmer}.} \bibinfo{year}{[n.\,d.]}\natexlab{}.
\newblock \bibinfo{title}{Taylor \& {{Francis AI Deal Sets}} `{{Worrying Precedent}}' for {{Academic Publishing}}}.
\newblock
\urldef\tempurl%
\url{https://www.insidehighered.com/news/faculty-issues/research/2024/07/29/taylor-francis-ai-deal-sets-worrying-precedent}
\showURL{%
\tempurl}


\bibitem[Palmer(2024)]%
        {Palmer_2024_Prestige_Factor_Propping_Up_Academic_Publishers}
\bibfield{author}{\bibinfo{person}{Kathryn Palmer}.} \bibinfo{year}{2024}\natexlab{}.
\newblock \bibinfo{title}{The {{Prestige Factor Propping Up Academic Publishers}}}.
\newblock
\urldef\tempurl%
\url{https://www.insidehighered.com/news/faculty-issues/research/2024/09/23/lawsuit-highlights-how-prestige-drives-academic-publishing}
\showURL{%
\tempurl}


\bibitem[Pandit et~al\mbox{.}(2026)]%
        {pandit2026termsabuseanalysisgenai}
\bibfield{author}{\bibinfo{person}{Harshvardhan~J. Pandit}, \bibinfo{person}{Dick A.~H. Blankvoort}, \bibinfo{person}{Adel Shaaban}, \bibinfo{person}{Sasha Luccioni}, {and} \bibinfo{person}{Abeba Birhane}.} \bibinfo{year}{2026}\natexlab{}.
\newblock \bibinfo{title}{Terms of (Ab)Use: An Analysis of GenAI Services}.
\newblock
\showeprint[arxiv]{2603.18964}~[cs.CY]
\urldef\tempurl%
\url{https://arxiv.org/abs/2603.18964}
\showURL{%
\tempurl}


\bibitem[Papineni et~al\mbox{.}(2002)]%
        {papineni-etal-2002-bleu}
\bibfield{author}{\bibinfo{person}{Kishore Papineni}, \bibinfo{person}{Salim Roukos}, \bibinfo{person}{Todd Ward}, {and} \bibinfo{person}{Wei-Jing Zhu}.} \bibinfo{year}{2002}\natexlab{}.
\newblock \showarticletitle{{B}leu: a Method for Automatic Evaluation of Machine Translation}. In \bibinfo{booktitle}{\emph{Proceedings of the 40th Annual Meeting of the Association for Computational Linguistics}}, \bibfield{editor}{\bibinfo{person}{Pierre Isabelle}, \bibinfo{person}{Eugene Charniak}, {and} \bibinfo{person}{Dekang Lin}} (Eds.). \bibinfo{publisher}{Association for Computational Linguistics}, \bibinfo{address}{Philadelphia, Pennsylvania, USA}, \bibinfo{pages}{311--318}.
\newblock
\href{https://doi.org/10.3115/1073083.1073135}{doi:\nolinkurl{10.3115/1073083.1073135}}


\bibitem[Paredes et~al\mbox{.}(2024)]%
        {ParedesSmithEtAl_2024_More_Articles_Are_Now_Created_by_AI_Than_Humans}
\bibfield{author}{\bibinfo{person}{Jose~Luis Paredes}, \bibinfo{person}{Ethan Smith}, \bibinfo{person}{Gregory Druck}, {and} \bibinfo{person}{Bevin Benson}.} \bibinfo{year}{2024}\natexlab{}.
\newblock \bibinfo{booktitle}{\emph{More {{Articles Are Now Created}} by {{AI Than Humans}}}}.
\newblock Graphite.
\newblock
\urldef\tempurl%
\url{https://graphite.io/five-percent/more-articles-are-now-created-by-ai-than-humans}
\showURL{%
\tempurl}


\bibitem[Park et~al\mbox{.}(2023)]%
        {park2023trakattributingmodelbehavior}
\bibfield{author}{\bibinfo{person}{Sung~Min Park}, \bibinfo{person}{Kristian Georgiev}, \bibinfo{person}{Andrew Ilyas}, \bibinfo{person}{Guillaume Leclerc}, {and} \bibinfo{person}{Aleksander Madry}.} \bibinfo{year}{2023}\natexlab{}.
\newblock \bibinfo{title}{TRAK: Attributing Model Behavior at Scale}.
\newblock
\showeprint[arxiv]{2303.14186}~[stat.ML]
\urldef\tempurl%
\url{https://arxiv.org/abs/2303.14186}
\showURL{%
\tempurl}


\bibitem[Peukert et~al\mbox{.}(2025)]%
        {PeukertAbeillonEtAl_2025_AI_and_Dynamic_Supply_of_Training_Data}
\bibfield{author}{\bibinfo{person}{Christian Peukert}, \bibinfo{person}{Florian Abeillon}, \bibinfo{person}{Jérémie Haese}, \bibinfo{person}{Franziska Kaiser}, {and} \bibinfo{person}{Alexander Staub}.} \bibinfo{year}{2025}\natexlab{}.
\newblock \bibinfo{booktitle}{\emph{{{AI}} and the {{Dynamic Supply}} of {{Training Data}}}}.
\newblock
\showeprint[arXiv]{2404.18445}~[econ]
\href{https://doi.org/10.48550/arXiv.2404.18445}{doi:\nolinkurl{10.48550/arXiv.2404.18445}}


\bibitem[Phelps(2022)]%
        {Phelps_2022_Challenging_Academic_Publisher_Oligopoly}
\bibfield{author}{\bibinfo{person}{Richard~P. Phelps}.} \bibinfo{year}{2022}\natexlab{}.
\newblock \bibinfo{title}{Challenging the {{Academic Publisher Oligopoly}}}.
\newblock
\urldef\tempurl%
\url{https://mindingthecampus.org/2022/11/18/challenging-the-academic-publisher-oligopoly/}
\showURL{%
\tempurl}


\bibitem[Piktus et~al\mbox{.}(2023)]%
        {PiktusAkikiEtAl_2023_ROOTS_Search_Tool_Data_Transparency_for_LLMs}
\bibfield{author}{\bibinfo{person}{Aleksandra Piktus}, \bibinfo{person}{Christopher Akiki}, \bibinfo{person}{Paulo Villegas}, \bibinfo{person}{Hugo Lauren{\c c}on}, \bibinfo{person}{G{\'e}rard Dupont}, \bibinfo{person}{Alexandra~Sasha Luccioni}, \bibinfo{person}{Yacine Jernite}, {and} \bibinfo{person}{Anna Rogers}.} \bibinfo{year}{2023}\natexlab{}.
\newblock \showarticletitle{The {{ROOTS Search Tool}}: {{Data Transparency}} for {{LLMs}}}. In \bibinfo{booktitle}{\emph{To Appearin {{ACL}} 2023 ({{Demo}} Track)}}. \bibinfo{publisher}{arXiv}.
\newblock
\showeprint[arxiv]{2302.14035}~[cs]
\urldef\tempurl%
\url{http://arxiv.org/abs/2302.14035}
\showURL{%
\tempurl}


\bibitem[ProRataAI(2025)]%
        {ProRataAI_2025_ProRata_Partners_with_Danish_Publishers_Group_DPCMO_to_Launch_First_Decentralized_Sovereign_AI_Answer_Engine}
\bibfield{author}{\bibinfo{person}{ProRataAI}.} \bibinfo{year}{2025}\natexlab{}.
\newblock \bibinfo{title}{{{ProRata Partners}} with {{Danish Publishers Group DPCMO}} to {{Launch}} the {{First Decentralized Sovereign AI Answer Engine}}}.
\newblock
\urldef\tempurl%
\url{https://www.prnewswire.com/news-releases/prorata-partners-with-danish-publishers-group-dpcmo-to-launch-the-first-decentralized-sovereign-ai-answer-engine-302637182.html}
\showURL{%
\tempurl}


\bibitem[Rafaeli and Ariel(2008)]%
        {RafaeliAriel_2008_Online_Motivational_Factors_Incentives_for_Participation_and_Contribution_in_Wikipedia}
\bibfield{author}{\bibinfo{person}{Sheizaf Rafaeli} {and} \bibinfo{person}{Yaron Ariel}.} \bibinfo{year}{2008}\natexlab{}.
\newblock \showarticletitle{Online {{Motivational Factors}}: {{Incentives}} for {{Participation}} and {{Contribution}} in {{Wikipedia}}}.
\newblock In \bibinfo{booktitle}{\emph{Psychological {{Aspects}} of {{Cyberspace}}: {{Theory}}, {{Research}}, {{Applications}}}}, \bibfield{editor}{\bibinfo{person}{Azy Barak}} (Ed.). \bibinfo{publisher}{Cambridge University Press}, \bibinfo{address}{Cambridge}, \bibinfo{pages}{243--267}.
\newblock
\showISBNx{978-0-521-87301-7}
\href{https://doi.org/10.1017/CBO9780511813740.012}{doi:\nolinkurl{10.1017/CBO9780511813740.012}}


\bibitem[Reimers and Gurevych(2019)]%
        {reimers-gurevych-2019-sentence}
\bibfield{author}{\bibinfo{person}{Nils Reimers} {and} \bibinfo{person}{Iryna Gurevych}.} \bibinfo{year}{2019}\natexlab{}.
\newblock \showarticletitle{Sentence-{BERT}: Sentence Embeddings using {S}iamese {BERT}-Networks}. In \bibinfo{booktitle}{\emph{Proceedings of the 2019 Conference on Empirical Methods in Natural Language Processing and the 9th International Joint Conference on Natural Language Processing (EMNLP-IJCNLP)}}, \bibfield{editor}{\bibinfo{person}{Kentaro Inui}, \bibinfo{person}{Jing Jiang}, \bibinfo{person}{Vincent Ng}, {and} \bibinfo{person}{Xiaojun Wan}} (Eds.). \bibinfo{publisher}{Association for Computational Linguistics}, \bibinfo{address}{Hong Kong, China}, \bibinfo{pages}{3982--3992}.
\newblock
\href{https://doi.org/10.18653/v1/D19-1410}{doi:\nolinkurl{10.18653/v1/D19-1410}}


\bibitem[Reisner(2025)]%
        {Reisner_2025_Company_Quietly_Funneling_Paywalled_Articles_to_AI_Developers}
\bibfield{author}{\bibinfo{person}{Alex Reisner}.} \bibinfo{year}{2025}\natexlab{}.
\newblock \bibinfo{title}{The {{Company Quietly Funneling Paywalled Articles}} to {{AI Developers}}}.
\newblock
\urldef\tempurl%
\url{https://www.theatlantic.com/technology/2025/11/common-crawl-ai-training-data/684567/}
\showURL{%
\tempurl}


\bibitem[Rendleman(2020)]%
        {Rendleman_2020_Copy_That_What_is_plagiarism_in_practice_of_law}
\bibfield{author}{\bibinfo{person}{Dennis~A. Rendleman}.} \bibinfo{year}{2020}\natexlab{}.
\newblock \bibinfo{title}{``{{Copy That}}!'': {{What}} Is Plagiarism in the Practice of Law?}
\newblock
\urldef\tempurl%
\url{https://www.americanbar.org/news/abanews/publications/youraba/2020/youraba-march-2020/\_copy-that-\_--what-is-plagiarism-in-the-practice-of-law-/}
\showURL{%
\tempurl}


\bibitem[Ringstad and L{\o}yland(2006)]%
        {RingstadLoyland_2006_Demand_for_Books_Estimated_by_Means_of_Consumer_Survey_Data}
\bibfield{author}{\bibinfo{person}{Vidar Ringstad} {and} \bibinfo{person}{Knut L{\o}yland}.} \bibinfo{year}{2006}\natexlab{}.
\newblock \showarticletitle{The {{Demand}} for {{Books Estimated}} by {{Means}} of {{Consumer Survey Data}}}.
\newblock \bibinfo{journal}{\emph{Journal of Cultural Economics}} \bibinfo{volume}{30}, \bibinfo{number}{2} (\bibinfo{date}{sep} \bibinfo{year}{2006}), \bibinfo{pages}{141--155}.
\newblock
\showISSN{0885-2545, 1573-6997}
\href{https://doi.org/10.1007/s10824-006-9006-7}{doi:\nolinkurl{10.1007/s10824-006-9006-7}}


\bibitem[Rivas(2025)]%
        {Rivas_2025_From_Sweatshops_to_Standards_History_of_US_Labor_Laws}
\bibfield{author}{\bibinfo{person}{Adelaida Rivas}.} \bibinfo{year}{2025}\natexlab{}.
\newblock \bibinfo{title}{From {{Sweatshops}} to {{Standards}}: {{The History}} of {{U}}.{{S}}. {{Labor Laws}}}.
\newblock
\urldef\tempurl%
\url{https://www.davisbaconsolutions.com/blog/history-us-labor-laws}
\showURL{%
\tempurl}


\bibitem[Robertson(2025)]%
        {Robertson_2025_How_Authors_Are_Thinking_About_AI_Survey_of_1200_Authors}
\bibfield{author}{\bibinfo{person}{Carlyn Robertson}.} \bibinfo{year}{2025}\natexlab{}.
\newblock \bibinfo{title}{How {{Authors Are Thinking About AI}} ({{Survey}} of 1,200+ {{Authors}})}.
\newblock
\urldef\tempurl%
\url{https://insights.bookbub.com/how-authors-are-thinking-about-ai-survey/}
\showURL{%
\tempurl}


\bibitem[Rogers(2026)]%
        {rogers2026the}
\bibfield{author}{\bibinfo{person}{Anna Rogers}.} \bibinfo{year}{2026}\natexlab{}.
\newblock \showarticletitle{The human knowledge loophole in the 'bitter lesson' for LLMs}. In \bibinfo{booktitle}{\emph{The Fifth Blogpost Track at ICLR 2026}}.
\newblock
\urldef\tempurl%
\url{https://iclr-blogposts.github.io/2026/blog/2026/llm-bitter-lesson/}
\showURL{%
\tempurl}


\bibitem[Rogers(2024)]%
        {Rogers_2024_Google_Search_Ranks_AI_Spam_Above_Original_Reporting_in_News_Results}
\bibfield{author}{\bibinfo{person}{Reece Rogers}.} \bibinfo{year}{2024}\natexlab{}.
\newblock \showarticletitle{Google {{Search Ranks AI Spam Above Original Reporting}} in {{News Results}}}.
\newblock  (\bibinfo{year}{2024}).
\newblock
\showISSN{1059-1028}
\urldef\tempurl%
\url{https://www.wired.com/story/google-search-ai-spam-original-reporting-news-results/}
\showURL{%
\tempurl}


\bibitem[Rosenblat et~al\mbox{.}(2025)]%
        {rosenblat2025beyond}
\bibfield{author}{\bibinfo{person}{Sruly Rosenblat}, \bibinfo{person}{Tim O’Reilly}, {and} \bibinfo{person}{Ilan Strauss}.} \bibinfo{year}{2025}\natexlab{}.
\newblock \showarticletitle{Beyond public access in LLM pre-training data: Non-public book content in OpenAI’s models}.
\newblock \bibinfo{journal}{\emph{SSRC AI Disclosures Project Working Paper Series}} \bibinfo{number}{1} (\bibinfo{year}{2025}).
\newblock


\bibitem[Roth(2024)]%
        {Roth_2024_Googles_AI_search_summaries_officially_have_ads}
\bibfield{author}{\bibinfo{person}{Emma Roth}.} \bibinfo{year}{2024}\natexlab{}.
\newblock \bibinfo{title}{Google's {{AI}} Search Summaries Officially Have Ads}.
\newblock
\urldef\tempurl%
\url{https://www.theverge.com/2024/10/3/24260637/googles-ai-overview-ads-launch}
\showURL{%
\tempurl}


\bibitem[Rønde(2026a)]%
        {Ronde_2026_DPCMO_Partners_with_TollBit_to_Offer_Solutions_to_Support_its_Members_DPCMO}
\bibfield{author}{\bibinfo{person}{Karen Rønde}.} \bibinfo{year}{2026}\natexlab{a}.
\newblock \bibinfo{booktitle}{\emph{DPCMO Partners with TollBit to Offer Solutions to Support its Members – DPCMO}}.
\newblock DPCMO.
\newblock
\urldef\tempurl%
\url{https://dpcmo.dk/dpcmo-partners-with-tollbit-to-offer-solutions-to-support-its-members/}
\showURL{%
\tempurl}


\bibitem[Rønde(2026b)]%
        {Ronde_2026_united_Danish_media_industry_takes_OpenAI_to_court}
\bibfield{author}{\bibinfo{person}{Karen Rønde}.} \bibinfo{year}{2026}\natexlab{b}.
\newblock \bibinfo{booktitle}{\emph{A united Danish media industry takes OpenAI to court}}.
\newblock DPCMO.
\newblock
\urldef\tempurl%
\url{https://dpcmo.dk/a-united-danish-media-industry-takes-openai-to-court/}
\showURL{%
\tempurl}


\bibitem[Salmons(2024)]%
        {Salmons_2024_Routledge_Sells_Out_Authors_to_AI}
\bibfield{author}{\bibinfo{person}{Janet Salmons}.} \bibinfo{year}{2024}\natexlab{}.
\newblock \bibinfo{title}{Routledge {{Sells Out Authors}} to {{AI}}}.
\newblock
\urldef\tempurl%
\url{https://blog.taaonline.net/2024/08/routledge-sells-out-authors-to-ai/}
\showURL{%
\tempurl}


\bibitem[Sanford(2025)]%
        {Sanford_2025_As_WA_government_officials_embrace_AI_policies_are_still_catching_up}
\bibfield{author}{\bibinfo{person}{Nate Sanford}.} \bibinfo{year}{2025}\natexlab{}.
\newblock \bibinfo{title}{As {{WA}} Government Officials Embrace {{AI}}, Policies Are Still Catching Up}.
\newblock
\urldef\tempurl%
\url{https://www.knkx.org/government/2025-08-27/washington-state-everett-bellingham-government-officials-embrace-artificial-intelligence-chatgpt-policies-catching-up}
\showURL{%
\tempurl}


\bibitem[Sankaran(2024)]%
        {Sankaran_2024_OpenAI_says_it_is_impossible_to_train_AI_without_using_copyrighted_works_for_free}
\bibfield{author}{\bibinfo{person}{Vishwam Sankaran}.} \bibinfo{year}{2024}\natexlab{}.
\newblock \showarticletitle{{{OpenAI}} Says It Is `Impossible' to Train {{AI}} without Using Copyrighted Works for Free}.
\newblock \bibinfo{journal}{\emph{The Independent}} (\bibinfo{date}{jan} \bibinfo{year}{2024}).
\newblock
\urldef\tempurl%
\url{https://www.independent.co.uk/tech/openai-chatgpt-copyrighted-work-use-b2475386.html}
\showURL{%
\tempurl}


\bibitem[Sauer(2024)]%
        {Sauer_2024_OpenAI_CEO_Sam_Altman_You_could_get_paid_one_day_for_AI_training_data_we_use}
\bibfield{author}{\bibinfo{person}{Megan Sauer}.} \bibinfo{year}{2024}\natexlab{}.
\newblock \bibinfo{title}{{{OpenAI CEO Sam Altman}}: {{You}} Could Get Paid One Day for the {{AI}} Training Data We Use}.
\newblock
\urldef\tempurl%
\url{https://www.cnbc.com/2024/12/06/openai-ceo-sam-altman-you-could-get-paid-one-day-for-ai-training-data.html}
\showURL{%
\tempurl}


\bibitem[Scao et~al\mbox{.}(2022)]%
        {ScaoFanEtAl_2022_BLOOM_176BParameter_OpenAccess_Multilingual_Language_Model}
\bibfield{author}{\bibinfo{person}{Teven~Le Scao}, \bibinfo{person}{Angela Fan}, \bibinfo{person}{Christopher Akiki}, \bibinfo{person}{Ellie Pavlick}, \bibinfo{person}{Suzana Ili{\'c}}, \bibinfo{person}{Daniel Hesslow}, \bibinfo{person}{Roman Castagn{\'e}}, \bibinfo{person}{Alexandra~Sasha Luccioni}, \bibinfo{person}{Fran{\c c}ois Yvon}, \bibinfo{person}{Matthias Gall{\'e}}, \bibinfo{person}{Jonathan Tow}, \bibinfo{person}{Alexander~M. Rush}, \bibinfo{person}{Stella Biderman}, \bibinfo{person}{Albert Webson}, \bibinfo{person}{Pawan~Sasanka Ammanamanchi}, \bibinfo{person}{Thomas Wang}, \bibinfo{person}{Beno{\^i}t Sagot}, \bibinfo{person}{Niklas Muennighoff}, \bibinfo{person}{Albert~Villanova {del Moral}}, \bibinfo{person}{Olatunji Ruwase}, \bibinfo{person}{Rachel Bawden}, \bibinfo{person}{Stas Bekman}, \bibinfo{person}{Angelina {McMillan-Major}}, \bibinfo{person}{Iz Beltagy}, \bibinfo{person}{Huu Nguyen}, \bibinfo{person}{Lucile Saulnier}, \bibinfo{person}{Samson Tan}, \bibinfo{person}{Pedro~Ortiz Suarez},
  \bibinfo{person}{Victor Sanh}, \bibinfo{person}{Hugo Lauren{\c c}on}, \bibinfo{person}{Yacine Jernite}, \bibinfo{person}{Julien Launay}, \bibinfo{person}{Margaret Mitchell}, \bibinfo{person}{Colin Raffel}, \bibinfo{person}{Aaron Gokaslan}, \bibinfo{person}{Adi Simhi}, \bibinfo{person}{Aitor Soroa}, \bibinfo{person}{Alham~Fikri Aji}, \bibinfo{person}{Amit Alfassy}, \bibinfo{person}{Anna Rogers}, \bibinfo{person}{Ariel~Kreisberg Nitzav}, \bibinfo{person}{Canwen Xu}, \bibinfo{person}{Chenghao Mou}, \bibinfo{person}{Chris Emezue}, \bibinfo{person}{Christopher Klamm}, \bibinfo{person}{Colin Leong}, \bibinfo{person}{Daniel {van Strien}}, \bibinfo{person}{David~Ifeoluwa Adelani}, \bibinfo{person}{Dragomir Radev}, \bibinfo{person}{Eduardo~Gonz{\'a}lez Ponferrada}, \bibinfo{person}{Efrat Levkovizh}, \bibinfo{person}{Ethan Kim}, \bibinfo{person}{Eyal~Bar Natan}, \bibinfo{person}{Francesco De~Toni}, \bibinfo{person}{G{\'e}rard Dupont}, \bibinfo{person}{Germ{\'a}n Kruszewski}, \bibinfo{person}{Giada Pistilli},
  \bibinfo{person}{Hady Elsahar}, \bibinfo{person}{Hamza Benyamina}, \bibinfo{person}{Hieu Tran}, \bibinfo{person}{Ian Yu}, \bibinfo{person}{Idris Abdulmumin}, \bibinfo{person}{Isaac Johnson}, \bibinfo{person}{Itziar {Gonzalez-Dios}}, \bibinfo{person}{Javier {de la Rosa}}, \bibinfo{person}{Jenny Chim}, \bibinfo{person}{Jesse Dodge}, \bibinfo{person}{Jian Zhu}, \bibinfo{person}{Jonathan Chang}, \bibinfo{person}{J{\"o}rg Frohberg}, \bibinfo{person}{Joseph Tobing}, \bibinfo{person}{Joydeep Bhattacharjee}, \bibinfo{person}{Khalid Almubarak}, \bibinfo{person}{Kimbo Chen}, \bibinfo{person}{Kyle Lo}, \bibinfo{person}{Leandro Von~Werra}, \bibinfo{person}{Leon Weber}, \bibinfo{person}{Long Phan}, \bibinfo{person}{Loubna~Ben {allal}}, \bibinfo{person}{Ludovic Tanguy}, \bibinfo{person}{Manan Dey}, \bibinfo{person}{Manuel~Romero Mu{\\textasciitilde n}oz}, \bibinfo{person}{Maraim Masoud}, \bibinfo{person}{Mar{\'i}a Grandury}, \bibinfo{person}{Mario {\v S}a{\v s}ko}, \bibinfo{person}{Max Huang}, \bibinfo{person}{Maximin
  Coavoux}, \bibinfo{person}{Mayank Singh}, \bibinfo{person}{Mike Tian-Jian Jiang}, \bibinfo{person}{Minh~Chien Vu}, \bibinfo{person}{Mohammad~A. Jauhar}, \bibinfo{person}{Mustafa Ghaleb}, \bibinfo{person}{Nishant Subramani}, \bibinfo{person}{Nora Kassner}, \bibinfo{person}{Nurulaqilla Khamis}, \bibinfo{person}{Olivier Nguyen}, \bibinfo{person}{Omar Espejel}, \bibinfo{person}{Ona {de Gibert}}, \bibinfo{person}{Paulo Villegas}, \bibinfo{person}{Peter Henderson}, \bibinfo{person}{Pierre Colombo}, \bibinfo{person}{Priscilla Amuok}, \bibinfo{person}{Quentin Lhoest}, \bibinfo{person}{Rheza Harliman}, \bibinfo{person}{Rishi Bommasani}, \bibinfo{person}{Roberto~Luis L{\'o}pez}, \bibinfo{person}{Rui Ribeiro}, \bibinfo{person}{Salomey Osei}, \bibinfo{person}{Sampo Pyysalo}, \bibinfo{person}{Sebastian Nagel}, \bibinfo{person}{Shamik Bose}, \bibinfo{person}{Shamsuddeen~Hassan Muhammad}, \bibinfo{person}{Shanya Sharma}, \bibinfo{person}{Shayne Longpre}, \bibinfo{person}{Somaieh Nikpoor}, \bibinfo{person}{Stanislav
  Silberberg}, \bibinfo{person}{Suhas Pai}, \bibinfo{person}{Sydney Zink}, \bibinfo{person}{Tiago~Timponi Torrent}, \bibinfo{person}{Timo Schick}, \bibinfo{person}{Tristan Thrush}, \bibinfo{person}{Valentin Danchev}, \bibinfo{person}{Vassilina Nikoulina}, \bibinfo{person}{Veronika Laippala}, \bibinfo{person}{Violette Lepercq}, \bibinfo{person}{Vrinda Prabhu}, \bibinfo{person}{Zaid Alyafeai}, \bibinfo{person}{Zeerak Talat}, \bibinfo{person}{Arun Raja}, \bibinfo{person}{Benjamin Heinzerling}, \bibinfo{person}{Chenglei Si}, \bibinfo{person}{Davut~Emre Ta{\c s}ar}, \bibinfo{person}{Elizabeth Salesky}, \bibinfo{person}{Sabrina~J. Mielke}, \bibinfo{person}{Wilson~Y. Lee}, \bibinfo{person}{Abheesht Sharma}, \bibinfo{person}{Andrea Santilli}, \bibinfo{person}{Antoine Chaffin}, \bibinfo{person}{Arnaud Stiegler}, \bibinfo{person}{Debajyoti Datta}, \bibinfo{person}{Eliza Szczechla}, \bibinfo{person}{Gunjan Chhablani}, \bibinfo{person}{Han Wang}, \bibinfo{person}{Harshit Pandey}, \bibinfo{person}{Hendrik Strobelt},
  \bibinfo{person}{Jason~Alan Fries}, \bibinfo{person}{Jos Rozen}, \bibinfo{person}{Leo Gao}, \bibinfo{person}{Lintang Sutawika}, \bibinfo{person}{M.~Saiful Bari}, \bibinfo{person}{Maged~S. {Al-shaibani}}, \bibinfo{person}{Matteo Manica}, \bibinfo{person}{Nihal Nayak}, \bibinfo{person}{Ryan Teehan}, \bibinfo{person}{Samuel Albanie}, \bibinfo{person}{Sheng Shen}, \bibinfo{person}{Srulik {Ben-David}}, \bibinfo{person}{Stephen~H. Bach}, \bibinfo{person}{Taewoon Kim}, \bibinfo{person}{Tali Bers}, \bibinfo{person}{Thibault Fevry}, \bibinfo{person}{Trishala Neeraj}, \bibinfo{person}{Urmish Thakker}, \bibinfo{person}{Vikas Raunak}, \bibinfo{person}{Xiangru Tang}, \bibinfo{person}{Zheng-Xin Yong}, \bibinfo{person}{Zhiqing Sun}, \bibinfo{person}{Shaked Brody}, \bibinfo{person}{Yallow Uri}, \bibinfo{person}{Hadar Tojarieh}, \bibinfo{person}{Adam Roberts}, \bibinfo{person}{Hyung~Won Chung}, \bibinfo{person}{Jaesung Tae}, \bibinfo{person}{Jason Phang}, \bibinfo{person}{Ofir Press}, \bibinfo{person}{Conglong Li},
  \bibinfo{person}{Deepak Narayanan}, \bibinfo{person}{Hatim Bourfoune}, \bibinfo{person}{Jared Casper}, \bibinfo{person}{Jeff Rasley}, \bibinfo{person}{Max Ryabinin}, \bibinfo{person}{Mayank Mishra}, \bibinfo{person}{Minjia Zhang}, \bibinfo{person}{Mohammad Shoeybi}, \bibinfo{person}{Myriam Peyrounette}, \bibinfo{person}{Nicolas Patry}, \bibinfo{person}{Nouamane Tazi}, \bibinfo{person}{Omar Sanseviero}, \bibinfo{person}{Patrick {von Platen}}, \bibinfo{person}{Pierre Cornette}, \bibinfo{person}{Pierre~Fran{\c c}ois Lavall{\'e}e}, \bibinfo{person}{R{\'e}mi Lacroix}, \bibinfo{person}{Samyam Rajbhandari}, \bibinfo{person}{Sanchit Gandhi}, \bibinfo{person}{Shaden Smith}, \bibinfo{person}{St{\'e}phane Requena}, \bibinfo{person}{Suraj Patil}, \bibinfo{person}{Tim Dettmers}, \bibinfo{person}{Ahmed Baruwa}, \bibinfo{person}{Amanpreet Singh}, \bibinfo{person}{Anastasia Cheveleva}, \bibinfo{person}{Anne-Laure Ligozat}, \bibinfo{person}{Arjun Subramonian}, \bibinfo{person}{Aur{\'e}lie N{\'e}v{\'e}ol},
  \bibinfo{person}{Charles Lovering}, \bibinfo{person}{Dan Garrette}, \bibinfo{person}{Deepak Tunuguntla}, \bibinfo{person}{Ehud Reiter}, \bibinfo{person}{Ekaterina Taktasheva}, \bibinfo{person}{Ekaterina Voloshina}, \bibinfo{person}{Eli Bogdanov}, \bibinfo{person}{Genta~Indra Winata}, \bibinfo{person}{Hailey Schoelkopf}, \bibinfo{person}{Jan-Christoph Kalo}, \bibinfo{person}{Jekaterina Novikova}, \bibinfo{person}{Jessica~Zosa Forde}, \bibinfo{person}{Jordan Clive}, \bibinfo{person}{Jungo Kasai}, \bibinfo{person}{Ken Kawamura}, \bibinfo{person}{Liam Hazan}, \bibinfo{person}{Marine Carpuat}, \bibinfo{person}{Miruna Clinciu}, \bibinfo{person}{Najoung Kim}, \bibinfo{person}{Newton Cheng}, \bibinfo{person}{Oleg Serikov}, \bibinfo{person}{Omer Antverg}, \bibinfo{person}{Oskar {van der Wal}}, \bibinfo{person}{Rui Zhang}, \bibinfo{person}{Ruochen Zhang}, \bibinfo{person}{Sebastian Gehrmann}, \bibinfo{person}{Shachar Mirkin}, \bibinfo{person}{Shani Pais}, \bibinfo{person}{Tatiana Shavrina}, \bibinfo{person}{Thomas
  Scialom}, \bibinfo{person}{Tian Yun}, \bibinfo{person}{Tomasz Limisiewicz}, \bibinfo{person}{Verena Rieser}, \bibinfo{person}{Vitaly Protasov}, \bibinfo{person}{Vladislav Mikhailov}, \bibinfo{person}{Yada Pruksachatkun}, \bibinfo{person}{Yonatan Belinkov}, \bibinfo{person}{Zachary Bamberger}, \bibinfo{person}{Zden{\v e}k Kasner}, \bibinfo{person}{Alice Rueda}, \bibinfo{person}{Amanda Pestana}, \bibinfo{person}{Amir Feizpour}, \bibinfo{person}{Ammar Khan}, \bibinfo{person}{Amy Faranak}, \bibinfo{person}{Ana Santos}, \bibinfo{person}{Anthony Hevia}, \bibinfo{person}{Antigona Unldreaj}, \bibinfo{person}{Arash Aghagol}, \bibinfo{person}{Arezoo Abdollahi}, \bibinfo{person}{Aycha Tammour}, \bibinfo{person}{Azadeh HajiHosseini}, \bibinfo{person}{Bahareh Behroozi}, \bibinfo{person}{Benjamin Ajibade}, \bibinfo{person}{Bharat Saxena}, \bibinfo{person}{Carlos~Mu{\\textasciitilde n}oz Ferrandis}, \bibinfo{person}{Danish Contractor}, \bibinfo{person}{David Lansky}, \bibinfo{person}{Davis David}, \bibinfo{person}{Douwe
  Kiela}, \bibinfo{person}{Duong~A. Nguyen}, \bibinfo{person}{Edward Tan}, \bibinfo{person}{Emi Baylor}, \bibinfo{person}{Ezinwanne Ozoani}, \bibinfo{person}{Fatima Mirza}, \bibinfo{person}{Frankline Ononiwu}, \bibinfo{person}{Habib Rezanejad}, \bibinfo{person}{Hessie Jones}, \bibinfo{person}{Indrani Bhattacharya}, \bibinfo{person}{Irene Solaiman}, \bibinfo{person}{Irina Sedenko}, \bibinfo{person}{Isar Nejadgholi}, \bibinfo{person}{Jesse Passmore}, \bibinfo{person}{Josh Seltzer}, \bibinfo{person}{Julio~Bonis Sanz}, \bibinfo{person}{Livia Dutra}, \bibinfo{person}{Mairon Samagaio}, \bibinfo{person}{Maraim Elbadri}, \bibinfo{person}{Margot Mieskes}, \bibinfo{person}{Marissa Gerchick}, \bibinfo{person}{Martha Akinlolu}, \bibinfo{person}{Michael McKenna}, \bibinfo{person}{Mike Qiu}, \bibinfo{person}{Muhammed Ghauri}, \bibinfo{person}{Mykola Burynok}, \bibinfo{person}{Nafis Abrar}, \bibinfo{person}{Nazneen Rajani}, \bibinfo{person}{Nour Elkott}, \bibinfo{person}{Nour Fahmy}, \bibinfo{person}{Olanrewaju Samuel},
  \bibinfo{person}{Ran An}, \bibinfo{person}{Rasmus Kromann}, \bibinfo{person}{Ryan Hao}, \bibinfo{person}{Samira Alizadeh}, \bibinfo{person}{Sarmad Shubber}, \bibinfo{person}{Silas Wang}, \bibinfo{person}{Sourav Roy}, \bibinfo{person}{Sylvain Viguier}, \bibinfo{person}{Thanh Le}, \bibinfo{person}{Tobi Oyebade}, \bibinfo{person}{Trieu Le}, \bibinfo{person}{Yoyo Yang}, \bibinfo{person}{Zach Nguyen}, \bibinfo{person}{Abhinav~Ramesh Kashyap}, \bibinfo{person}{Alfredo Palasciano}, \bibinfo{person}{Alison Callahan}, \bibinfo{person}{Anima Shukla}, \bibinfo{person}{Antonio {Miranda-Escalada}}, \bibinfo{person}{Ayush Singh}, \bibinfo{person}{Benjamin Beilharz}, \bibinfo{person}{Bo Wang}, \bibinfo{person}{Caio Brito}, \bibinfo{person}{Chenxi Zhou}, \bibinfo{person}{Chirag Jain}, \bibinfo{person}{Chuxin Xu}, \bibinfo{person}{Cl{\'e}mentine Fourrier}, \bibinfo{person}{Daniel~Le{\'o}n Peri{\\textasciitilde n}{\'a}n}, \bibinfo{person}{Daniel Molano}, \bibinfo{person}{Dian Yu}, \bibinfo{person}{Enrique Manjavacas},
  \bibinfo{person}{Fabio Barth}, \bibinfo{person}{Florian Fuhrimann}, \bibinfo{person}{Gabriel Altay}, \bibinfo{person}{Giyaseddin Bayrak}, \bibinfo{person}{Gully Burns}, \bibinfo{person}{Helena~U. Vrabec}, \bibinfo{person}{Imane Bello}, \bibinfo{person}{Ishani Dash}, \bibinfo{person}{Jihyun Kang}, \bibinfo{person}{John Giorgi}, \bibinfo{person}{Jonas Golde}, \bibinfo{person}{Jose~David Posada}, \bibinfo{person}{Karthik~Rangasai Sivaraman}, \bibinfo{person}{Lokesh Bulchandani}, \bibinfo{person}{Lu Liu}, \bibinfo{person}{Luisa Shinzato}, \bibinfo{person}{Madeleine~Hahn {de Bykhovetz}}, \bibinfo{person}{Maiko Takeuchi}, \bibinfo{person}{Marc P{\`a}mies}, \bibinfo{person}{Maria~A. Castillo}, \bibinfo{person}{Marianna Nezhurina}, \bibinfo{person}{Mario S{\"a}nger}, \bibinfo{person}{Matthias Samwald}, \bibinfo{person}{Michael Cullan}, \bibinfo{person}{Michael Weinberg}, \bibinfo{person}{Michiel De~Wolf}, \bibinfo{person}{Mina Mihaljcic}, \bibinfo{person}{Minna Liu}, \bibinfo{person}{Moritz Freidank},
  \bibinfo{person}{Myungsun Kang}, \bibinfo{person}{Natasha Seelam}, \bibinfo{person}{Nathan Dahlberg}, \bibinfo{person}{Nicholas~Michio Broad}, \bibinfo{person}{Nikolaus Muellner}, \bibinfo{person}{Pascale Fung}, \bibinfo{person}{Patrick Haller}, \bibinfo{person}{Ramya Chandrasekhar}, \bibinfo{person}{Renata Eisenberg}, \bibinfo{person}{Robert Martin}, \bibinfo{person}{Rodrigo Canalli}, \bibinfo{person}{Rosaline Su}, \bibinfo{person}{Ruisi Su}, \bibinfo{person}{Samuel Cahyawijaya}, \bibinfo{person}{Samuele Garda}, \bibinfo{person}{Shlok~S. Deshmukh}, \bibinfo{person}{Shubhanshu Mishra}, \bibinfo{person}{Sid Kiblawi}, \bibinfo{person}{Simon Ott}, \bibinfo{person}{Sinee {Sang-aroonsiri}}, \bibinfo{person}{Srishti Kumar}, \bibinfo{person}{Stefan Schweter}, \bibinfo{person}{Sushil Bharati}, \bibinfo{person}{Tanmay Laud}, \bibinfo{person}{Th{\'e}o Gigant}, \bibinfo{person}{Tomoya Kainuma}, \bibinfo{person}{Wojciech Kusa}, \bibinfo{person}{Yanis Labrak}, \bibinfo{person}{Yash~Shailesh Bajaj}, \bibinfo{person}{Yash
  Venkatraman}, \bibinfo{person}{Yifan Xu}, \bibinfo{person}{Yingxin Xu}, \bibinfo{person}{Yu Xu}, \bibinfo{person}{Zhe Tan}, \bibinfo{person}{Zhongli Xie}, \bibinfo{person}{Zifan Ye}, \bibinfo{person}{Mathilde Bras}, \bibinfo{person}{Younes Belkada}, {and} \bibinfo{person}{Thomas Wolf}.} \bibinfo{year}{2022}\natexlab{}.
\newblock \bibinfo{title}{{{BLOOM}}: {{A 176B-Parameter Open-Access Multilingual Language Model}}}.
\newblock
\showeprint[arxiv]{2211.05100}~[cs]
\href{https://doi.org/10.48550/arXiv.2211.05100}{doi:\nolinkurl{10.48550/arXiv.2211.05100}}


\bibitem[Shazeer et~al\mbox{.}(2017)]%
        {shazeerOutrageouslyLargeNeural2017}
\bibfield{author}{\bibinfo{person}{Noam Shazeer}, \bibinfo{person}{*Azalia Mirhoseini}, \bibinfo{person}{*Krzysztof Maziarz}, \bibinfo{person}{Andy Davis}, \bibinfo{person}{Quoc Le}, \bibinfo{person}{Geoffrey Hinton}, {and} \bibinfo{person}{Jeff Dean}.} \bibinfo{year}{2017}\natexlab{}.
\newblock \showarticletitle{Outrageously {{Large Neural Networks}}: {{The Sparsely-Gated Mixture-of-Experts Layer}}}. In \bibinfo{booktitle}{\emph{International {{Conference}} on {{Learning Representations}}}}.
\newblock


\bibitem[Shi et~al\mbox{.}(2025)]%
        {shi2025flexolmoopenlanguagemodels}
\bibfield{author}{\bibinfo{person}{Weijia Shi}, \bibinfo{person}{Akshita Bhagia}, \bibinfo{person}{Kevin Farhat}, \bibinfo{person}{Niklas Muennighoff}, \bibinfo{person}{Pete Walsh}, \bibinfo{person}{Jacob Morrison}, \bibinfo{person}{Dustin Schwenk}, \bibinfo{person}{Shayne Longpre}, \bibinfo{person}{Jake Poznanski}, \bibinfo{person}{Allyson Ettinger}, \bibinfo{person}{Daogao Liu}, \bibinfo{person}{Margaret Li}, \bibinfo{person}{Dirk Groeneveld}, \bibinfo{person}{Mike Lewis}, \bibinfo{person}{Wen tau Yih}, \bibinfo{person}{Luca Soldaini}, \bibinfo{person}{Kyle Lo}, \bibinfo{person}{Noah~A. Smith}, \bibinfo{person}{Luke Zettlemoyer}, \bibinfo{person}{Pang~Wei Koh}, \bibinfo{person}{Hannaneh Hajishirzi}, \bibinfo{person}{Ali Farhadi}, {and} \bibinfo{person}{Sewon Min}.} \bibinfo{year}{2025}\natexlab{}.
\newblock \bibinfo{title}{FlexOlmo: Open Language Models for Flexible Data Use}.
\newblock
\showeprint[arxiv]{2507.07024}~[cs.CL]
\urldef\tempurl%
\url{https://arxiv.org/abs/2507.07024}
\showURL{%
\tempurl}


\bibitem[Shimabucoro et~al\mbox{.}(2024)]%
        {shimabucoro-etal-2024-llm}
\bibfield{author}{\bibinfo{person}{Lu{\'i}sa Shimabucoro}, \bibinfo{person}{Sebastian Ruder}, \bibinfo{person}{Julia Kreutzer}, \bibinfo{person}{Marzieh Fadaee}, {and} \bibinfo{person}{Sara Hooker}.} \bibinfo{year}{2024}\natexlab{}.
\newblock \showarticletitle{{LLM} See, {LLM} Do: Leveraging Active Inheritance to Target Non-Differentiable Objectives}. In \bibinfo{booktitle}{\emph{Proceedings of the 2024 Conference on Empirical Methods in Natural Language Processing}}, \bibfield{editor}{\bibinfo{person}{Yaser Al-Onaizan}, \bibinfo{person}{Mohit Bansal}, {and} \bibinfo{person}{Yun-Nung Chen}} (Eds.). \bibinfo{publisher}{Association for Computational Linguistics}, \bibinfo{address}{Miami, Florida, USA}, \bibinfo{pages}{9243--9267}.
\newblock
\href{https://doi.org/10.18653/v1/2024.emnlp-main.521}{doi:\nolinkurl{10.18653/v1/2024.emnlp-main.521}}


\bibitem[Sim et~al\mbox{.}(2022)]%
        {SimXuEtAl_2022_Data_Valuation_in_Machine_Learning_Ingredients_Strategies_and_Open_Challenges}
\bibfield{author}{\bibinfo{person}{Rachael Hwee~Ling Sim}, \bibinfo{person}{Xinyi Xu}, {and} \bibinfo{person}{Bryan Kian~Hsiang Low}.} \bibinfo{year}{2022}\natexlab{}.
\newblock \showarticletitle{Data {{Valuation}} in {{Machine Learning}}: "{{Ingredients}}", {{Strategies}}, and {{Open Challenges}}}. In \bibinfo{booktitle}{\emph{Proceedings of the {{Thirty-First International Joint Conference}} on {{Artificial Intelligence}}}}. \bibinfo{publisher}{International Joint Conferences on Artificial Intelligence Organization}, \bibinfo{address}{Vienna, Austria}, \bibinfo{pages}{5607--5614}.
\newblock
\showISBNx{978-1-956792-00-3}
\href{https://doi.org/10.24963/ijcai.2022/782}{doi:\nolinkurl{10.24963/ijcai.2022/782}}


\bibitem[Skorup and Huddleston(2019)]%
        {SkorupHuddleston_2019_Erosion_of_Publisher_Liability_in_American_Law_Section_230_and_Future_of_Online_Curation}
\bibfield{author}{\bibinfo{person}{Brent Skorup} {and} \bibinfo{person}{Jennifer Huddleston}.} \bibinfo{year}{2019}\natexlab{}.
\newblock \showarticletitle{The {{Erosion}} of {{Publisher Liability}} in {{American Law}}, {{Section}} 230, and the {{Future}} of {{Online Curation}}}.
\newblock \bibinfo{journal}{\emph{SSRN Electronic Journal}} (\bibinfo{year}{2019}).
\newblock
\showISSN{1556-5068}
\href{https://doi.org/10.2139/ssrn.3420304}{doi:\nolinkurl{10.2139/ssrn.3420304}}


\bibitem[Smith(2021)]%
        {Smith_2021_13400_Artists_Out_of_7_Million_Earn_50k_or_More_From_Spotify_Yearly}
\bibfield{author}{\bibinfo{person}{Dylan Smith}.} \bibinfo{year}{2021}\natexlab{}.
\newblock \bibinfo{title}{13,400 {{Artists}} ({{Out}} of 7 {{Million}}) {{Earn}} \$50k or {{More From Spotify Yearly}}}.
\newblock
\urldef\tempurl%
\url{https://www.digitalmusicnews.com/2021/03/18/spotify-artist-earnings-figures/}
\showURL{%
\tempurl}


\bibitem[Sommer(2025)]%
        {Sommer_2025_AI-Generated_Books_on_Amazon_Are_Hurting_Authors_and_Publishing_Industry}
\bibfield{author}{\bibinfo{person}{Joanna Sommer}.} \bibinfo{year}{2025}\natexlab{}.
\newblock \bibinfo{booktitle}{\emph{{{AI-Generated Books}} on {{Amazon Are Hurting Authors}} and the {{Publishing Industry}}}}.
\newblock InsideHook.
\newblock
\urldef\tempurl%
\url{https://www.insidehook.com/books/ai-generated-books-amazon-authors-publishing-industry}
\showURL{%
\tempurl}


\bibitem[Su et~al\mbox{.}(2023)]%
        {su-etal-2023-one}
\bibfield{author}{\bibinfo{person}{Hongjin Su}, \bibinfo{person}{Weijia Shi}, \bibinfo{person}{Jungo Kasai}, \bibinfo{person}{Yizhong Wang}, \bibinfo{person}{Yushi Hu}, \bibinfo{person}{Mari Ostendorf}, \bibinfo{person}{Wen-tau Yih}, \bibinfo{person}{Noah~A. Smith}, \bibinfo{person}{Luke Zettlemoyer}, {and} \bibinfo{person}{Tao Yu}.} \bibinfo{year}{2023}\natexlab{}.
\newblock \showarticletitle{One Embedder, Any Task: Instruction-Finetuned Text Embeddings}. In \bibinfo{booktitle}{\emph{Findings of the Association for Computational Linguistics: ACL 2023}}, \bibfield{editor}{\bibinfo{person}{Anna Rogers}, \bibinfo{person}{Jordan Boyd-Graber}, {and} \bibinfo{person}{Naoaki Okazaki}} (Eds.). \bibinfo{publisher}{Association for Computational Linguistics}, \bibinfo{address}{Toronto, Canada}, \bibinfo{pages}{1102--1121}.
\newblock
\href{https://doi.org/10.18653/v1/2023.findings-acl.71}{doi:\nolinkurl{10.18653/v1/2023.findings-acl.71}}


\bibitem[{Sutherland-Smith}(2016)]%
        {Sutherland-Smith_2016_Authorship_Ownership_and_Plagiarism_in_Digital_Age}
\bibfield{author}{\bibinfo{person}{Wendy {Sutherland-Smith}}.} \bibinfo{year}{2016}\natexlab{}.
\newblock \showarticletitle{Authorship, {{Ownership}}, and {{Plagiarism}} in the {{Digital Age}}}.
\newblock In \bibinfo{booktitle}{\emph{Handbook of {{Academic Integrity}}}}. \bibinfo{publisher}{Springer, Singapore}, \bibinfo{pages}{575--589}.
\newblock
\showISBNx{978-981-287-098-8}
\href{https://doi.org/10.1007/978-981-287-098-8\_14}{doi:\nolinkurl{10.1007/978-981-287-098-8\_14}}


\bibitem[Temple(2019)]%
        {Temple_2019_Authors_Attribution_and_Integrity_Examining_Moral_Rights_in_United_States_Report_of_Register_of_Copyrights}
\bibfield{author}{\bibinfo{person}{Karyn~A. Temple}.} \bibinfo{year}{2019}\natexlab{}.
\newblock \bibinfo{booktitle}{\emph{Authors, {{Attribution}}, and {{Integrity}}: {{Examining Moral Rights}} in the {{United States}} -- {{A Report}} of the {{Register}} of {{Copyrights}}, {{April}} 2019}}.
\newblock \bibinfo{type}{{T}echnical {R}eport}. \bibinfo{institution}{US Copyright Office}.
\newblock
\urldef\tempurl%
\url{https://www.copyright.gov/policy/moralrights/full-report.pdf}
\showURL{%
\tempurl}


\bibitem[Thompson(2025)]%
        {Thompson_2025_They_Criticized_Musk_on_Then_Their_Reach_Collapsed}
\bibfield{author}{\bibinfo{person}{Stuart~A. Thompson}.} \bibinfo{year}{2025}\natexlab{}.
\newblock \bibinfo{title}{They {{Criticized Musk}} on {{X}}. {{Then Their Reach Collapsed}}.}
\newblock
\urldef\tempurl%
\url{https://www.nytimes.com/interactive/2025/04/23/business/elon-musk-x-suppression-laura-loomer.html}
\showURL{%
\tempurl}


\bibitem[Ulea(2025)]%
        {Ulea_2025_OpenAI_cannot_use_song_lyrics_without_paying_German_court_rules}
\bibfield{author}{\bibinfo{person}{Anca Ulea}.} \bibinfo{year}{2025}\natexlab{}.
\newblock \showarticletitle{{{OpenAI}} Cannot Use Song Lyrics without Paying, {{German}} Court Rules}.
\newblock \bibinfo{journal}{\emph{Euronews}} (\bibinfo{date}{nov} \bibinfo{year}{2025}).
\newblock
\urldef\tempurl%
\url{http://www.euronews.com/next/2025/11/11/openai-chatbots-cannot-use-song-lyrics-without-paying-german-court-rules-in-landmark-trial}
\showURL{%
\tempurl}


\bibitem[Van~Noorden(2013)]%
        {VanNoorden_2013_Open_access_true_cost_of_science_publishing}
\bibfield{author}{\bibinfo{person}{Richard Van~Noorden}.} \bibinfo{year}{2013}\natexlab{}.
\newblock \showarticletitle{Open Access: {{The}} True Cost of Science Publishing}.
\newblock \bibinfo{journal}{\emph{Nature}} \bibinfo{volume}{495}, \bibinfo{number}{7442} (\bibinfo{date}{mar} \bibinfo{year}{2013}), \bibinfo{pages}{426--429}.
\newblock
\showISSN{1476-4687}
\href{https://doi.org/10.1038/495426a}{doi:\nolinkurl{10.1038/495426a}}


\bibitem[Vanian(2025)]%
        {Vanian_2025_Meta_greenlights_Facebook_Instagram_ads_based_on_your_AI_chats}
\bibfield{author}{\bibinfo{person}{Jonathan Vanian}.} \bibinfo{year}{2025}\natexlab{}.
\newblock \bibinfo{title}{Meta Greenlights {{Facebook}}, {{Instagram}} Ads Based on Your {{AI}} Chats}.
\newblock
\urldef\tempurl%
\url{https://www.cnbc.com/2025/10/01/meta-facebook-instagram-ads-ai-chat.html}
\showURL{%
\tempurl}


\bibitem[Vera et~al\mbox{.}(2025)]%
        {vera2025embeddinggemmapowerfullightweighttext}
\bibfield{author}{\bibinfo{person}{Henrique~Schechter Vera}, \bibinfo{person}{Sahil Dua}, \bibinfo{person}{Biao Zhang}, \bibinfo{person}{Daniel Salz}, \bibinfo{person}{Ryan Mullins}, \bibinfo{person}{Sindhu~Raghuram Panyam}, \bibinfo{person}{Sara Smoot}, \bibinfo{person}{Iftekhar Naim}, \bibinfo{person}{Joe Zou}, \bibinfo{person}{Feiyang Chen}, \bibinfo{person}{Daniel Cer}, \bibinfo{person}{Alice Lisak}, \bibinfo{person}{Min Choi}, \bibinfo{person}{Lucas Gonzalez}, \bibinfo{person}{Omar Sanseviero}, \bibinfo{person}{Glenn Cameron}, \bibinfo{person}{Ian Ballantyne}, \bibinfo{person}{Kat Black}, \bibinfo{person}{Kaifeng Chen}, \bibinfo{person}{Weiyi Wang}, \bibinfo{person}{Zhe Li}, \bibinfo{person}{Gus Martins}, \bibinfo{person}{Jinhyuk Lee}, \bibinfo{person}{Mark Sherwood}, \bibinfo{person}{Juyeong Ji}, \bibinfo{person}{Renjie Wu}, \bibinfo{person}{Jingxiao Zheng}, \bibinfo{person}{Jyotinder Singh}, \bibinfo{person}{Abheesht Sharma}, \bibinfo{person}{Divyashree Sreepathihalli}, \bibinfo{person}{Aashi Jain},
  \bibinfo{person}{Adham Elarabawy}, \bibinfo{person}{AJ Co}, \bibinfo{person}{Andreas Doumanoglou}, \bibinfo{person}{Babak Samari}, \bibinfo{person}{Ben Hora}, \bibinfo{person}{Brian Potetz}, \bibinfo{person}{Dahun Kim}, \bibinfo{person}{Enrique Alfonseca}, \bibinfo{person}{Fedor Moiseev}, \bibinfo{person}{Feng Han}, \bibinfo{person}{Frank~Palma Gomez}, \bibinfo{person}{Gustavo~Hernández Ábrego}, \bibinfo{person}{Hesen Zhang}, \bibinfo{person}{Hui Hui}, \bibinfo{person}{Jay Han}, \bibinfo{person}{Karan Gill}, \bibinfo{person}{Ke Chen}, \bibinfo{person}{Koert Chen}, \bibinfo{person}{Madhuri Shanbhogue}, \bibinfo{person}{Michael Boratko}, \bibinfo{person}{Paul Suganthan}, \bibinfo{person}{Sai Meher~Karthik Duddu}, \bibinfo{person}{Sandeep Mariserla}, \bibinfo{person}{Setareh Ariafar}, \bibinfo{person}{Shanfeng Zhang}, \bibinfo{person}{Shijie Zhang}, \bibinfo{person}{Simon Baumgartner}, \bibinfo{person}{Sonam Goenka}, \bibinfo{person}{Steve Qiu}, \bibinfo{person}{Tanmaya Dabral}, \bibinfo{person}{Trevor
  Walker}, \bibinfo{person}{Vikram Rao}, \bibinfo{person}{Waleed Khawaja}, \bibinfo{person}{Wenlei Zhou}, \bibinfo{person}{Xiaoqi Ren}, \bibinfo{person}{Ye Xia}, \bibinfo{person}{Yichang Chen}, \bibinfo{person}{Yi-Ting Chen}, \bibinfo{person}{Zhe Dong}, \bibinfo{person}{Zhongli Ding}, \bibinfo{person}{Francesco Visin}, \bibinfo{person}{Gaël Liu}, \bibinfo{person}{Jiageng Zhang}, \bibinfo{person}{Kathleen Kenealy}, \bibinfo{person}{Michelle Casbon}, \bibinfo{person}{Ravin Kumar}, \bibinfo{person}{Thomas Mesnard}, \bibinfo{person}{Zach Gleicher}, \bibinfo{person}{Cormac Brick}, \bibinfo{person}{Olivier Lacombe}, \bibinfo{person}{Adam Roberts}, \bibinfo{person}{Qin Yin}, \bibinfo{person}{Yunhsuan Sung}, \bibinfo{person}{Raphael Hoffmann}, \bibinfo{person}{Tris Warkentin}, \bibinfo{person}{Armand Joulin}, \bibinfo{person}{Tom Duerig}, {and} \bibinfo{person}{Mojtaba Seyedhosseini}.} \bibinfo{year}{2025}\natexlab{}.
\newblock \bibinfo{title}{EmbeddingGemma: Powerful and Lightweight Text Representations}.
\newblock
\showeprint[arxiv]{2509.20354}~[cs.CL]
\urldef\tempurl%
\url{https://arxiv.org/abs/2509.20354}
\showURL{%
\tempurl}


\bibitem[Vincent et~al\mbox{.}(2025)]%
        {vincentCollectiveBargainingInformation2025}
\bibfield{author}{\bibinfo{person}{Nicholas Vincent}, \bibinfo{person}{Matthew Prewitt}, {and} \bibinfo{person}{Hanlin Li}.} \bibinfo{year}{2025}\natexlab{}.
\newblock \bibinfo{title}{Collective {{Bargaining}} in the {{Information Economy Can Address AI-Driven Power Concentration}}}.
\newblock
\showeprint[arxiv]{2506.10272}~[cs]
\href{https://doi.org/10.48550/arXiv.2506.10272}{doi:\nolinkurl{10.48550/arXiv.2506.10272}}


\bibitem[Viswanathan et~al\mbox{.}(2025)]%
        {viswanathan-etal-2025-synthetic}
\bibfield{author}{\bibinfo{person}{Vijay Viswanathan}, \bibinfo{person}{Xiang Yue}, \bibinfo{person}{Alisa Liu}, \bibinfo{person}{Yizhong Wang}, {and} \bibinfo{person}{Graham Neubig}.} \bibinfo{year}{2025}\natexlab{}.
\newblock \showarticletitle{Synthetic Data in the Era of Large Language Models}. In \bibinfo{booktitle}{\emph{Proceedings of the 63rd Annual Meeting of the Association for Computational Linguistics (Volume 5: Tutorial Abstracts)}}, \bibfield{editor}{\bibinfo{person}{Yuki Arase}, \bibinfo{person}{David Jurgens}, {and} \bibinfo{person}{Fei Xia}} (Eds.). \bibinfo{publisher}{Association for Computational Linguistics}, \bibinfo{address}{Vienna, Austria}, \bibinfo{pages}{11--12}.
\newblock
\showISBNx{979-8-89176-255-8}
\href{https://doi.org/10.18653/v1/2025.acl-tutorials.7}{doi:\nolinkurl{10.18653/v1/2025.acl-tutorials.7}}


\bibitem[Walsh(2025)]%
        {Walsh_2025_YouTube_error_that_couldve_cost_thousands}
\bibfield{author}{\bibinfo{person}{Marcus Walsh}.} \bibinfo{year}{2025}\natexlab{}.
\newblock \bibinfo{title}{{{YouTube}} Error That Could've Cost Thousands}.
\newblock
\urldef\tempurl%
\url{https://cybernews.com/news/youtube-monetization-influencer-error/}
\showURL{%
\tempurl}


\bibitem[Wang et~al\mbox{.}(2024)]%
        {wang2024economicsolutioncopyrightchallenges}
\bibfield{author}{\bibinfo{person}{Jiachen~T. Wang}, \bibinfo{person}{Zhun Deng}, \bibinfo{person}{Hiroaki Chiba-Okabe}, \bibinfo{person}{Boaz Barak}, {and} \bibinfo{person}{Weijie~J. Su}.} \bibinfo{year}{2024}\natexlab{}.
\newblock \bibinfo{title}{An Economic Solution to Copyright Challenges of Generative AI}.
\newblock
\showeprint[arxiv]{2404.13964}~[cs.LG]
\urldef\tempurl%
\url{https://arxiv.org/abs/2404.13964}
\showURL{%
\tempurl}


\bibitem[Weller et~al\mbox{.}(2025)]%
        {WellerChangEtAl_2025_FollowIR_Evaluating_and_Teaching_Information_Retrieval_Models_to_Follow_Instructions}
\bibfield{author}{\bibinfo{person}{Orion Weller}, \bibinfo{person}{Benjamin Chang}, \bibinfo{person}{Sean MacAvaney}, \bibinfo{person}{Kyle Lo}, \bibinfo{person}{Arman Cohan}, \bibinfo{person}{Benjamin Van~Durme}, \bibinfo{person}{Dawn Lawrie}, {and} \bibinfo{person}{Luca Soldaini}.} \bibinfo{year}{2025}\natexlab{}.
\newblock \showarticletitle{{{FollowIR}}: {{Evaluating}} and {{Teaching Information Retrieval Models}} to {{Follow Instructions}}}. In \bibinfo{booktitle}{\emph{Proceedings of the 2025 {{Conference}} of the {{Nations}} of the {{Americas Chapter}} of the {{Association}} for {{Computational Linguistics}}: {{Human Language Technologies}} ({{Volume}} 1: {{Long Papers}})}}, \bibfield{editor}{\bibinfo{person}{Luis Chiruzzo}, \bibinfo{person}{Alan Ritter}, {and} \bibinfo{person}{Lu~Wang}} (Eds.). \bibinfo{publisher}{Association for Computational Linguistics}, \bibinfo{address}{Albuquerque, New Mexico}, \bibinfo{pages}{11926--11942}.
\newblock
\showISBNx{979-8-89176-189-6}
\href{https://doi.org/10.18653/v1/2025.naacl-long.597}{doi:\nolinkurl{10.18653/v1/2025.naacl-long.597}}


\bibitem[Wettig et~al\mbox{.}(2025)]%
        {wettig-et-al-2025_organizewebconstructingdomains}
\bibfield{author}{\bibinfo{person}{Alexander Wettig}, \bibinfo{person}{Kyle Lo}, \bibinfo{person}{Sewon Min}, \bibinfo{person}{Hannaneh Hajishirzi}, \bibinfo{person}{Danqi Chen}, {and} \bibinfo{person}{Luca Soldaini}.} \bibinfo{year}{2025}\natexlab{}.
\newblock \bibinfo{title}{Organize the Web: Constructing Domains Enhances Pre-Training Data Curation}.
\newblock
\showeprint[arxiv]{2502.10341}~[cs.CL]
\urldef\tempurl%
\url{https://arxiv.org/abs/2502.10341}
\showURL{%
\tempurl}


\bibitem[Widder et~al\mbox{.}(2024)]%
        {WidderWhittakerEtAl_2024_Why_open_AI_systems_are_actually_closed_and_why_this_matters}
\bibfield{author}{\bibinfo{person}{David~Gray Widder}, \bibinfo{person}{Meredith Whittaker}, {and} \bibinfo{person}{Sarah~Myers West}.} \bibinfo{year}{2024}\natexlab{}.
\newblock \showarticletitle{Why `Open' {{AI}} Systems Are Actually Closed, and Why This Matters}.
\newblock \bibinfo{journal}{\emph{Nature}} \bibinfo{volume}{635}, \bibinfo{number}{8040} (\bibinfo{date}{nov} \bibinfo{year}{2024}), \bibinfo{pages}{827--833}.
\newblock
\showISSN{1476-4687}
\href{https://doi.org/10.1038/s41586-024-08141-1}{doi:\nolinkurl{10.1038/s41586-024-08141-1}}


\bibitem[Wiggers(2024)]%
        {Wiggers_2024_OpenAI_inks_deal_to_train_AI_on_Reddit_data}
\bibfield{author}{\bibinfo{person}{Kyle Wiggers}.} \bibinfo{year}{2024}\natexlab{}.
\newblock \bibinfo{title}{{{OpenAI}} Inks Deal to Train {{AI}} on {{Reddit}} Data}.
\newblock
\urldef\tempurl%
\url{https://techcrunch.com/2024/05/16/openai-inks-deal-to-train-ai-on-reddit-data/}
\showURL{%
\tempurl}


\bibitem[Wiggers(2025)]%
        {Wiggers_2025_Mark_Zuckerberg_gave_Metas_Llama_team_OK_to_train_on_copyrighted_works_filing_claims}
\bibfield{author}{\bibinfo{person}{Kyle Wiggers}.} \bibinfo{year}{2025}\natexlab{}.
\newblock \bibinfo{title}{Mark {{Zuckerberg}} Gave {{Meta}}'s {{Llama}} Team the {{OK}} to Train on Copyrighted Works, Filing Claims}.
\newblock
\urldef\tempurl%
\url{https://techcrunch.com/2025/01/09/mark-zuckerberg-gave-metas-llama-team-the-ok-to-train-on-copyrighted-works-filing-claims/}
\showURL{%
\tempurl}


\bibitem[Wilkins(2026)]%
        {Wilkins_2026_Furious_AI_Users_Say_Their_Prompts_Are_Being_Plagiarized}
\bibfield{author}{\bibinfo{person}{Joe Wilkins}.} \bibinfo{year}{2026}\natexlab{}.
\newblock \bibinfo{title}{Furious {{AI Users Say Their Prompts Are Being Plagiarized}}}.
\newblock
\urldef\tempurl%
\url{https://futurism.com/artificial-intelligence/ai-prompt-plagiarism-art}
\showURL{%
\tempurl}


\bibitem[Worledge et~al\mbox{.}(2024)]%
        {worledge2024unifying}
\bibfield{author}{\bibinfo{person}{Theodora Worledge}, \bibinfo{person}{Judy~Hanwen Shen}, \bibinfo{person}{Nicole Meister}, \bibinfo{person}{Caleb Winston}, {and} \bibinfo{person}{Carlos Guestrin}.} \bibinfo{year}{2024}\natexlab{}.
\newblock \showarticletitle{Unifying corroborative and contributive attributions in large language models}. In \bibinfo{booktitle}{\emph{2024 IEEE Conference on Secure and Trustworthy Machine Learning (SaTML)}}. IEEE, \bibinfo{publisher}{IEEE Computer Society}, \bibinfo{address}{Los Alamitos, CA, USA}, \bibinfo{pages}{665--683}.
\newblock


\bibitem[Wu et~al\mbox{.}(2026)]%
        {WuLiuEtAl_2026_Ads_in_AI_Chatbots_Analysis_of_How_Large_Language_Models_Navigate_Conflicts_of_Interest}
\bibfield{author}{\bibinfo{person}{Addison~J. Wu}, \bibinfo{person}{Ryan Liu}, \bibinfo{person}{Shuyue~Stella Li}, \bibinfo{person}{Yulia Tsvetkov}, {and} \bibinfo{person}{Thomas~L. Griffiths}.} \bibinfo{year}{2026}\natexlab{}.
\newblock \bibinfo{booktitle}{\emph{Ads in {{AI Chatbots}}? {{An Analysis}} of {{How Large Language Models Navigate Conflicts}} of {{Interest}}}}.
\newblock
\showeprint[arXiv]{2604.08525}~[cs]
\href{https://doi.org/10.48550/arXiv.2604.08525}{doi:\nolinkurl{10.48550/arXiv.2604.08525}}


\bibitem[Wu(2016)]%
        {Wu_2016_attention_merchants_from_daily_newspaper_to_social_media_how_our_time_and_attention_is_harvested_and_sold}
\bibfield{author}{\bibinfo{person}{Tim Wu}.} \bibinfo{year}{2016}\natexlab{}.
\newblock \bibinfo{booktitle}{\emph{The Attention Merchants: From the Daily Newspaper to Social Media, How Our Time and Attention Is Harvested and Sold}}.
\newblock \bibinfo{publisher}{Atlantic Books}, \bibinfo{address}{London}.
\newblock
\showISBNx{978-1-78239-483-9 978-1-78239-482-2}


\bibitem[Yoon et~al\mbox{.}(2019)]%
        {yoon2019datavaluationusingreinforcement}
\bibfield{author}{\bibinfo{person}{Jinsung Yoon}, \bibinfo{person}{Sercan~O. Arik}, {and} \bibinfo{person}{Tomas Pfister}.} \bibinfo{year}{2019}\natexlab{}.
\newblock \bibinfo{title}{Data Valuation using Reinforcement Learning}.
\newblock
\showeprint[arxiv]{1909.11671}~[cs.LG]
\urldef\tempurl%
\url{https://arxiv.org/abs/1909.11671}
\showURL{%
\tempurl}


\bibitem[Zandbergen(2023)]%
        {Zandbergen_2023_Canadian_media_trained_audiences_to_use_Facebook_With_Meta_blocking_news_whats_next}
\bibfield{author}{\bibinfo{person}{Rebecca Zandbergen}.} \bibinfo{year}{2023}\natexlab{}.
\newblock \showarticletitle{Canadian Media Trained Audiences to Use {{Facebook}}. {{With Meta}} Blocking News, What's Next?}
\newblock \bibinfo{journal}{\emph{CBC Radio}} (\bibinfo{date}{aug} \bibinfo{year}{2023}).
\newblock
\urldef\tempurl%
\url{https://www.cbc.ca/radio/sunday/canadian-media-news-meta-facebook-1.6939274}
\showURL{%
\tempurl}


\bibitem[Zhang(2026)]%
        {Like_Company_Google_CJEU_Holds_First-Ever_Hearing_on_Generative_AI_and_Copyright_on_10_March_2026_Bird_Bird}
\bibfield{author}{\bibinfo{person}{Cen Zhang}.} \bibinfo{year}{2026}\natexlab{}.
\newblock \bibinfo{booktitle}{\emph{Like {{Company}} v {{Google CJEU Holds First-Ever Hearing}} on {{Generative AI}} and {{Copyright}} on 10 {{March}} 2026 - {{Bird}} \& {{Bird}}}}.
\newblock Bird \& Bird.
\newblock
\urldef\tempurl%
\url{https://www.twobirds.com/en/insights/2026/like-company-v-google-cjeu-holds-first-ever-hearing-on-generative-ai-and-copyright-on-10-march-2026}
\showURL{%
\tempurl}


\bibitem[Zhang et~al\mbox{.}(2025)]%
        {ZhangJiaoEtAl_2025_Fairshare_Data_Pricing_via_Data_Valuation_for_Large_Language_Models}
\bibfield{author}{\bibinfo{person}{Luyang Zhang}, \bibinfo{person}{Cathy Jiao}, \bibinfo{person}{Beibei Li}, {and} \bibinfo{person}{Chenyan Xiong}.} \bibinfo{year}{2025}\natexlab{}.
\newblock \bibinfo{title}{Fairshare {{Data Pricing}} via {{Data Valuation}} for {{Large Language Models}}}.
\newblock
\showeprint[arxiv]{2502.00198}~[cs]
\href{https://doi.org/10.48550/arXiv.2502.00198}{doi:\nolinkurl{10.48550/arXiv.2502.00198}}


\bibitem[Zuboff(2019)]%
        {Zuboff_2019_age_of_surveillance_capitalism_fight_for_human_future_at_new_frontier_of_power}
\bibfield{author}{\bibinfo{person}{Shoshana Zuboff}.} \bibinfo{year}{2019}\natexlab{}.
\newblock \bibinfo{booktitle}{\emph{The Age of Surveillance Capitalism: The Fight for a Human Future at the New Frontier of Power} (\bibinfo{edition}{first edition} ed.)}.
\newblock \bibinfo{publisher}{PublicAffairs}, \bibinfo{address}{New York}.
\newblock
\showISBNx{978-1-61039-569-4}
\showLCCN{HF5415.32 .Z83 2019}


\end{thebibliography}
